\documentclass [12pt]{article}
\usepackage{amsmath,amssymb,cite}
\usepackage{appendix}
\usepackage{feynmp}
\usepackage[vcentermath,enableskew]{youngtab}
\usepackage{tabularx}
\usepackage[english]{babel}
\usepackage[dvips]{graphicx}
\usepackage{indentfirst}
\usepackage{epsfig}
\usepackage{physics}
\usepackage{pdfpages}
\usepackage{cases}
\usepackage{appendix}
\usepackage{tabularx}
\usepackage[english]{babel}
\usepackage[dvips]{graphicx}
\usepackage{indentfirst}
\usepackage{epsfig}
\usepackage{slashed}
\usepackage{fancyhdr}
\usepackage{cases}
\usepackage{tcolorbox}
\usepackage{amsmath,amssymb,cite}
\usepackage{amsfonts, geometry, hyperref}
\usepackage{booktabs,longtable}
\usepackage{appendix}
\usepackage{feynmp}
\usepackage{float}
\usepackage{subfigure}
\usepackage[vcentermath,enableskew]{youngtab}
\usepackage{tabularx}
\usepackage[english]{babel}
\usepackage{graphicx}
\usepackage{indentfirst}
\usepackage{epsfig}
\usepackage{color}
\usepackage{epstopdf}
\usepackage[]{hyperref}
\usepackage[section]{placeins}
\usepackage[stable]{footmisc}
\usepackage{appendix}
\usepackage{adjustbox}
\usepackage{tabularx}
\usepackage[english]{babel}
\usepackage{graphicx}
\usepackage{indentfirst}
\usepackage{epsfig}
\usepackage{slashed}
\usepackage{fancyhdr}
\usepackage{url}
\usepackage{pgfplots}
\usepackage{amsmath}

\pgfplotsset{compat=1.18}
\numberwithin{equation}{section}

\makeatletter
\patchcmd{\beamer@sectionintoc}
  {\vfill}
  {\vskip\itemsep}
  {}
  {}

\long\def\beamer@section[#1]#2{%
  \beamer@savemode%
  \mode<all>%
  \ifbeamer@inlecture
    \refstepcounter{section}%
    \beamer@ifempty{#2}%
    {\long\def\secname{#1}\long\def\lastsection{#1}}%
    {\global\advance\beamer@tocsectionnumber by 1\relax%
      \long\def\secname{#2}%
      \long\def\lastsection{#1}%
      \addtocontents{toc}{\protect\beamer@sectionintoc{\the\c@section}{#2\hfill\the\c@page}{\the\c@page}{\the\c@part}%
        {\the\beamer@tocsectionnumber}}}%
    {\let\\=\relax\xdef\sectionlink{{Navigation\the\c@page}{\noexpand\secname}}}%
    \beamer@tempcount=\c@page\advance\beamer@tempcount by -1%
    \beamer@ifempty{#1}{}{%
      \addtocontents{nav}{\protect\headcommand{\protect\sectionentry{\the\c@section}{#1}{\the\c@page}{\secname}{\the\c@part}}}%
      \addtocontents{nav}{\protect\headcommand{\protect\beamer@sectionpages{\the\beamer@sectionstartpage}{\the\beamer@tempcount}}}%
      \addtocontents{nav}{\protect\headcommand{\protect\beamer@subsectionpages{\the\beamer@subsectionstartpage}{\the\beamer@tempcount}}}%
    }%
    \beamer@sectionstartpage=\c@page%
    \beamer@subsectionstartpage=\c@page%
    \def\insertsection{\expandafter\hyperlink\sectionlink}%
    \def\insertsubsection{}%
    \def\insertsubsubsection{}%
    \def\insertsectionhead{\hyperlink{Navigation\the\c@page}{#1}}%
    \def\insertsubsectionhead{}%
    \def\insertsubsubsectionhead{}%
    \def\lastsubsection{}%
    \Hy@writebookmark{\the\c@section}{\secname}{Outline\the\c@part.\the\c@section}{2}{toc}%
    \hyper@anchorstart{Outline\the\c@part.\the\c@section}\hyper@anchorend%
    \beamer@ifempty{#2}{\beamer@atbeginsections}{\beamer@atbeginsection}%
  \fi%
  \beamer@resumemode}%

\def\beamer@subsection[#1]#2{%
  \beamer@savemode%
  \mode<all>%
  \ifbeamer@inlecture%
    \refstepcounter{subsection}%
    \beamer@ifempty{#2}{\long\def\subsecname{#1}\long\def\lastsubsection{#1}}
    {%
      \long\def\subsecname{#2}%
      \long\def\lastsubsection{#1}%
      \addtocontents{toc}{\protect\beamer@subsectionintoc{\the\c@section}{\the\c@subsection}{#2\hfill\the\c@page}{\the\c@page}{\the\c@part}{\the\beamer@tocsectionnumber}}%
    }%
    \beamer@tempcount=\c@page\advance\beamer@tempcount by -1%
    \addtocontents{nav}{%
      \protect\headcommand{\protect\beamer@subsectionentry{\the\c@part}{\the\c@section}{\the\c@subsection}{\the\c@page}{\lastsubsection}}%
      \protect\headcommand{\protect\beamer@subsectionpages{\the\beamer@subsectionstartpage}{\the\beamer@tempcount}}%
    }%
    \beamer@subsectionstartpage=\c@page%
    \edef\subsectionlink{{Navigation\the\c@page}{\noexpand\subsecname}}%
    \def\insertsubsection{\expandafter\hyperlink\subsectionlink}%
    \def\insertsubsubsection{}%
    \def\insertsubsectionhead{\hyperlink{Navigation\the\c@page}{#1}}%
    \def\insertsubsubsectionhead{}%
    \Hy@writebookmark{\the\c@subsection}{#2}{Outline\the\c@part.\the\c@section.\the\c@subsection.\the\c@page}{3}{toc}%
    \hyper@anchorstart{Outline\the\c@part.\the\c@section.\the\c@subsection.\the\c@page}\hyper@anchorend%
    \beamer@ifempty{#2}{\beamer@atbeginsubsections}{\beamer@atbeginsubsection}%
  \fi%
  \beamer@resumemode}

\makeatother

\begin{document}

\title{A Double--Scaling Large--\(d\) Saddle of BFSS/BMN Matrix Quantum Mechanics\\
\large\emph{BFSS/BMN Matrix Quantum Mechanics I}
}

\author{
Badis Ydri\\[2mm]
Department of Physics, Badji Mokhtar Annaba University, Algeria\\
}

\maketitle

\begin{abstract}

We study the large--\(d\) dynamics of the mass--deformed bosonic
\(\mathrm{BFSS}_{d+1}\) matrix quantum mechanics using a Hubbard--Stratonovich
localization of the Yang--Mills interaction. After integrating out the matrix
coordinates, the theory reduces to a holonomy--dependent effective action for an
auxiliary adjoint kernel. We introduce a commuting--symmetric saddle and its
maximally symmetric specialization, in which the interaction is encoded in a
single dynamically generated mass shift \(k_0\). The resulting large--\(d\)
description is a gauged matrix harmonic oscillator with self--consistent
frequency \(s^2=m+k_0\), fixed by a gap equation.

We analyze the low--temperature \(X\)-space physics, the holonomy effective
action, the Yang--Mills observable, and the associated phase structure. We then
identify a correlated double--scaling limit in which \(d\to\infty\),
\(m\to\infty\), and \(\kappa=m^{3/2}/d\) is held fixed. In this limit the
Yang--Mills interaction and the explicit mass deformation remain parametrically
balanced: the theory interpolates between the commutator--dominated BFSS regime
and the mass--dominated Gaussian regime.

The double--scaled theory exhibits two complementary large--\(d\) regimes. At
low temperature, the enhanced gap pushes the deconfinement scale upward and opens
a parametrically large uniform--holonomy region, where the bulk dynamics behaves
as weakly coupled \(\mathrm{BFSS}_2\)--type gauged harmonic--oscillator sectors.
At the same time, the high--temperature branch reveals an overlap window in
which the Gaussian description remains self--consistent while the commutator
contribution per matrix pair is parametrically suppressed. The resulting
dynamics is therefore \(\mathrm{BFSS}_2\)--like in its enlarged
uniform--holonomy sector and IKKT--like in its almost--commuting matrix behavior.

\end{abstract}


\tableofcontents

\section{Introduction and Goal}
\subsection{Generalities:BFSS/BMN systems}

\medskip
\noindent
The BFSS$_{d+1}$ models form a central class of matrix quantum mechanics:
one--dimensional \(U(N)\) gauge theories with \(d\) adjoint matrix coordinates
\(X_a\), obtained by dimensional reduction of ten--dimensional
\(\mathcal N=1\) super Yang--Mills theory~\cite{Brink:1976bc}. Their Euclidean action may be written schematically as
\begin{eqnarray}
S_{\rm BFSS}^{\rm E}
=
\frac{1}{g^2}
\int_{0}^{\beta}\!dt\,
\Tr\left[
\frac12(D_tX_a)^2
-\frac14[X_a,X_b]^2
\right]
+
S_{\rm BFSS,F}^{\rm E},
\qquad
D_t=\partial_t-i[A_t,\,\cdot\,],
\label{BFSSIntro}
\end{eqnarray}
where \(S_{\rm BFSS,F}^{\rm E}\) contains the fermionic kinetic and Yukawa
terms required by supersymmetry.

\medskip
\noindent
Supersymmetry restricts the allowed matrix quantum mechanics to
\(D_{\rm YM}=d+1=10,6,4,3,2\), by the Fierz identity analysis of
Baake, Reinicke and Rittenberg~\cite{Baake:1984ie}; the corresponding
supermembrane or M--theory dimensions are \(D_{\rm M}=d+2=11,7,5,4,3\).

\medskip
\noindent
The holographic regime is obtained in the planar 't~Hooft limit~\cite{tHooft1974},
\begin{eqnarray}
N\to\infty,
\qquad
g^2\to0,
\qquad
\lambda=g^2N=\hbox{fixed}.
\end{eqnarray}
This is the natural large--\(N\) setting for holography~\cite{tHooft1993,Susskind1995},
and at strong coupling it is expected to admit a dual weakly curved classical
supergravity description.

\medskip
\noindent
The central example is the \(d=9\) BFSS$_{10}$ model, or M-(atrix) theory,
which gives a large-\(N\) nonperturbative formulation of M-theory in the
infinite--momentum frame~\cite{BanksFischlerShenkerSusskind1997}. It describes the low--energy worldvolume dynamics of \(N\) coincident
D\(0\)-branes~\cite{Witten1996,Polchinski1995}. At strong coupling and large
\(N\), its thermodynamics is dual to the near--horizon black 0--brane geometry of
type IIA supergravity~\cite{Itzhaki1998}, which is the low--energy limit of type
IIA superstring theory. Equivalently, type IIA supergravity
is obtained from eleven--dimensional supergravity~\cite{Cremmer1978} by
compactification on \(S^1\)~\cite{Witten1995}, and reducing the eleven-dimensional
M-wave along the compact direction gives the ten-dimensional black 0-brane
solution~\cite{Hyakutake:2014maa,Hyakutake:2006aq}. This places BFSS$_{10}$ at
the intersection of D\(0\)-brane dynamics, black 0-brane supergravity,
light-cone supermembranes~\cite{Hoppe1982,Hoppe1988,deWitHoppeNicolai1988}, and
superparticles in maximally supersymmetric pp-wave backgrounds
\cite{Kowalski-Glikman:1984qtj,Blau:2001ne}.

\medskip
\noindent
In this interpretation, the matrices \(X_a\) are noncommutative transverse
coordinates of the D\(0\)-branes: their diagonal components encode brane
positions, while their off-diagonal components describe open strings stretched
between different branes, becoming light when the branes coincide. See, for
example,~\cite{Azeyanagi2009} and the pedagogical accounts
\cite{Zwiebach2009,Becker2006}.

\medskip
\noindent
In the decoupling limit, the BFSS$_{10}$ matrix quantum mechanics is related by
gauge/gravity duality to type IIA string theory on the near--horizon black
0--brane background~\cite{Maldacena1999,Gubser1998,Witten1998}. In the
large--\(N\), strong--coupling regime, this becomes a weakly curved type IIA
supergravity description, while the gauge theory side remains a nonperturbative
matrix quantum mechanics which admits a nonperturbative lattice regularization~\cite{Wilson:1974sk}.
Thus the black 0--brane correspondence provides a concrete nonperturbative
setting for quantum gravity and black--hole thermodynamics.

\medskip
\noindent
This duality has been tested extensively by Monte Carlo simulations
\cite{Catterall2008,Anagnostopoulos2008,Hanada2014,Hanada2016b,Filev:2015hiaF}
and analytic methods~\cite{Kabat2001,Hanada2009,Hyakutake2014}. In particular,
BFSS$_{10}$/black 0--brane thermodynamics relates quantum--gravity corrections
on the gravity side to finite--\(N\) and finite--coupling effects in the matrix
model. For a pedagogical review, see~\cite{Hanada2016}.

\medskip
\noindent
More generally, one may consider the lower--dimensional BFSS$_{d+1}$ matrix
quantum mechanical models associated with the other allowed supersymmetric
Yang--Mills dimensions, besides the maximally supersymmetric BFSS$_{10}$ case.
These models describe lower--dimensional DLCQ dynamics of D\(0\)-branes and the
associated gauge/gravity dualities. They also provide the corresponding matrix regularizations
of light--cone supermembranes and are naturally related to the light--cone
quantization of superparticles in maximally supersymmetric pp--wave backgrounds.

\medskip
\noindent
Although simpler than BFSS$_{10}$, these lower--dimensional models retain many
of its essential dynamical features: confinement--deconfinement behavior,
holonomy eigenvalue condensation, gauge/gravity--inspired matrix dynamics, and
emergent geometric phases. They therefore provide analytically and numerically
more tractable laboratories for studying the mechanisms underlying BFSS
holography and matrix quantum geometry.

\medskip
\noindent
A second natural generalization is obtained by adding maximally supersymmetric
mass deformations to the BFSS$_{d+1}$ models. The prototype is the BMN
plane--wave matrix model~\cite{BerensteinMaldacenaNastase2002}, with the general
classification given in~\cite{Kim:2006,Park:2005}. These deformations add quadratic matrix masses, Myers--type cubic
interactions~\cite{Myers}, and the corresponding fermionic mass terms
required by supersymmetry.  Schematically,
\begin{eqnarray}
S_{\rm BMN}^{\rm E}
=
S_{\rm BFSS}^{\rm E}
+
\frac{1}{g^2}
\int_{0}^{\beta}\!dt\,
\Tr\left[
\mu_1 X_a^2
+
\mu_2\epsilon_{ijk}X_iX_jX_k
\right]
+
S_{\rm BMN,F}^{\rm def},
\label{BMNIntro}
\end{eqnarray}
where \(S_{\rm BMN,F}^{\rm def}\) denotes the fermionic part of the mass
deformation.

\medskip
\noindent
The allowed cases are \(d=2,3,5,9\), corresponding to
BMN$_{3,4,6,10}\), while the special \(d=1\) case, BMN$_2\), is discussed
in~\cite{Park:2005,Ydri2025}. These deformations generally break the transverse
\(SO(d)\) symmetry into smaller rotational factors: the quadratic masses may
split the matrix directions, while the Myers term selects an \(SO(3)\) sector.

\medskip
\noindent
The mass term may be viewed as a cosmological, or curvature, scale: it lifts the
flat directions of BFSS and replaces the flat target-space interpretation by a
curved pp-wave background. On the gravity side, this is the distinction between
the undeformed black 0-brane/type IIA setting and the deformed pp-wave setting.
In the maximally supersymmetric case, the half-BPS sectors are related to LLM
bubbling geometries~\cite{Lin:2004nb}. Some Monte Carlo studies of the BMN model can
be found in~\cite{Asano:2018nol,Asano:2020yry} and references therein. For related recent developments, see also the preprints
\cite{OConnor:2026zlf,Lei:2026fep}.

In summary, each BFSS\(_{d+1}\) model admits a corresponding BMN deformation preserving maximal supersymmetry. These deformations describe supermembranes and superparticles in maximally supersymmetric pp-wave backgrounds. The corresponding classification is summarized in Table~\ref{so3}.

\begin{table}[h]
\centering
\begin{tabular}{@{}lllll@{}}
\toprule
\textbf{Model} & \(D_{\rm YM}\) & Splitting of \(SO(D_{\rm YM}-1)\) & Superalgebra & Deformation parameter \\
\midrule
\(\mathcal N=16\) & 10 & \(SO(6)\times SO(3)\) & \(\mathfrak{su}(2|4)\) & \(\mu\) \\
\(\mathcal N=8\) type I & 6 & \(SO(3)\times SO(2)\) & \(\mathfrak{su}(2|2)\) & \(\mu\) \\
\(\mathcal N=8\) type II & 6 & \(SO(4)\) & \(\mathfrak{su}(2|1)\oplus\mathfrak{su}(2|1)\) & \(\mu\) \\
\(\mathcal N=4\) type I & 4 & \(SO(3)\) & \(\mathfrak{su}(2|1)\) & \(\mu_1,\mu_2\) \\
\(\mathcal N=4\) type II & 4 & \(SO(2)\) & \(\mathrm{Clifford}_{4}(\mathbb R)\) & \(\mu\) \\
\(\mathcal N=2\) & 3 & \(SO(2)\) & \(\mathrm{Clifford}_{2}(\mathbb R)\) & \(\mu\) \\
\(\mathcal N=1+1\) & 2 & \(SO(1,2)\) & \(\mathfrak{osp}(1|2,\mathbb R)\) & \(\Lambda(t),\rho(t)\) \\
\bottomrule
\end{tabular}
\caption{Classification of massive supersymmetric Yang--Mills quantum mechanics models and their deformation parameters.}
\label{so3}
\end{table}

 \subsection{Goal and overview}

 \medskip
\noindent
 We study here the mass--deformed BFSS$_{d+1}$ matrix model using a combined \emph{large-$d$} and \emph{saddle--point} analysis, with the aim of isolating the dominant low--energy degrees of freedom and clarifying their geometric and gravitational interpretation. Our approach is motivated by earlier large-$d$ studies of matrix quantum mechanics, which show that the dynamics simplifies dramatically and becomes governed by a small set of collective variables, most notably the holonomy around the Euclidean time circle \cite{Mandal:2009vzN,Mandal:2011hbN,Kabat:2000zv,Kabat:2001ve}.

\medskip
\noindent
The first step of our analysis is to integrate out the matrix degrees of freedom $X_a$ at large $d$ using a Hubbard--Stratonovich localization and a saddle--point approximation. This leads to an effective description in terms of a dynamically generated mass scale determined self--consistently by a gap equation. At leading order in $1/d$, we show that the interacting BFSS model reduces to a \emph{gauged matrix harmonic oscillator}. This observation builds on earlier results demonstrating that the matrix harmonic oscillator already captures the essential thermodynamic and spectral features of the BFSS model in the low--temperature regime \cite{Furuuchi:2003syNe,Filev:2015hiaF}.

\medskip
\noindent
After integrating out the $X_a$ matrices, the remaining dynamics is encoded in an effective action for the holonomy eigenvalues. This holonomy effective action consists of a universal Vandermonde repulsion term together with an attractive contribution generated by the massive adjoint fields. Its structure is closely related to unitary matrix models of the Gross--Witten--Wadia type \cite{Gross:1980heF,Wadia:1980cpF}, and can be derived using standard methods for gauge theories on a circle, following Polyakov’s treatment of eigenvalue measures and holonomies \cite{Polyakov1987}.

\medskip
\noindent
Within this effective description we compute physical observables such as the free energy and the \emph{extent of space}, defined by $\langle \mathrm{Tr}\,X_a^2\rangle$. In the Gaussian saddle, these observables are controlled by the dynamically generated mass and can be evaluated explicitly. We show that the saddle--point expectation values reproduce the classical constraints implied by the Hubbard--Stratonovich equations of motion, providing a nontrivial consistency check of the large-$d$ approximation.

\medskip
\noindent
The resulting phase structure is governed by the competition between eigenvalue repulsion and attraction. At low temperatures the holonomy eigenvalues are uniformly distributed, corresponding to a confining phase in the gauge--theory sense. As the temperature increases, the uniform distribution becomes unstable and gives way to a non--uniform or gapped distribution, signaling a deconfinement transition. This transition is closely related to the Hagedorn phenomenon in string theory \cite{Aharony:2003sxF,Aharony:2004igF,Harmark:2004ws,AlvarezGaume:2005fvF} and to the Gregory--Laflamme instability of black strings
\cite{Gregory:1993vy,Mandal:2011ws}.

\medskip
\noindent
A central result here is that a positive mass deformation pushes the deconfinement transition to parametrically higher temperatures. Combined with the large-$d$ limit, this allows one to keep the system deep in the uniform, confining phase even for moderately large $d$. In this regime, BFSS$_{d+1}$ behaves effectively as a collection of BFSS$_2$--like sectors governed by matrix quantum mechanics with an AdS$_2$ dual description. This provides a controlled setting in which the Yang--Mills phase of the IKKT model can be identified with a confining, string--like phase encoding emergent geometry in a nontrivial way.

\medskip
\noindent
From the gravitational perspective, the uniform holonomy phase corresponds to a
black--string--like configuration, while the gapped phase corresponds to a
localized black hole in the dual description~\cite{Susskind:1997drNe}. Thus our
analysis shows how mass deformations and large--\(d\) dynamics can suppress the
black--hole phase and stabilize the stringy, confining phase over a wide range of
parameters. In this way, BFSS$_2$--like physics dominates the low--energy sector
of higher--dimensional BFSS models, and geometry emerges through AdS$_2$ quantum
mechanics rather than higher--dimensional black--hole dynamics.

\medskip
\noindent
At the same time, the double--scaled saddle has a complementary IKKT--like
aspect. The low--temperature analysis already shows that the matrices are
localized near the origin and that the commutator contribution per fixed matrix
pair is parametrically suppressed. This interpretation becomes sharper after
combining the low--temperature and high--temperature saddle descriptions: there
is an overlap window in which both Gaussian descriptions are self--consistent,
while the matrices remain localized and approximately commuting. Thus the system
is not driven by strong noncommutativity, even though the original Yang--Mills
interaction is retained self--consistently through the dynamically generated
mass. Together with the enlarged BFSS$_2$--like uniform-holonomy regime, this
shows that the double--scaling limit interpolates between BFSS$_2$--like
AdS$_2$ quantum mechanics and IKKT--like commuting-matrix geometry.

\medskip
\noindent
The present work is part of a broader program on one--dimensional
BFSS/BMN matrix gauge theories with an arbitrary number \(d\) of
noncommuting matrix coordinates.  The program combines large--\(d\)
Gaussian reduction, Molien--Weyl singlet counting, endpoint and
Wishart--Stiefel formulations, and Monte Carlo methods, with the broader
aim of clarifying matrix quantum mechanics and its possible
interpretation as matrix quantum gravity.  Related aspects are developed
in
\cite{Ydri:2026smo,YdriUnpublished0,Ydri:2026rke,
Ydri:2026zlj,YdriUnpublished2,YdriUnpublished3}.

\medskip
\noindent
For completeness, the first arXiv version of the present work, which
contains additional material not retained in this manuscript, is available in Ref.~\cite{Ydri:2026gsy}.

\subsection{Organization of the paper}

\medskip
\noindent
The paper is organized as follows.  Section~\ref{section2} develops the
localized large--\(d\) formulation of the mass--deformed bosonic
BFSS$_{d+1}$ model, including the holonomy variables, the adjoint
projector, and the Hubbard--Stratonovich kernel.  Section~\ref{section3}
derives the saddle--point equation and introduces the
commuting--symmetric and maximally symmetric ansätze, together with a
discussion of the scope of the latter.  Section~\ref{section4} studies
the low--temperature \(X\)-space physics of the saddle, including the
self--consistent matrix harmonic oscillator, the auxiliary-field
fluctuations, the free energy, and the extent of space.

\medskip
\noindent
Section~\ref{section5} derives the low--temperature holonomy effective
action, analyzes the associated phase structure, and discusses its
black--string interpretation.  Section~\ref{section6} evaluates the
Yang--Mills observable at the large--\(d\) saddle and studies its
small-- and large--mass limits.  Section~\ref{section7} introduces the
double--scaling limit in which the Yang--Mills interaction and the
explicit mass deformation remain parametrically balanced.  It then
analyzes the nearly commuting regime, the high--temperature Gaussian
branch, and the size of the auxiliary-field fluctuations in that
regime.

\medskip
\noindent
Section~\ref{section8} discusses split--mass saddles, Gaussian
Molien--Weyl approximations, and the supersymmetric completion of the
large--\(d\) bosonic Gaussian theory, together with its purely
holonomic and bosonized lattice formulations.  The main results and
their physical implications are summarized in
Section~\ref{conclusion}.  Finally,
Appendix~\ref{appendix-holonomy-effective-action} gives the Fourier
derivation of the holonomy effective action.

\section{Hubbard--Stratonovich localization of the BFSS/BMN action}\label{section2}
\medskip
\noindent
For earlier large--\(d\) treatments of bosonic matrix quantum mechanics,
including the auxiliary--field reduction, the dynamically generated mass gap,
and the associated thermal phase structure, see
Refs.~\cite{Mandal:2009vzN,Mandal:2011hbN,Filev:2015hiaF,Kawahara:2007fnF}
and references therein.

\subsection{Action and observables}

\medskip
\noindent We consider the Euclidean mass--deformed bosonic BFSS$_{d+1}$ model
\begin{eqnarray}
S
&=&
N\int_{0}^{\beta}\! dt\ \mathrm{Tr}\bigg[\frac{1}{2}(D_tX_a)^2+\frac{m}{2}X_a^2-\frac{\alpha}{4}[X_a,X_b]^2\bigg],
\qquad (\alpha=1\ \text{at the end}),
\end{eqnarray}
with partition function and free energy
\begin{eqnarray}
Z(\alpha)\;=\;\int\!\mathcal{D}X\,e^{-S(\alpha)},
\qquad
F(\alpha)\;=\;-\frac{1}{\beta}\log Z(\alpha).
\end{eqnarray}
A primary observable is the quartic Yang--Mills term in this \emph{original} BFSS theory,
\begin{eqnarray}
\Big\langle \frac{N}{4}\int_{0}^{\beta}\! dt\ \mathrm{Tr}[X_a,X_b]^2\Big\rangle
&=&
\left.\frac{\partial \ln Z(\alpha)}{\partial \alpha}\right|_{\alpha=1}
\;=\;
-\beta\,\left.\frac{\partial F(\alpha)}{\partial \alpha}\right|_{\alpha=1}.
\label{YM-from-alpha}
\end{eqnarray}
A second key observable is the extent of space,
\begin{eqnarray}
R^2
&\equiv&
\frac{1}{N\beta}\Big\langle \int_{0}^{\beta}\!dt\ \mathrm{Tr}\,X_a(t)\,X_a(t)\Big\rangle
\;.
\label{def-R2}
\end{eqnarray}
In the large--$d$ saddle, the mass--deformed BFSS$_{d+1}$ is approximated by a gauged matrix harmonic oscillator with
a dynamically generated effective mass $m_{\rm eff}$ determined self-consistently by the gap equation.
In this Gaussian saddle, $R^2$ can be obtained either from the holonomy--dependent propagator, or
equivalently from the MHO free energy through
\begin{eqnarray}
\frac{\partial F_{\rm MHO}}{\partial m_{\rm eff}}
&=&
\frac{N^2}{2}\,R^2,
\label{R2-from-F}
\end{eqnarray}
(up to the same normalization conventions used in the quadratic action).

\medskip
\noindent
The internal energy of the system can be expressed as a linear combination of the Yang--Mills observable and the extent of space~\cite{Kawahara:2007fnF}. The internal energy is defined by 
\begin{eqnarray}
E
=
-\frac{\partial}{\partial\beta}\ln Z,
\qquad
Z(\beta)=\int\!\mathcal{D}X\ e^{-S}.
\label{E-def}
\end{eqnarray}
To isolate the explicit $\beta$--dependence, rescale
\begin{eqnarray}
t=\beta\tau,\qquad X_a(t)=\sqrt{\beta}\,\widetilde X_a(\tau),
\qquad \tau\in[0,1].
\end{eqnarray}
Then the action becomes
\begin{eqnarray}
S
=
N\int_{0}^{1}\!d\tau\ \Tr\Big[
\frac{1}{2}(D_\tau\widetilde X_a)^2
+\frac{\beta^2 m}{2}\,\widetilde X_a^2
-\frac{\beta^3\alpha}{4}\,[\widetilde X_a,\widetilde X_b]^2
\Big],
\label{S-rescaled}
\end{eqnarray}
so that (up to $\beta$--independent measure factors)
\begin{eqnarray}
-\frac{\partial}{\partial\beta}\ln Z
=
\Big\langle \frac{\partial S}{\partial\beta}\Big\rangle
=
N\int_{0}^{1}\!d\tau\ \Tr\Big[
\beta m\,\widetilde X_a^2
-\frac{3\beta^2\alpha}{4}\,[\widetilde X_a,\widetilde X_b]^2
\Big].
\end{eqnarray}
Returning to the original variables gives the compact identity
\begin{eqnarray}
E
=
-\frac{3\alpha}{\beta}\Big\langle \frac{N}{4}\int_{0}^{\beta}\!dt\ \Tr[X_a,X_b]^2\Big\rangle
\;+\;
N^2\,m\,R^2.
\label{E-YM-R2}
\end{eqnarray}
\medskip
\noindent An alternative formula follows from the Ward identity. Indeed,  on a lattice with spacing $a$ and $\Lambda={\beta}/{a}$ sites, the path integral is invariant under
the infinitesimal rescaling $X_a\to (1+\varepsilon)X_a$.
Since the Jacobian contributes $\varepsilon\,d\,\Lambda\,(N^2-1)$, while the action varies as
$\delta S = 2\varepsilon\,S - 2\varepsilon\,\alpha\hat {\cal O}_{\rm YM}$ with
\begin{eqnarray}
\hat{\cal O}_{\rm YM}
\equiv
\Big\langle \frac{N}{4}\int_{0}^{\beta}\!dt\ \Tr[X_a,X_b]^2\Big\rangle,
\end{eqnarray}
one obtains
\begin{eqnarray}
2\langle S\rangle
-2\alpha\Big\langle \frac{N}{4}\int_{0}^{\beta}\!dt\ \Tr[X_a,X_b]^2\Big\rangle
=
d\big(\Lambda (N^2-1)\big).
\label{Ward-lattice}
\end{eqnarray}
(In a continuum treatment, the right-hand side is replaced by the corresponding regulated
counting of bosonic degrees of freedom.)

\medskip
\noindent In the following, we will compute the free energy $F$ and extent of space $R^2$ explicitly in the presence of holonomy, and then evaluate the
Yang--Mills expectation value via \eqref{YM-from-alpha} within the large--$d$ saddle approximation.
These observables determine the holonomy effective action and the phase structure, and quantify the
regime in which BFSS$_{d+1}$ is accurately captured by the effective gauged matrix harmonic oscillator.

\subsection{Action in terms of large--$d$ effective variables}

\medskip
\noindent The computation proceeds, as we will discuss in detail, in several steps. First, we rewrite the action in a form suitable for the large--$d$ analysis as follows:

\begin{itemize}
\item \textbf{Kinetic term}: Since we are on a Euclidean time circle of circumference $\beta$, all bosonic fields are periodic: $X_a(t+\beta)=X_a(t)$. This periodicity quantizes the temporal momentum, so the Fourier expansion is
\begin{eqnarray}
X_a^{ij}(t)&=&
\frac{1}{\sqrt{\beta}}\sum_{n\in\mathbb{Z}} X_a^{ij}(n)\,e^{i\omega_n t},
\qquad
\omega_n=\frac{2\pi n}{\beta}.
\end{eqnarray}
$X_a$ are Hermitian matrices, $X_a^{ji}(-n)=[X_a^{ij}(n)]^\ast$.

\noindent The gauge field enters only through the temporal component $A_0(t)$. In one-dimensional gauge theory, $A_0$ has no dynamics of its own—it is an auxiliary field enforcing Gauss' law. However, the path-ordered exponential
\begin{eqnarray}
U \;=\; \mathcal{P}\exp\!\Big(-i\int_{0}^{\beta}\!dt\,A_0(t)\Big)
\end{eqnarray}
around the circle (the Polyakov loop or holonomy) is gauge-invariant up to conjugation. Indeed, the holonomy matrix $U$ transforms by conjugation under a periodic gauge transformation $g(t)$ with $g(0)=g(\beta)$:
\begin{eqnarray}
U \;\longrightarrow\; g(0)\,U\,g^{-1}(0).
\end{eqnarray}
Thus the matrix $U$ itself is not gauge-invariant, but its physical information---namely its spectrum, or equivalently the set of holonomy angles $\{\theta_i\}$---remains unchanged.

However, by a gauge transformation that is periodic in Euclidean time, one can always bring $A_0(t)$ to a time-independent, diagonal form:
\begin{eqnarray}
A_0=-\frac{1}{\beta}\mathrm{diag}\big(\theta_1,\ldots,\theta_N\big),\quad U=\mathrm{diag}\big(e^{i\theta_1},\ldots,e^{i\theta_N}\big),
\quad
\sum_{i=1}^{N}\theta_i=0\ \ \ (\text{for }SU(N)).\nonumber\\
\end{eqnarray}
This is the static diagonal (Polyakov) gauge. The only residual information in $A_0$ is the set of eigenvalues $\{\theta_i\}$, i.e. the holonomy angles; all fluctuations of $A_0(t)$ away from this constant diagonal form can be gauged away.

\noindent 
With this gauge choice, the covariant derivative $D_t X_a=\partial_t X_a-i[A_0,X_a]$ acting on Fourier modes shifts the Matsubara frequencies by color-dependent phases:
\begin{eqnarray}
(D_t X_a)^{ij}(n)
&=&
\frac{i}{\sqrt{\beta}}\sum_{n,ij}\Big(\omega_n+\frac{\theta_i-\theta_j}{\beta}\Big)\,X_a^{ij}(n)\,e^{i\omega_n t}.
\end{eqnarray}
\noindent Thus, in the kinetic term one always encounters the combination 
$\omega_n+(\theta_i-\theta_j)/\beta$. The kinetic term then reads
\begin{eqnarray}
  \frac{N}{2}\int_{0}^{\beta}\! dt\ \mathrm{Tr}(D_tX_a)^2
  &=&
  \frac{N}{2}\sum_{a=1}^d\sum_{n\in\mathbb{Z}}\sum_{i,j=1}^N
  \Big(\frac{2\pi n+\theta_i-\theta_j}{\beta}\Big)^2
  X_a^{ij}(n)\,X_a^{ji}(-n).
\end{eqnarray}

\noindent Each adjoint mode $(i,j)$ thus experiences the shifted Matsubara frequency 
$\omega_n\mapsto \omega_n+(\theta_i-\theta_j)/\beta$. 
In particular, the trace component ($i=j$) corresponds to the overall $U(1)$ mode 
and can be removed by inserting the adjoint projector
\begin{eqnarray}
P_{ij,kl}=\delta_{il}\delta_{jk}-\frac{1}{N}\delta_{ij}\delta_{kl},
\end{eqnarray}
which projects onto traceless $SU(N)$ matrices. The projected kinetic term reads
\begin{eqnarray}
S_{\rm kin}
&=&
\frac{N}{2}\sum_{a=1}^{d}\sum_{n\in\mathbb{Z}}\sum_{i,j=1}^N
X_a^{ij}(-n)\,
\Bigg(\Big(\omega_n+\frac{\theta_i-\theta_j}{\beta}\Big)^2\,
P_{ij,kl}\Bigg)\,
X_a^{kl}(n).
\end{eqnarray}

\item \textbf{Mass term}:
\noindent We expand the matrix coordinates as
\begin{eqnarray}
&& X_a=\sum_{\mu=1}^{N^2-1} x_a^{\mu}\,\Lambda_{\mu},
\nonumber\\
&& [\Lambda_{\mu},\Lambda_{\nu}]=2i\,f_{\mu\nu\lambda}\,\Lambda_{\lambda},
\qquad
\mathrm{Tr}\big(\Lambda_{\mu}\Lambda_{\nu}\big)=2\,\delta_{\mu\nu}.
\end{eqnarray}
\noindent The quadratic term then takes the form
\begin{eqnarray}
  \frac{Nm}{2}\int_{0}^{\beta}\! dt\ \mathrm{Tr}\,X_a^2
  &=& Nm\int_{0}^{\beta}\! dt\ (x_a^{\mu}x_a^{\mu}),
  \qquad
  x_a^{\mu}=\frac{1}{2}\,\mathrm{Tr}\,(X_a\Lambda_{\mu})
  \nonumber\\
  &=& \frac{Nm}{2}\int_{0}^{\beta}\! dt\ X_a^{ij} X_a^{kl}\,P_{ij,kl},
  \qquad
  P_{ij,kl}=\frac{1}{2}\sum_{\mu}\Lambda_{\mu}^{\,ji}\Lambda_{\mu}^{\,lk}
  =\delta_{il}\delta_{jk}-\frac{1}{N}\delta_{ij}\delta_{kl}
  \nonumber\\
  &=& \frac{Nm}{2}\sum_{a=1}^{d}\sum_{n\in\mathbb{Z}}
      \sum_{i,j,k,l=1}^{N}
      X_a^{ij}(-n)\,P_{ij,kl}\,X_a^{kl}(n).
\end{eqnarray}
\noindent Again $P_{ij,kl}$ is the adjoint projector, which removes the trace ($i=j$) and enforces the $SU(N)$ sector.

\item \textbf{Quartic coupling}: \noindent Similarly, the quartic action takes the form
\begin{eqnarray}
  -\frac{N\alpha }{4}\int_{0}^{\beta}\! dt\ \mathrm{Tr}[X_a,X_b]^2&=&
-\frac{N\alpha}{4}\,\lambda_{\mu\rho\nu\sigma}\int_{0}^{\beta}\! dt\ (x_a^{\mu}x_a^{\rho})(x_b^{\nu}x_b^{\sigma}),
\end{eqnarray}
with the coefficient
\begin{eqnarray}
\lambda_{\mu\rho\nu\sigma}
&=&
\frac{1}{2}\,\mathrm{Tr}\bigg([\Lambda_{\mu},\Lambda_{\nu}][\Lambda_{\rho},\Lambda_{\sigma}]
+[\Lambda_{\mu},\Lambda_{\sigma}][\Lambda_{\rho},\Lambda_{\nu}]\bigg).
\end{eqnarray}
\noindent Let $\mu_{\mu\rho\nu\sigma}$ denote the inverse of $\lambda_{\mu\rho\nu\sigma}$. Then we must have the pairwise identities
\begin{eqnarray}
\mu_{\mu\rho\nu\sigma}\,\lambda_{\nu\sigma\alpha\beta}
=\delta_{\mu\alpha}\,\delta_{\rho\beta},
\qquad
\lambda_{\mu\rho\nu\sigma}\,\mu_{\nu\sigma\alpha\beta}
=\delta_{\mu\alpha}\,\delta_{\rho\beta}.\label{ik}
\end{eqnarray}

\noindent For $N=2$ (so $f_{\mu\nu\kappa}=\varepsilon_{\mu\nu\kappa}$), we compute
\begin{eqnarray}
\lambda_{\mu\rho\nu\sigma}
&=&
-4\left(f_{\mu\nu\kappa}f_{\rho\sigma\kappa}+f_{\mu\sigma\kappa}f_{\rho\nu\kappa}\right)
\nonumber\\
&=&
-4\big(2\,\delta_{\mu\rho}\delta_{\nu\sigma}-\delta_{\mu\sigma}\delta_{\nu\rho}-\delta_{\mu\nu}\delta_{\rho\sigma}\big),
\qquad (SU(2)).
\end{eqnarray}

\noindent An ansatz for the inverse kernel $\mu$ is given by
\begin{eqnarray}
\mu_{\mu\rho\nu\sigma}
= a\,\delta_{\mu\rho}\delta_{\nu\sigma}
+ b\,\delta_{\mu\sigma}\delta_{\nu\rho}
+ c\,\delta_{\mu\nu}\delta_{\rho\sigma}.
\end{eqnarray}
Solving (\ref{ik}) gives $a=-b=-c=-1/16$, so that
\begin{eqnarray}
\mu_{\mu\rho\nu\sigma}
=\frac{1}{16}\Big(-\delta_{\mu\rho}\delta_{\nu\sigma}
+\delta_{\mu\sigma}\delta_{\nu\rho}
+\delta_{\mu\nu}\delta_{\rho\sigma}\Big).\ \qquad (SU(2)).
\end{eqnarray}

\noindent Moreover, we can check the two contractions
\begin{eqnarray}
\lambda_{\mu\rho\nu\sigma}\,\delta_{\nu\sigma}
= -16\,\delta_{\mu\rho},\qquad \mu_{\mu\rho\nu\sigma}\,\delta_{\nu\sigma}
= -\frac{1}{16}\,\delta_{\mu\rho}.
\end{eqnarray}

\noindent For generic $N$, we will have
\begin{eqnarray}
\lambda_{\mu\rho\nu\sigma}\,\delta_{\nu\sigma}
= -8N\,\delta_{\mu\rho},\qquad \mu_{\mu\rho\nu\sigma}\,\delta_{\nu\sigma}
= -\frac{1}{8N}\,\delta_{\mu\rho}.
\end{eqnarray}
\item \textbf{Total action}: The total action can then be written as
  
  \begin{eqnarray}
S[X]
&=&
\frac{N}{2}\sum_{a=1}^{d}\sum_{n\in\mathbb{Z}}\sum_{i,j=1}^N
X_a^{ij}(-n)\,
\Bigg(\Big(\omega_n+\frac{\theta_i-\theta_j}{\beta}\Big)^2\,
P_{ij,kl}+mP_{ij,kl}\Bigg)\,
X_a^{kl}(n)\nonumber\\
&-&\frac{N\alpha}{4}\,\lambda_{\mu\rho\nu\sigma}\int_{0}^{\beta}\! dt \,Y_{\mu\rho}\,Y_{\nu\sigma}.
  \end{eqnarray}
  Here we have introduced the bilinear

\begin{eqnarray}
Y_{\mu\rho}(t)\;=\;\sum_{a=1}^{d}x_a^{\mu}(t)\,x_a^{\rho}(t).
\end{eqnarray}
\end{itemize}
\subsection{Localization: Hubbard–Stratonovich decoupling} 
Next, we introduce an auxiliary symmetric kernel $k_{\mu\rho}=k_{\rho\mu}$ and consider the following Gaussian identity (a completion of the square)

\begin{eqnarray}
0&=&\frac{N}{4\alpha}\int_{0}^{\beta}\!dt\ \mu_{\mu\rho\nu\sigma}\,
\Big(k_{\mu\rho}+\alpha\,\lambda_{\mu\rho\alpha\beta}Y_{\alpha\beta}\Big)
\Big(k_{\nu\sigma}+\alpha\,\lambda_{\nu\sigma\gamma\delta}Y_{\gamma\delta}\Big)
\nonumber\\
&&\hspace{1.6cm}
-\;\frac{N\alpha}{4}\int_{0}^{\beta}\!dt\ \lambda_{\mu\rho\nu\sigma}\,Y_{\mu\rho}\,Y_{\nu\sigma}
\;-\;\frac{N}{2}\int_{0}^{\beta}\!dt\ k_{\mu\rho}Y_{\mu\rho}
\;-\;\frac{N}{4\alpha}\int_{0}^{\beta}\!dt\ \mu_{\mu\rho\nu\sigma}\,k_{\mu\rho}k_{\nu\sigma}.\nonumber\\
\end{eqnarray}

\noindent In other words, the path integral is extended to include the auxiliary field $k_{\mu\rho}$:
\begin{eqnarray}
Z=\int\!\mathcal{D}X\,e^{-S[X]}
=\int\!\mathcal{D}X\,\mathcal{D}k\,e^{-S_{\rm loc}[X,k]},
\end{eqnarray}
so that we are now integrating over both the original matrices $X_a$ and the new auxiliary variable $k_{\mu\rho}$. The localization action is given explicitly by 
\begin{eqnarray}
  S_{\rm loc}[X,k]=S[X]+\frac{N}{4\alpha}\int_{0}^{\beta}\!dt\ \mu_{\mu\rho\nu\sigma}\,
\Big(k_{\mu\rho}+\alpha\,\lambda_{\mu\rho\alpha\beta}Y_{\alpha\beta}\Big)
\Big(k_{\nu\sigma}+\alpha\,\lambda_{\nu\sigma\gamma\delta}Y_{\gamma\delta}\Big).
\end{eqnarray}
Explicitly, we have
\begin{eqnarray}
S_{\rm loc}
&=&
\frac{N}{2}\sum_{a=1}^{d}\sum_{n\in\mathbb{Z}}\sum_{i,j=1}^N
X_a^{ij}(-n)\,
\Bigg[
\Big(\omega_n+\frac{\theta_i-\theta_j}{\beta}\Big)^2\,P_{ij,kl}
\;+\;m\,P_{ij,kl}
\Bigg]\,
X_a^{kl}(n)
\nonumber\\
&&\qquad \;+\;\frac{N}{2}\int_{0}^{\beta}\!dt\ k_{\mu\rho}Y_{\mu\rho}
+\;\frac{N}{4\alpha}\mu_{\mu\rho\nu\sigma}\,\int_{0}^{\beta}\!dt k_{\mu\rho}k_{\nu\sigma}.
\end{eqnarray}
\noindent Strictly speaking, the auxiliary kernel introduced in the Hubbard–Stratonovich step
is a function of Euclidean time, since it couples directly to the time–dependent bilinear $Y_{\mu\rho}(t)$. 
Thus the auxiliary field is not \emph{a priori} constant in $t$. 
However, in the large–$d$ limit (or at the saddle point of the effective action after integrating out $X_a$) the dominant configurations of $k_{\mu\rho}(t)$ are time–independent.
This justifies the use of a constant $k_{\mu\rho}(t)$ in the action, which then reads

\begin{eqnarray}
S_{\rm loc}
&=&
\frac{N}{2}\sum_{a=1}^{d}\sum_{n\in\mathbb{Z}}\sum_{i,j=1}^N
X_a^{ij}(-n)\,
\Bigg[
\Big(\omega_n+\frac{\theta_i-\theta_j}{\beta}\Big)^2\,P_{ij,kl}
\;+\;m\,P_{ij,kl}+k_{ij,kl}
\Bigg]\,
X_a^{kl}(n)
\nonumber\\
&&\qquad 
+\;\frac{N}{4\alpha}\,\beta\;\mu_{\mu\rho\nu\sigma}\,k_{\mu\rho}k_{\nu\sigma},
\label{S-loc}
\end{eqnarray}
with the adjoint–to–matrix map
\begin{eqnarray}
k_{ij,kl} \;=\;\frac{1}{4}\,k_{\mu\rho}\,\Lambda_{\mu}^{\,ji}\,\Lambda_{\rho}^{\,lk}.
\end{eqnarray}
\noindent Thus the interaction has been \emph{localized} into a linear coupling of $k$ to the bilinear $Y_{\mu\rho}$ plus a Gaussian weight for $k$. Integrating out $X_a$ is now Gaussian and yields an effective action for $k$; the saddle point of that effective action is  the real starting point of  the large–$d$ analysis.

\noindent Moreover, note that $k_{ij,kl}$ is traceless in each index pair:
\begin{eqnarray}
\delta_{ij}\,k_{ij,kl}
&=& \frac{1}{4}\,k_{\mu\rho}\,(\delta_{ij}\Lambda_{\mu}^{\,ji})\,\Lambda_{\rho}^{\,lk}
=\frac{1}{4}\,k_{\mu\rho}\,\mathrm{Tr}(\Lambda_\mu)\,\Lambda_{\rho}^{\,lk}=0,
\nonumber\\
\delta_{kl}\,k_{ij,kl}
&=& \frac{1}{4}\,k_{\mu\rho}\,\Lambda_{\mu}^{\,ji}\,(\delta_{kl}\Lambda_{\rho}^{\,lk})
=\frac{1}{4}\,k_{\mu\rho}\,\Lambda_{\mu}^{\,ji}\,\mathrm{Tr}(\Lambda_\rho)=0.
\end{eqnarray}
\noindent Now apply the projector from the right:
\begin{eqnarray}
(kP)_{ij,kl}
&=& k_{ij,rs}\,P_{rs,kl}
\nonumber\\
&=& k_{ij,rs}\Big(\delta_{rk}\delta_{sl}-\frac{1}{N}\delta_{rs}\delta_{kl}\Big)
= k_{ij,kl} - \frac{1}{N}\,(\delta_{rs}k_{ij,rs})\,\delta_{kl}=k_{ij,kl}.
\end{eqnarray}
\noindent Similarly, applying the projector from the left gives
\begin{eqnarray}
(Pk)_{ij,kl}
&=& P_{ij,rs}\,k_{rs,kl}
=\Big(\delta_{ir}\delta_{js}-\frac{1}{N}\delta_{ij}\delta_{rs}\Big)\,k_{rs,kl}
= k_{ij,kl}-\frac{1}{N}\,\delta_{ij}\,(\delta_{rs}k_{rs,kl})=k_{ij,kl}.\nonumber\\
\end{eqnarray}
\noindent Combining both results, we obtain
\begin{eqnarray}
(PkP)_{ij,kl}=k_{ij,kl}.
\end{eqnarray}

\subsection{Effective action}

\medskip 
 \noindent Integrating out the Gaussian matrices $X_a$ in \eqref{S-loc} yields the effective action for the auxiliary kernel $k$:
\begin{eqnarray}
S_{\rm eff}[k]
&=&
\frac{d}{2}\sum_n
\operatorname{Tr}\!
\log (P{\cal W}(n)P)
\;+\;\frac{N\beta}{4\alpha}\,\mu_{\mu\rho\nu\sigma}\,k_{\mu\rho}k_{\nu\sigma},
\label{S-eff-general}
\end{eqnarray}
with
\begin{eqnarray}
{\cal W}_{ij,kl}(n)=\Big(\omega_n+\frac{\theta_i-\theta_j}{\beta}\Big)^2 P_{ij,kl}
\;+\;m\,P_{ij,kl}
\;+\;k_{ij,kl}.
\end{eqnarray}
\noindent In the free part, we have also used the fact that $P$ commutes with the quadratic kernel 
${\cal W}^{(0)}_{ij}(n)$ defined by 
\begin{eqnarray}
{\cal W}^{(0)}_{ij,kl}(n)
\;=\;
A_{ij}\,P_{ij,kl},
\qquad
A_{ij}\equiv\Big(\omega_n+\frac{\theta_i-\theta_j}{\beta}\Big)^2+m,
\qquad
P_{ij,kl}=\delta_{il}\delta_{jk}-\frac{1}{N}\delta_{ij}\delta_{kl}.\nonumber\\
\end{eqnarray}
Indeed, we have
\begin{eqnarray}
{\cal W}^{(0)}(n)\,P\;=\; P\,{\cal W}^{(0)}(n).
\end{eqnarray}
The proof goes as follows. First, we compute
\begin{eqnarray}
(P{\cal W}^{(0)})_{ij,rs}
&=& P_{ij,kl}{\cal W}^{(0)}_{kl,rs}=P_{ij,kl}\,A_{kl}\,P_{kl,rs}
=\Big(\delta_{ik}\delta_{jl}-\frac{1}{N}\delta_{ij}\delta_{kl}\Big)\,A_{kl}\,
\Big(\delta_{kr}\delta_{ls}-\frac{1}{N}\delta_{kl}\delta_{rs}\Big)\nonumber\\
&=& T_1+T_2+T_3+T_4,
\end{eqnarray}
with
\begin{align}
T_1&=\delta_{ik}\delta_{jl}\,A_{kl}\,\delta_{kr}\delta_{ls}
= A_{ij}\,\delta_{ir}\delta_{js},\\
T_2&=-\delta_{ik}\delta_{jl}\,A_{kl}\,\frac{1}{N}\delta_{kl}\delta_{rs}
= -\frac{1}{N}\,\delta_{ij}\delta_{rs}\,A_{ii}
= -\frac{1}{N}(\omega_n^2+m)\,\delta_{ij}\delta_{rs},\\
T_3&=-\frac{1}{N}\delta_{ij}\delta_{kl}\,A_{kl}\,\delta_{kr}\delta_{ls}
= -\frac{1}{N}\,\delta_{ij}\delta_{rs}\,A_{rs},\\
T_4&=+\frac{1}{N^2}\delta_{ij}\delta_{kl}\,A_{kl}\,\delta_{kl}\delta_{rs}
= \frac{1}{N}\,\delta_{ij}\delta_{rs}\,(\omega_n^2+m),
\end{align}
where we used $A_{ii}=\omega_n^2+m$ and $\sum_k A_{kk}=N(\omega_n^2+m)$. Hence $T_2+T_4=0$ and
\begin{eqnarray}
(P{\cal W}^{(0)})_{ij,rs}
= A_{ij}\,\delta_{ir}\delta_{js} - \frac{1}{N}\,\delta_{ij}\delta_{rs}\,A_{rs}.
\end{eqnarray}
On the other hand, we compute
\begin{eqnarray}
({\cal W}^{(0)}P)_{ij,rs}={\cal W}^{(0)}_{ij,kl}P_{kl,rs}
= A_{ij}\,P_{ij,rs}
= A_{ij}\Big(\delta_{ir}\delta_{js}-\frac{1}{N}\delta_{ij}\delta_{rs}\Big).
\end{eqnarray}
Therefore $(P{\cal W}^{(0)})_{ij,rs}=({\cal W}^{(0)}P)_{ij,rs}$.

\section{Saddle-point equation and the maximally symmetric ansatz}\label{section3}
\medskip
\noindent
The saddle--point and self--consistent Gaussian methods used in this section
are closely related to the large--\(d\) analyses of
Refs.~\cite{Mandal:2009vzN,Mandal:2011hbN,Filev:2015hiaF,Kawahara:2007fnF}
and references therein.
\subsection{Saddle point equation}

\medskip
\noindent
Motivated by the fact that the free quadratic kernel is compatible with the adjoint
projector,
\begin{eqnarray}
P\Delta^{(0)}=\Delta^{(0)}P,
\end{eqnarray}
we impose the analogous requirement on the Hubbard--Stratonovich interaction
kernel. We therefore adopt the factorized ansatz
\begin{eqnarray}
k_{ij,kl}
\;\equiv\;
k_{ij}\,P_{ij,kl}
\;=\;
P_{ij,kl}\,k_{kl}.
\label{ansatz-k}
\end{eqnarray}
This choice makes the interaction kernel act as an effective mass term in the
adjoint sector, rather than as a new kinetic operator. For consistency of the two
factorizations in \eqref{ansatz-k}, one must have
\begin{eqnarray}
k_{ij}=k_{ji},
\qquad
k_{ii}=k_{jj}\qquad \forall\,i,j.
\end{eqnarray}
Indeed,
\begin{eqnarray}
(kP)_{ij,rs}
=
k_{ij}P_{ij,mn}P_{mn,rs}
=
k_{ij}
\left(
\delta_{ir}\delta_{js}
-\frac{1}{N}\delta_{ij}\delta_{rs}
\right),
\end{eqnarray}
while
\begin{eqnarray}
(Pk)_{ij,rs}
  =
P_{ij,mn}k_{mn}P_{mn,rs}=
k_{ji}\delta_{ir}\delta_{js}
-\frac{1}{N}\delta_{ij}\delta_{rs}k_{ii}
-\frac{1}{N}\delta_{ij}\delta_{rs}k_{rr}
+\frac{1}{N^2}\delta_{ij}\delta_{rs}\sum_m k_{mm}.
\end{eqnarray}
Using \(k_{ij}=k_{ji}\) and the equality of all diagonal entries, this becomes
\begin{eqnarray}
(Pk)_{ij,rs}
&=&
k_{ij}
\left(
\delta_{ir}\delta_{js}
-\frac{1}{N}\delta_{ij}\delta_{rs}
\right)
=
(kP)_{ij,rs}.
\end{eqnarray}
Thus
\begin{eqnarray}
Pk=kP.
\end{eqnarray}
Since the free quadratic kernel also commutes with \(P\), the full quadratic
operator satisfies
\begin{eqnarray}
P{\cal W}(n)={\cal W}(n)P.
\label{coi}
\end{eqnarray}

\medskip
\noindent
Therefore, on the projected adjoint subspace, \(P\) acts as the identity and
\begin{eqnarray}
P{\cal W}(n)P
=
P{\cal W}(n)
=
{\cal W}(n)P.
\end{eqnarray}
Hence the trace-log in \eqref{S-eff-general} is understood as a trace over the
projected \(SU(N)\) sector:
\begin{eqnarray}
\Tr\log\!\big(P{\cal W}(n)P\big)
=
\Tr_{\rm adj}\log{\cal W}(n).
\end{eqnarray}

\medskip
\noindent
The ansatz \eqref{ansatz-k} also gives the factorized form
\begin{eqnarray}
{\cal W}_{ij,kl}(n)
=
\bar A_{ij}(n)\,P_{ij,kl}
=
P_{ij,kl}\,\bar A_{lk}(n),
\qquad
\bar A_{ij}(n)\equiv A_{ij}(n)+k_{ij}.
\label{W-Delta}
\end{eqnarray}
The second equality contains \(\bar A_{lk}\), not \(\bar A_{kl}\), because the
projector \(P_{ij,kl}\) reverses the bi--index pair. The projected trace therefore
reduces to the sum over adjoint bi--index modes,
\begin{eqnarray}
\Tr_{\rm adj}\log{\cal W}(n)
=
\sum_{i,j=1}^{N}
P_{ij,ji}\,
\log\bar A_{ij}(n).
\end{eqnarray}
Thus the effective action becomes
\begin{eqnarray}
S_{\rm eff}[k]
&=&
\frac{d}{2}
\sum_{n\in\mathbb Z}
\sum_{i,j=1}^{N}
P_{ij,ji}\,
\log\bar A_{ij}(n)
+
\frac{N\beta}{4\alpha}
\mu_{\mu\rho\nu\sigma}k_{\mu\rho}k_{\nu\sigma}.
\label{Seff-simplified}
\end{eqnarray}

\medskip
\noindent
Recall now the adjoint–to–matrix map
\begin{eqnarray}
k_{ij,kl}=\frac{1}{4}\,k_{\mu\rho}\,\Lambda_{\mu}^{\,ji}\,\Lambda_{\rho}^{\,lk},
\qquad
\mathrm{Tr}(\Lambda_\mu\Lambda_\nu)=2\,\delta_{\mu\nu}.
\end{eqnarray}
\noindent Varying $S_{\rm eff}$ with respect to the \emph{matrix} variables $k_{ij}$ gives
\begin{eqnarray}
0=\frac{\partial S_{\rm eff}}{\partial k_{ij}}
&=&
\frac{d}{2}\sum_{n\in\mathbb{Z}}\,P_{ij,ji}\,\frac{1}{\bar A_{ij}(n)}
\;+\;\frac{N\beta}{2\alpha}\,\mu_{\mu\rho\nu\sigma}\,k_{\nu\sigma}\,
\frac{\partial k_{\mu\rho}}{\partial k_{ij}}.
\label{var-kij}
\end{eqnarray}
We use the inverse of the above map (obtained by contracting with the generators and using the commuting-symmetric ansatz $k_{ij,kl}=P_{ij,kl}\,k_{ij}$),
\begin{eqnarray}
k_{\mu\rho} \;=\; \Lambda_{\mu}^{\,ij}\,k_{ij,kl}\,\Lambda_{\rho}^{\,kl}=\; \Lambda_{\mu}^{\,ij}\,P_{ij,kl}\,k_{ij}\,\Lambda_{\rho}^{\,kl},
\end{eqnarray}
to obtain the variation 
\begin{eqnarray}
\frac{\partial k_{\mu\rho}}{\partial k_{i_1j_1}}
= \Lambda_{\mu}^{\,ij}\,P_{ij,kl}\,\Lambda_{\rho}^{\,kl}\ \delta_{ii_1}\delta_{jj_1}
= \Lambda_{\mu}^{\,i_1j_1}\,P_{i_1j_1,kl}\,\Lambda_{\rho}^{\,kl}.
\end{eqnarray}
Hence \eqref{var-kij} becomes
\begin{eqnarray}
0
&=&
\frac{d}{2}\sum_{n\in\mathbb{Z}}\,P_{ij,ji}\,\frac{1}{\bar A_{ij}(n)}
\;+\;\frac{N\beta}{2\alpha}\,\mu_{\mu\rho\nu\sigma}\,k_{\nu\sigma}\,
\Lambda_{\mu}^{\,ij}\,P_{ij,kl}\,\Lambda_{\rho}^{\,kl}.
\label{gap-raw}
\end{eqnarray}
Since $P$ projects onto traceless matrices and each $\Lambda_\rho$ is traceless, we have
$P_{ij,kl}\,\Lambda_{\rho}^{\,kl}=\Lambda_{\rho}^{\,ji}$, and therefore the saddle equation simplifies to
\begin{eqnarray}
0
&=&
\frac{d}{2}\sum_{n\in\mathbb{Z}}\,P_{ij,ji}\,\frac{1}{\bar A_{ij}(n)}
\;+\;\frac{N\beta}{2\alpha}\,\mu_{\mu\rho\nu\sigma}\,k_{\nu\sigma}\,
\Lambda_{\mu}^{\,ij}\,\Lambda_{\rho}^{\,ji}.
\label{gap-pre-sum}
\end{eqnarray}
\noindent The Matsubara sum is standard \cite{Kapusta:2006pm,LeBellac:1996}:
\begin{eqnarray}
\sum_{n\in\mathbb{Z}}
\frac{1}{\Big(\omega_n+\frac{\theta_i-\theta_j}{\beta}\Big)^2+\tilde{k}_{ij}}
\;=\;
\beta\,\frac{1}{2\sqrt{\tilde{k}_{ij}}}\,
\frac{\sinh\!\big(\beta\sqrt{\tilde{k}_{ij}}\big)}
{\cosh\!\big(\beta\sqrt{\tilde{k}_{ij}}\big)-\cos(\theta_i-\theta_j)},\quad \tilde{k}_{ij}=k_{ij}+m.\label{Matsu}
\end{eqnarray}
Substituting this back we obtain the saddle–point (gap) equation in the commuting-symmetric ansatz:
\begin{eqnarray}
0
&=&
\frac{d\,\beta}{4}\,P_{ij,ji}\,
\frac{1}{\sqrt{\tilde{k}_{ij}}}\,
\frac{\sinh\!\big(\beta\sqrt{\tilde{k}_{ij}}\big)}
{\cosh\!\big(\beta\sqrt{\tilde{k}_{ij}}\big)-\cos(\theta_i-\theta_j)}
\;+\;\frac{N\beta}{2\alpha}\,\mu_{\mu\rho\nu\sigma}\,k_{\nu\sigma}\,
\Lambda_{\mu}^{\,ij}\,\Lambda_{\rho}^{\,ji}.\nonumber\\
\label{gap-final}
\end{eqnarray}
Equations \eqref{ansatz-k}, \eqref{Seff-simplified}, and \eqref{gap-final} encapsulate the commuting-symmetric ansatz, the corresponding simplification of the effective action in the constant commuting sector, and the resulting saddle–point (gap) condition obtained after performing the Matsubara summation.

\subsection{Maximally symmetric ansatz}

\medskip
\noindent
We now pass from the commuting--symmetric saddle used above to the maximally
symmetric ansatz.  The constancy of \(k_{\mu\rho}(t)\) in Euclidean time and the
factorization of the interaction kernel through the adjoint projector have already
been used in deriving \eqref{Seff-simplified} and \eqref{gap-final}.  The new
ingredient in the present subsection is the final strengthening of this ansatz:
uniformity in color space, \(k_{ij}=k_0\).

\medskip
\noindent
Therefore, we package together several consistency requirements:

\medskip
\noindent\emph{(i) Constancy in Euclidean time (commutativity).}
\begin{eqnarray}
k_{\mu\rho}(t)=k_{\mu\rho},\qquad k_{\mu\rho} \;=\; \Lambda_{\mu}^{\,ij}\,k_{ij,kl}\,\Lambda_{\rho}^{\,kl} \quad\text{(time–independent).}
\end{eqnarray}
This condition has already been assumed in passing to the constant large--\(d\)
saddle and in deriving the effective action.  Also, since \(k_{\mu\rho}\) is an
adjoint kernel, its matrix representative lies in the \(SU(N)\) sector; thus
\(PkP=k\) follows from adjoint support and is not an additional assumption.

\noindent\emph{(ii) Factorization with the adjoint projector (symmetric properties).}
\begin{eqnarray}
k_{ij,kl}=k_{ij}\,P_{ij,kl}=P_{ij,kl}\,k_{kl},\qquad k_{ij}=k_{ji},
\end{eqnarray}
From this one obtains 
\begin{eqnarray}
k_{ij}=k_{ji}\quad (\text{symmetry}),\qquad
k_{ii}=k_{jj}\ \ \forall\,i,j \quad (\text{all diagonals equal}).
\end{eqnarray}
This is precisely the symmetric ansatz already used above: it makes the
interaction kernel compatible with the same adjoint projector structure as the
free quadratic kernel.  In turn, the four--index structure collapses to a
two--index one, and \(k_{ij}\) behaves as a holonomy--dependent effective mass
term.

\noindent\emph{(iii) Uniformity in color space.}
\begin{eqnarray}
k_{ij}=k_0\quad\forall\,i,j.
\end{eqnarray}
This is the genuinely new step of the maximally symmetric ansatz.  Consequently
the remaining two--index structure collapses to a single scalar mass parameter,
\begin{eqnarray}
k_{ij,kl}=k_0\,P_{ij,kl}\qquad\text{with }k_0\in\mathbb{R}.
\end{eqnarray}

\medskip
\noindent
Thus, in the terminology used here, the commuting--symmetric saddle consists of
(i) and (ii), while the maximally symmetric ansatz is obtained by adding the
uniformity condition (iii).

\medskip
\noindent In the literature this is often termed simply the “symmetric ansatz”: symmetric in Euclidean time (constant),
symmetric in indices ($k_{ij}=k_{ji}$), and effectively uniform in matrix (color) space.

\medskip
\noindent
Starting now from
\begin{eqnarray}
0
&=&
\frac{d\,\beta}{4}\,P_{ij,ji}\,
\frac{1}{\sqrt{\tilde{k}_{ij}}}\,
\frac{\sinh\!\big(\beta\sqrt{\tilde{k}_{ij}}\big)}
{\cosh\!\big(\beta\sqrt{\tilde{k}_{ij}}\big)-\cos(\theta_i-\theta_j)}
\;+\;\frac{N\beta}{2\alpha}\,\mu_{\mu\rho\nu\sigma}\,k_{\nu\sigma}\,
\Lambda_{\mu}^{\,ij}\,\Lambda_{\rho}^{\,ji},
\label{eq:start-gap}
\end{eqnarray}
we now impose the maximally symmetric ansatz \(k_{ij}=k_0\). At this stage we keep
the \((i,j)\) dependence explicit; the final scalar gap equation is obtained only
after projecting or summing over the color indices, or after specifying the
holonomy saddle. Using
\begin{eqnarray}
\mathrm{Tr} \Lambda_{\mu}\Lambda_{\rho}=2\,\delta_{\mu\rho},
\qquad
k_{ij,kl}=k_0\,P_{ij,kl}\;\Longleftrightarrow\;k_{\nu\sigma}=2k_0\,\delta_{\nu\sigma},
\nonumber\\
\end{eqnarray}
and the identity
\begin{eqnarray}
\mu_{\mu\rho\nu\sigma}\,\delta_{\nu\sigma}=-\frac{1}{8N}\,\delta_{\mu\rho},
\end{eqnarray}
the second term in \eqref{eq:start-gap} gives
\begin{eqnarray}
\frac{N\beta}{2\alpha}\,\mu_{\mu\rho\nu\sigma}\,k_{\nu\sigma}\,
\Lambda_{\mu}^{\,ij}\Lambda_{\rho}^{\,ji}
&=&
\frac{N\beta}{2\alpha}\,\mu_{\mu\rho\nu\sigma}\,(2k_0\,\delta_{\nu\sigma})\,
\Lambda_{\mu}^{\,ij}\Lambda_{\rho}^{\,ji}
\nonumber\\
&=&
\frac{N\beta}{2\alpha}\,
\left(-\frac{1}{8N}\delta_{\mu\rho}\right)(2k_0)\,
\Lambda_{\mu}^{\,ij}\Lambda_{\rho}^{\,ji}
\nonumber\\
&=&
\frac{N\beta}{2\alpha}\,
\left(-\frac{1}{8N}\right)(2k_0)(2P_{ij,ji})
\nonumber\\
&=&-\frac{\beta k_0}{4\alpha}P_{ij,ji}.
\end{eqnarray}
In the third line, we used
\begin{eqnarray}
\sum_{\mu}\Lambda_{\mu}^{\,ij}\Lambda_{\mu}^{\,ji}=2\,P_{ij,ji}.
\end{eqnarray}
Hence the maximally symmetric ansatz gives the pairwise saddle condition
\begin{eqnarray}
0
&=&
\frac{\beta P_{ij,ji}}{4}
\left[
\frac{d}{\sqrt{\tilde{k}_0}}\,
\frac{\sinh(\beta\sqrt{\tilde{k}_0})}
{\cosh(\beta\sqrt{\tilde{k}_0})-\cos(\theta_i-\theta_j)}
-\frac{k_0}{\alpha}
\right].
\end{eqnarray}
For the components surviving the adjoint projection, this is equivalently written as
\begin{eqnarray}
0
&=&
\frac{d}{\sqrt{\tilde{k}_0}}\,
\frac{\sinh(\beta\sqrt{\tilde{k}_0})}
{\cosh(\beta\sqrt{\tilde{k}_0})-\cos(\theta_i-\theta_j)}
-\frac{k_0}{\alpha}.
\label{eq:gap-exact}
\end{eqnarray}
The uniform ansatz removes thus the color dependence of the auxiliary mass \(k_{ij}\),
but it does not remove the holonomy dependence carried by
\(\theta_i-\theta_j\). Thus the final scalar gap equation for \(k_0\) is obtained
only after a further treatment of the holonomy sector, for example by summing over
\(i,j\), averaging over the holonomy distribution, or evaluating at a chosen
holonomy saddle.

\medskip
\noindent
For example, in the center--symmetric, or confining, holonomy saddle the
eigenvalues are uniformly distributed on the unit circle,
\begin{eqnarray}
\theta_i
=
\frac{2\pi}{N}\left(i-\frac{N+1}{2}\right),
\qquad
i=1,\ldots,N,
\end{eqnarray}
and therefore
\begin{eqnarray}
\theta_i-\theta_j
=
\frac{2\pi}{N}(i-j).
\end{eqnarray}
In this case the residual holonomy dependence is converted into a discrete color
average, giving
\begin{eqnarray}
\frac{k_0}{\alpha}
&=&
\frac{d}{N^2-1}
\sum_{i,j=1}^{N}
P_{ij,ji}\,
\frac{1}{\sqrt{\tilde{k}_0}}\,
\frac{\sinh(\beta\sqrt{\tilde{k}_0})}
{\cosh(\beta\sqrt{\tilde{k}_0})
-\cos\!\left(\frac{2\pi(i-j)}{N}\right)}.
\label{eq:gap-center-symmetric}
\end{eqnarray}
By contrast, setting all \(\theta_i=0\) corresponds to the collapsed, or
deconfined, holonomy saddle.

\subsection{Scope of the maximally symmetric ansatz}
\label{subsection:scope-maximal-saddle}

\medskip
\noindent
It is useful to derive the scalar gap equation once more by imposing the
maximally symmetric ansatz directly in the effective action.  This provides
a complementary derivation of the result above and clarifies the precise
scope of the ansatz.

\medskip
\noindent
Substituting
\begin{eqnarray}
k_{ij,kl}=k_0P_{ij,kl},
\qquad
k_{\mu\rho}=2k_0\delta_{\mu\rho},
\qquad
s^2=m+k_0,
\end{eqnarray}
into \eqref{Seff-simplified}, and using again \(\mu_{\mu\rho\nu\sigma}\delta_{\nu\sigma}=-\delta_{\mu\rho}/8N\), one obtains, up to terms independent of \(k_0\),
\begin{eqnarray}
S_{\rm eff}^{\rm max}(k_0;U)
=
\frac{d}{2}
\sum_{n\in{\mathbb Z}}
\sum_{i,j=1}^{N}
P_{ij,ji}
\log\left[
\left(
\omega_n+\frac{\theta_i-\theta_j}{\beta}
\right)^2+s^2
\right]-\frac{\beta(N^2-1)}{8\alpha}k_0^2.
\label{Seff-direct-k0}
\end{eqnarray}
Varying with respect to \(k_0\) gives
\begin{eqnarray}
0
=
\frac{d}{2}
\sum_{n\in{\mathbb Z}}
\sum_{i,j=1}^{N}
\frac{P_{ij,ji}}
{
\left(
\omega_n+\frac{\theta_i-\theta_j}{\beta}
\right)^2+s^2
}
-
\frac{\beta(N^2-1)}{4\alpha}k_0.
\end{eqnarray}
After performing the Matsubara sum, this becomes
\begin{eqnarray}
\frac{d}{s}{\cal R}(s;U)
=
\frac{k_0}{\alpha}
=
\frac{s^2-m}{\alpha},
\label{gap-direct-k0}
\end{eqnarray}
where
\begin{eqnarray}
{\cal R}(s;U)
=
\frac{1}{N^2-1}
\sum_{i,j=1}^{N}
P_{ij,ji}
\frac{\sinh(\beta s)}
{\cosh(\beta s)-\cos(\theta_i-\theta_j)}.
\label{R-direct-k0}
\end{eqnarray}
This is precisely the projector--weighted projection of the pairwise saddle
equation \eqref{eq:gap-exact} onto the maximally symmetric variation
\(\delta k_{ij}=\delta k_0\), which sums over the adjoint color modes with
weight \(P_{ij,ji}\).  The pairwise derivation and the direct scalar
variation are therefore complementary: the former retains the residual
\((i,j)\)-dependent holonomy structure, whereas the latter yields the
single equation obeyed within the restricted \(k_0\) sector.

\medskip
\noindent
The important point is that the maximally symmetric ansatz concerns the
Hubbard--Stratonovich kernel, not the Polyakov-loop holonomy.  It sets the
dynamically generated mass shift equal in all adjoint color directions, but
the eigenvalues \(\theta_i\) remain dynamical and enter the gap equation
through \({\cal R}(s;U)\).  The ansatz therefore does not impose a uniform
holonomy distribution.

Indeed, within the same maximally symmetric auxiliary sector, the leading
holonomy action derived below is
\begin{eqnarray}
\frac{S_{\rm hol}}{N^2}
=
\sum_{n\geq1}
\frac{1-de^{-n\beta s}}{n}|u_n|^2+\cdots,
\qquad
u_n=\frac{1}{N}\Tr U^n.
\label{holonomy-preview-k0}
\end{eqnarray}
The confined saddle is characterized by \(u_n=0\), whereas nonuniform and
gapped deconfined saddles have nonzero Polyakov moments.  The uniform
distribution loses stability when
\begin{eqnarray}
1-de^{-\beta s}=0.
\end{eqnarray}
Consequently, condition (iii) does not exclude deconfinement: the confined,
nonuniform, and gapped holonomy phases are all compatible with the same
scalar auxiliary saddle.

\medskip
\noindent
The motivation for the ansatz is particularly clear at low temperature.
Using
\begin{eqnarray}
\frac{\sinh(\beta s)}
{\cosh(\beta s)-\cos\phi}
=
1+2\sum_{n\geq1}e^{-n\beta s}\cos(n\phi),
\end{eqnarray}
one finds
\begin{eqnarray}
{\cal R}(s;U)
=
1+2\sum_{n\geq1}C_n(U)e^{-n\beta s},
\qquad
C_n(U)
=
\frac{|\Tr U^n|^2-1}{N^2-1}.
\end{eqnarray}
The leading zero-temperature saddle is therefore color independent, while
its holonomy dependence is exponentially suppressed.  This supports the
use of the symmetry-preserving scalar kernel as the leading large-\(d\),
low-temperature saddle, as in the large-\(d\) analysis of
Ref.~\cite{Filev:2015hiaF}.

\medskip
\noindent
The ansatz is nevertheless a restriction of the full auxiliary-field
saddle space.  More general color-dependent solutions, or nonsinglet
fluctuations of \(k_{\mu\rho}\), may affect subleading corrections and the
detailed nonlinear structure of the deconfined branch.  The present work
does not claim to classify such saddles; it shows instead that the
maximally symmetric auxiliary saddle itself is fully compatible with both
confined and deconfined holonomy configurations.

\section{Low--temperature \(X\)-space physics of the large--\(d\) saddle}\label{section4}

\subsection{Gap equation at low temperature}

\noindent Starting from the exact gap equation
\begin{eqnarray}
\frac{d}{\sqrt{\tilde{k}_0}}\,
\frac{\sinh(\beta\sqrt{\tilde{k}_0})}{\cosh(\beta\sqrt{\tilde{k}_0})-\cos\phi}
=\frac{k_0}{\alpha},
\qquad \phi\equiv \theta_i-\theta_j,\quad s\equiv\sqrt{\tilde{k}_0}>0,\quad \tilde{k}_0=k_0+m.
\label{eq:gap-exact-LT}
\end{eqnarray}
we expand the thermal factor for $\beta\to\infty$:
\begin{eqnarray}
\frac{\sinh(\beta s)}{\cosh(\beta s)-\cos\phi}
= \frac{\tfrac{1}{2}e^{\beta s}(1-e^{-2\beta s})}{\tfrac{1}{2}e^{\beta s}\big(1-2\cos\phi\,e^{-\beta s}+e^{-2\beta s}\big)}
= 1 + 2\cos\phi\,e^{-\beta s} + \mathcal{O}\!\big(e^{-2\beta s}\big).
\label{eq:ratio-expansion}
\end{eqnarray}
Inserting \eqref{eq:ratio-expansion} into \eqref{eq:gap-exact-LT} gives
\begin{eqnarray}
\frac{d}{s}\Big[1 + 2\cos\phi\,e^{-\beta s} + \mathcal{O}(e^{-2\beta s})\Big]
=\frac{s^{2}-m}{\alpha}.
\label{eq:gap-expand}
\end{eqnarray}
Let \(s=s_0+\delta s\), with \(s_0\) the \(T=0\) solution and
\(\delta s=\mathcal{O}(e^{-\beta s_0})\) the first thermal correction.
At \(T=0\) we have
\(d/s_0=(s_0^2-m)/\alpha\), i.e.
\begin{eqnarray}
s_0^3-ms_0=\alpha d.
\label{eq:zeroT}
\end{eqnarray}
Keeping only terms linear in \(\delta s\) and \(e^{-\beta s_0}\), and
neglecting
\(\delta s\,e^{-\beta s_0}
=\mathcal{O}(e^{-2\beta s_0})\),
Eq.~\eqref{eq:gap-expand} gives
\begin{eqnarray}
\underbrace{
\left(
\frac{d}{s_0}
-
\frac{s_0^2-m}{\alpha}
\right)
}_{=\,0}
+
\left(
-\frac{d}{s_0^2}
-\frac{2s_0}{\alpha}
\right)\delta s
+
\frac{2d}{s_0}
\cos\phi\,e^{-\beta s_0}
=
0.
\end{eqnarray}
Using
\begin{eqnarray}
-\frac{d}{s_0^2}
-\frac{2s_0}{\alpha}
=
-\frac{3s_0^2-m}{\alpha s_0},
\end{eqnarray}
we obtain
\begin{eqnarray}
\delta s
=
\frac{2s_0(s_0^2-m)}
{3s_0^2-m}
\cos\phi\,e^{-\beta s_0}.
\label{eq:delta-s}
\end{eqnarray}
Hence the mass gap and the auxiliary kernel are
\begin{align}
s=\sqrt{\widetilde k_0}
&=
s_0
\left[
1+
2\frac{s_0^2-m}{3s_0^2-m}
\cos\phi\,e^{-\beta s_0}
+
\mathcal{O}\!\left(e^{-2\beta s_0}\right)
\right],
\label{eq:gap-s}
\\[4pt]
\widetilde k_0
&=
s_0^2
\left[
1+
4\frac{s_0^2-m}{3s_0^2-m}
\cos\phi\,e^{-\beta s_0}
+
\mathcal{O}\!\left(e^{-2\beta s_0}\right)
\right].
\label{eq:gap-k}
\end{align}
In a center-symmetric (confining) holonomy one often has the projector-weighted average $\langle\cos\phi\rangle=0$, in which case the first thermal correction vanishes and the leading thermal effect is $\mathcal{O}(e^{-2\beta s_0})$. However, if the holonomy is not center-symmetric, the \(\cos\phi\) factor should be understood as the appropriate (projector-weighted) average over color pairs.

The large-d saddle point $s_0$ is determined by the cubic equation \eqref{eq:zeroT}
In Cardano form $s_0^3+ps_0+q=0$ we have $p=-m$ and $q=-a$. The discriminant is
\begin{eqnarray}
\Delta=\Big(\frac{q}{2}\Big)^2+\Big(\frac{p}{3}\Big)^3
=\frac{a^2}{4}-\frac{m^3}{27},
\end{eqnarray}
and the (real) root can be written as
\begin{eqnarray}
s_0
=
\Big(\frac{a}{2}+\sqrt{\Delta}\Big)^{\!\!1/3}
+
\Big(\frac{a}{2}-\sqrt{\Delta}\Big)^{\!\!1/3}.
\label{Cardano}
\end{eqnarray}
\medskip
\noindent We now expand \eqref{Cardano} at large $d$, i.e. at large $a=\alpha d$ with fixed $m$.
Introduce the small parameter
\begin{eqnarray}
\varepsilon \equiv \frac{m}{a^{2/3}}.
\label{eps}
\end{eqnarray}
First expand $\sqrt{\Delta}$:
\begin{eqnarray}
\sqrt{\Delta}
&=&
\frac{a}{2}\sqrt{1-\frac{4m^3}{27a^2}}
=\frac{a}{2}\left[1-\frac{1}{2}\frac{4m^3}{27a^2}-\frac{1}{8}\Big(\frac{4m^3}{27a^2}\Big)^2+\cdots\right]\nonumber\\
&=&
\frac{a}{2}
-\frac{m^3}{27a}
-\frac{m^6}{729a^3}
+\cdots.
\label{sqrtDelta}
\end{eqnarray}
Therefore
\begin{eqnarray}
\frac{a}{2}+\sqrt{\Delta}
&=&
a-\frac{m^3}{27a}-\frac{m^6}{729a^3}+\cdots
=a\left[1-\frac{m^3}{27a^2}-\frac{m^6}{729a^4}+\cdots\right],
\label{Ap}\\
\frac{a}{2}-\sqrt{\Delta}
&=&
\frac{m^3}{27a}+\frac{m^6}{729a^3}+\cdots
=\frac{m^3}{27a}\left[1+\frac{m^3}{27a^2}+\cdots\right].
\label{Am}
\end{eqnarray}
Taking cube roots and using $(1+u)^{1/3}=1+\frac{1}{3}u-\frac{1}{9}u^2+\cdots$ gives
\begin{eqnarray}
\Big(\frac{a}{2}+\sqrt{\Delta}\Big)^{\!\!1/3}
&=&
a^{1/3}\left[1-\frac{1}{3}\frac{m^3}{27a^2}+\mathcal{O}\Big(\frac{m^6}{a^4}\Big)\right]
=
a^{1/3}-\frac{m^3}{81\,a^{5/3}}+\cdots,
\label{root1}\\
\Big(\frac{a}{2}-\sqrt{\Delta}\Big)^{\!\!1/3}
&=&
\left(\frac{m^3}{27a}\right)^{\!1/3}\left[1+\frac{1}{3}\frac{m^3}{27a^2}+\cdots\right]
=
\frac{m}{3\,a^{1/3}}+\frac{m^4}{243\,a^{7/3}}+\cdots.
\label{root2}
\end{eqnarray}
Adding \eqref{root1} and \eqref{root2} yields the large-$d$ expansion of the positive root:
\begin{eqnarray}
s_0
&=&
a^{1/3}
+\frac{m}{3\,a^{1/3}}
-\frac{m^3}{81\,a^{5/3}}
+\frac{m^4}{243\,a^{7/3}}
+\mathcal{O}\!\Big(\frac{m^6}{a^{11/3}}\Big),
\qquad a=\alpha d.
\label{sd-expansion}
\end{eqnarray}
Equivalently, in terms of $\varepsilon=m/a^{2/3}$,
\begin{eqnarray}
s_0
=
a^{1/3}\left(1+\frac{\varepsilon}{3}-\frac{\varepsilon^3}{81}+\frac{\varepsilon^4}{243}+\cdots\right),
\qquad
\varepsilon=\frac{m}{(\alpha d)^{2/3}}.
\label{sd-eps}
\end{eqnarray}

\medskip
\noindent The domain of validity is $\varepsilon\ll1$, i.e.
\begin{eqnarray}
\frac{m}{(\alpha d)^{2/3}}\ll 1.
\end{eqnarray}

\subsection{Careful treatment of the holonomy}

\medskip
\noindent
Starting from the maximally symmetric pairwise gap condition, we can restore the
projector and average over the color indices as follows. With
\begin{eqnarray}
s\equiv\sqrt{\tilde k_0}>0,
\qquad
\tilde k_0=k_0+m,
\end{eqnarray}
the scalar gap equation is written as
\begin{eqnarray}
\frac{d}{s}\,{\cal R}(s;U)
=
\frac{k_0}{\alpha}
=
\frac{s^2-m}{\alpha},
\label{eq:gap-projector-averaged}
\end{eqnarray}
where the projector-weighted holonomy kernel is
\begin{eqnarray}
{\cal R}(s;U)
&=&
\frac{1}{N^2-1}
\sum_{i,j=1}^{N}
P_{ij,ji}\,
\frac{\sinh(\beta s)}
{\cosh(\beta s)-\cos(\theta_i-\theta_j)}.
\label{Rkernel_projected}
\end{eqnarray}
Here
\begin{eqnarray}
P_{ij,ji}=1-\frac{1}{N}\delta_{ij},
\qquad
\sum_{i,j}P_{ij,ji}=N^2-1.
\end{eqnarray}

\medskip
\noindent
The thermal kernel admits the standard low-temperature expansion
\begin{eqnarray}
\frac{\sinh(\beta s)}
{\cosh(\beta s)-\cos(\theta_i-\theta_j)}
=
1
+
2\sum_{p\geq 1}
e^{-p\beta s}\cos\big(p(\theta_i-\theta_j)\big).
\end{eqnarray}
Therefore
\begin{eqnarray}
{\cal R}(s;U)
=
1
+
2\sum_{p\geq 1}
{\cal C}_p(U)\,e^{-p\beta s},
\end{eqnarray}
with the projector-weighted holonomy moments
\begin{eqnarray}
{\cal C}_p(U)
&=&
\frac{1}{N^2-1}
\sum_{i,j=1}^{N}
P_{ij,ji}
\cos\big(p(\theta_i-\theta_j)\big)
\nonumber\\
&=&
\frac{|\Tr U^p|^2-1}{N^2-1}.
\label{Cp_def}
\end{eqnarray}
Thus the first correction is controlled not by a single \(\cos\phi\), but by the
projector-weighted moment \({\cal C}_1(U)\). To first thermal order,
\begin{eqnarray}
{\cal R}(s;U)
=
1+2{\cal C}_1(U)e^{-\beta s}
+
{\cal O}(e^{-2\beta s}).
\end{eqnarray}

\medskip
\noindent
The zero-temperature gap \(s_0\) is determined by \eqref{eq:zeroT}.
Writing \(s=s_0+\delta s\), with
\(\delta s={\cal O}(e^{-\beta s_0})\), we obtain
\begin{eqnarray}
\delta s
=
\frac{2s_0(s_0^2-m)}{3s_0^2-m}\,
{\cal C}_1(U)\,e^{-\beta s_0}
+
{\cal O}(e^{-2\beta s_0}).
\label{delta_s_projected}
\end{eqnarray}
Hence
\begin{eqnarray}
s
&=&
s_0\left[
1+
2\frac{s_0^2-m}{3s_0^2-m}\,
{\cal C}_1(U)e^{-\beta s_0}
+
{\cal O}(e^{-2\beta s_0})
\right],
\\
\tilde k_0
&=&
s_0^2\left[
1+
4\frac{s_0^2-m}{3s_0^2-m}\,
{\cal C}_1(U)e^{-\beta s_0}
+
{\cal O}(e^{-2\beta s_0})
\right].
\end{eqnarray}

\medskip
\noindent
This makes the role of the holonomy explicit. For the collapsed, deconfined
saddle \(U=\mathbf 1_N\), one has \({\cal C}_1=1\). For a center-symmetric
confining holonomy, \(\Tr U=0\), and therefore
\begin{eqnarray}
{\cal C}_1
=
-\frac{1}{N^2-1},
\end{eqnarray}
which vanishes only in the strict large-\(N\) limit. Thus the first thermal
correction is exactly absent at leading planar order in the center-symmetric
phase, but at finite \(N\) there remains a projector-induced correction of order
\(1/N^2\).

\subsection{Bosonic effective action}

\noindent
Let us start from the localized action:
\begin{eqnarray}
S_{\rm loc}[X,k]
&=&
\frac{N}{2}\sum_{a=1}^{d}\sum_{n\in\mathbb{Z}}\sum_{i,j=1}^N
X_a^{ij}(-n)\,
\Bigg[
\Big(\omega_n+\frac{\theta_i-\theta_j}{\beta}\Big)^2\,P_{ij,kl}
\;+\;m\,P_{ij,kl}
\Bigg]\,
X_a^{kl}(n)
\nonumber\\
&&\qquad
+\;\frac{N}{2}\int_{0}^{\beta}\!dt\ k_{\mu\rho}(t)\,Y_{\mu\rho}(t)
\;+\;\frac{N}{4\alpha}\int_{0}^{\beta}\!dt\
\mu_{\mu\rho\nu\sigma}\,k_{\mu\rho}(t)\,k_{\nu\sigma}(t),
\label{Sloc-start-dk}
\end{eqnarray}
with
\begin{eqnarray}
Y_{\mu\rho}(t)
=
\sum_{a=1}^{d}x_a^{\mu}(t)\,x_a^{\rho}(t),
\end{eqnarray}
and
\begin{eqnarray}
k_{\mu\rho}
=
\Lambda_{\mu}^{\,ij}k_{ij,kl}\Lambda_{\rho}^{\,kl},
\qquad
k_{ij,kl}(t)
=
\frac{1}{4}k_{\mu\rho}(t)
\Lambda_{\mu}^{\,ji}\Lambda_{\rho}^{\,lk},
\qquad
\mu_{\mu\rho\nu\sigma}\lambda_{\nu\sigma\alpha\beta}
=
\delta_{\mu\alpha}\delta_{\rho\beta}.
\end{eqnarray}

\medskip
\noindent
We expand around the time--independent saddle
\begin{eqnarray}
k_{\mu\rho}(t)
=
k_{0\,\mu\rho}
+
\delta k_{\mu\rho}(t),
\qquad
k_{0\,\mu\rho}
=
\hbox{const.}
\label{k-split}
\end{eqnarray}
It is important to remember that \(k_0\) is a stationary point of the auxiliary-field
effective action obtained after integrating out the coordinate matrices:
\begin{eqnarray}
e^{-S_{\rm eff}[k]}
=
\int{\cal D}X\,
e^{-S_{\rm loc}[X,k]},
\qquad
\left.
\frac{\delta S_{\rm eff}[k]}
{\delta k_{\mu\rho}(t)}
\right|_{k=k_0}
=
0.
\end{eqnarray}
It is not, in general, a stationary point of
\(S_{\rm loc}[X,k]\) for every fixed configuration \(X\).

\medskip
\noindent
In the maximally symmetric sector one has
\(k_{0ij,kl}=k_0P_{ij,kl}\), or equivalently
\(k_{0\mu\rho}=2k_0\delta_{\mu\rho}\).  Hence the \(k_0\)-piece in
\(\frac{N}{2}\int kY\) produces precisely a shift of the quadratic
kernel by \(k_0P\).  Therefore we rewrite
\begin{eqnarray}
S_{\rm loc}[X,k_0+\delta k]
&=&
S_{\rm MHO}[X;\theta,m_{\rm eff}]
\;+\;S_k[k_0]
\;+\;S_{\rm int}[X,\delta k]
\;+\;S_{\delta k}^{(2)}[\delta k],
\label{split-1}
\end{eqnarray}
with the \emph{effective mass}
\begin{eqnarray}
m_{\rm eff}
=
s^2
=
\tilde{k}_0
=
m+k_0,
\label{s0-def}
\end{eqnarray}
and
\begin{eqnarray}
S_{\rm MHO}[X;\theta,m_{\rm eff}]
&=&
\frac{N}{2}
\sum_{a=1}^{d}
\sum_{n\in\mathbb{Z}}
\sum_{i,j=1}^N
X_a^{ij}(-n)\,
\Bigg[
\Big(\omega_n+\frac{\theta_i-\theta_j}{\beta}\Big)^2P_{ij,kl}
\;+\;m_{\rm eff}P_{ij,kl}
\Bigg]\,
X_a^{kl}(n).
\nonumber\\
\end{eqnarray}
The three remaining contributions to \eqref{split-1} are:

\begin{itemize}

\item \textbf{Zeroth--order term.}
Evaluating the purely \(k_0\)--dependent piece gives
\begin{eqnarray}
S_k[k_0]
&\equiv&
\frac{N}{4\alpha}
\int_{0}^{\beta}\!dt\;
\mu_{\mu\rho\nu\sigma}\,
k_{0\,\mu\rho}\,
k_{0\,\nu\sigma}
\nonumber\\
&=&
-\frac{\beta}{8\alpha}\,
(N^2-1)\,
k_0^2,
\end{eqnarray}
where we used
\(\mu_{\mu\rho\nu\sigma}\delta_{\nu\sigma}
=-(8N)^{-1}\delta_{\mu\rho}\)
and the fact that
\(k_{0\,\mu\rho}\propto\delta_{\mu\rho}\)
lives in the adjoint sector.

\item \textbf{Linear order term.}
The terms linear in the fluctuation \(\delta k\) combine into
\begin{eqnarray}
S_{\rm int}[X,\delta k]
&\equiv&
\frac{N}{2}
\int_{0}^{\beta}\!dt\;
\delta k_{\mu\rho}(t)\,
Y_{\mu\rho}(t)
\nonumber\\
&&
+
\frac{N}{2\alpha}
\int_{0}^{\beta}\!dt\;
\mu_{\mu\rho\nu\sigma}\,
k_{0\,\nu\sigma}\,
\delta k_{\mu\rho}(t)
\nonumber\\
&=&
\frac{N}{2}
\int_{0}^{\beta}\!dt\;
\delta k_{\mu\rho}(t)\,
\tilde Y_{\mu\rho}(t),
\end{eqnarray}
with
\begin{eqnarray}
\tilde Y_{\mu\rho}
\;\equiv\;
Y_{\mu\rho}
+
\frac{1}{\alpha}
\mu_{\mu\rho\nu\sigma}
k_{0\,\nu\sigma}
\;=\;
Y_{\mu\rho}
-
\frac{k_0}{4\alpha N}
\delta_{\mu\rho}.
\label{lin-compact}
\end{eqnarray}
Since \(k_0\) is a saddle point of \(S_{\rm eff}[k]\), stationarity
requires the expectation value of the coefficient of \(\delta k\) to
vanish.  The resulting gap equation can therefore be written as the
self--consistency condition
\begin{eqnarray}
\left\langle
Y_{\mu\rho}
\right\rangle_{k_0}
=
-\frac{1}{\alpha}
\mu_{\mu\rho\nu\sigma}
k_{0\,\nu\sigma}
=
\frac{k_0}{4\alpha N}
\delta_{\mu\rho}.
\label{saddle-point-cond}
\end{eqnarray}
Accordingly,
\begin{eqnarray}
S_{\rm int}[X,\delta k]
=
\frac{N}{2}
\int_{0}^{\beta}\!dt\;
\delta k_{\mu\rho}(t)
\left[
Y_{\mu\rho}(t)
-
\left\langle
Y_{\mu\rho}
\right\rangle_{k_0}
\right].
\end{eqnarray}
The linear coupling does not vanish identically for each fluctuating
configuration \(X\).  Rather,
\begin{eqnarray}
\left\langle
\tilde Y_{\mu\rho}
\right\rangle_{k_0}
=
0,
\end{eqnarray}
so that the linear term vanishes after the coordinate matrices have
been integrated out, as required by the stationarity of
\(S_{\rm eff}[k]\).

\item \textbf{Second--order term.}
The bare quadratic fluctuation term in the localized action is
\begin{eqnarray}
S_{\delta k}^{(2)}[\delta k]
&\equiv&
\frac{N}{4\alpha}
\int_{0}^{\beta}\!dt\;
\mu_{\mu\rho\nu\sigma}\,
\delta k_{\mu\rho}(t)\,
\delta k_{\nu\sigma}(t).
\label{Sdk2}
\end{eqnarray}
After the \(X_a\) matrices have been integrated out, their connected
correlators contribute to the quadratic kernel of the auxiliary
effective action.  Expanding around \(k_0\), one obtains
\begin{eqnarray}
S_{\rm eff}[k_0+\delta k]
&=&
S_{\rm eff}[k_0]
+
\frac{1}{2}
\int_{0}^{\beta}\!dt
\int_{0}^{\beta}\!dt'\;
\delta k_{\mu\rho}(t)\,
\Gamma^{(2)}_{\mu\rho,\nu\sigma}(t,t')\,
\delta k_{\nu\sigma}(t')+
\mathcal{O}(\delta k^3),\nonumber\\
\end{eqnarray}
where
\begin{eqnarray}
\Gamma^{(2)}_{\mu\rho,\nu\sigma}(t,t')
&=&
\frac{N}{2\alpha}
\mu_{\mu\rho\nu\sigma}\,
\delta(t-t')-
\frac{N^2}{4}
\left\langle
\tilde Y_{\mu\rho}(t)
\tilde Y_{\nu\sigma}(t')
\right\rangle_{k_0,c}. \label{hessian}
\end{eqnarray}
The one--loop integration over the auxiliary fluctuation is therefore
controlled by the full Hessian \(\Gamma^{(2)}\):
\begin{eqnarray}
Z
\simeq
e^{-S_{\rm eff}[k_0]}
\left[
\det\Gamma^{(2)}
\right]^{-1/2}.
\end{eqnarray}
Thus, integrating out \(\delta k\) at fixed \(X\) would not yield only
an irrelevant constant: it would generate interactions involving
\(Y-\langle Y\rangle_{k_0}\).  These interactions describe the
subleading corrections to the Gaussian saddle.

\end{itemize}

\medskip
\noindent
We now estimate the two contributions to the auxiliary-field Hessian
\eqref{hessian}.  The first is the bare kernel inherited from the
quadratic \(k^2\) term in the localized action, while the second is the
polarization kernel generated by the connected fluctuations of the
coordinate matrices.

\medskip
\noindent
Consider a time--independent fluctuation in the maximally symmetric
channel,
\begin{eqnarray}
k=k_0+\delta k,
\qquad
s^2=m+k_0.
\end{eqnarray}
In this channel, and suppressing the holonomy dependence and common
adjoint multiplicity, the effective action takes the schematic form

\begin{eqnarray}
S_{\rm eff}(k)
=
(N^2-1)
\left[
-\frac{\beta}{8\alpha}k^2
+
\frac{d}{2}
\sum_{n\in\mathbb Z}
\log\!\left(\omega_n^2+m+k\right)
\right].
\end{eqnarray}
Expanding around \(k_0\) gives
\begin{eqnarray}
S_{\rm eff}(k_0+\delta k)
&=&
S_{\rm eff}(k_0)
+
(N^2-1)\delta k
\left[
-\frac{\beta k_0}{4\alpha}
+
\frac{d}{2}
\sum_{n\in\mathbb Z}
\frac{1}{\omega_n^2+s^2}
\right]
\nonumber\\
&&
-
(N^2-1)(\delta k)^2
\left[
\frac{\beta}{8\alpha}
+
\frac{d}{4}
\sum_{n\in\mathbb Z}
\frac{1}{(\omega_n^2+s^2)^2}
\right]
+
\mathcal{O}(\delta k^3).
\label{tr-log}
\end{eqnarray}
\medskip
\noindent
The coefficient of the linear fluctuation vanishes by the static
auxiliary-field saddle equation
\begin{eqnarray}
-\frac{\beta k_0}{4\alpha}
+
\frac{d}{2}
\sum_{n\in\mathbb Z}
\frac{1}{\omega_n^2+s^2}
=
0\qquad\Longleftrightarrow\qquad 
\frac{k_0}{\alpha}
=
\frac{d}{s}
\coth\!\left(\frac{\beta s}{2}\right).
\end{eqnarray}
In the zero--temperature limit, and upon using \(k_0=s^2-m\), this
saddle equation is precisely equivalent to the cubic gap equation
\eqref{eq:zeroT}.

\medskip
\noindent
The two terms in the second line of \eqref{tr-log} are precisely the two
contributions to the static Hessian \eqref{hessian}.  The coordinate
polarization is governed by
\begin{eqnarray}
\sum_{n\in\mathbb Z}
\frac{1}{(\omega_n^2+s^2)^2}
&=&
\frac{\beta}{4s^3}
\coth\!\left(\frac{\beta s}{2}\right)
+
\frac{\beta^2}{8s^2}
\operatorname{csch}^2\!\left(\frac{\beta s}{2}\right)
\nonumber\\
&=&
\frac{\beta}{4s^3}
\left[
1+\mathcal{O}\!\left(\beta s\,e^{-\beta s}\right)
\right],
\qquad
\beta s\gg1.
\end{eqnarray}
It therefore scales parametrically as
\begin{eqnarray}
\Gamma_{{\rm stat},X}^{(2)}
\sim
\frac{d\beta}{s^3},
\end{eqnarray}
up to its numerical coefficient, sign, and adjoint tensor eigenvalue.

\medskip
\noindent
The bare contribution scales as \(\beta/\alpha\).  To leading
low--temperature order, the cubic gap equation \eqref{eq:zeroT} gives
\begin{eqnarray}
\Gamma_{{\rm stat},{\rm bare}}^{(2)}
\sim
\frac{d\beta}{s^3}\,
\frac{s^2}{s^2-m},
\end{eqnarray}
again up to the corresponding numerical coefficient and tensor
eigenvalue.  Thus, provided \(s^2-m\) is of order \(s^2\), both parts of
the static Hessian are parametrically of order \(d\beta/s^3\).

\medskip
\noindent For a stable eigenmode of the complete tensor kernel, understood along
the appropriate steepest--descent contour of the Hubbard--Stratonovich
integral, one may therefore write

\begin{eqnarray}
\Gamma_{\rm stat}^{(2)}
\sim
\frac{d\beta}{s^3}\,
{\cal C}\!\left(\frac{m}{s^2}\right)
\left[
1+\mathcal{O}\!\left(\beta s\,e^{-\beta s}\right)
\right],
\end{eqnarray}
where the dimensionless coefficient \({\cal C}\) combines the bare and
polarization contributions.  This is a parametric estimate of an
eigenvalue of the full Hessian, rather than an equality for the complete
tensor-valued kernel.

\medskip
\noindent Provided that \({\cal C}\neq0\) and that no parametrically soft mode is
present, the variance of the corresponding static fluctuation along the
stable steepest--descent direction is controlled parametrically by the
inverse Hessian:

\begin{eqnarray}
\left\langle(\delta k)^2\right\rangle
\sim
\frac{s^3}{d\beta}\,
\frac{1}{{\cal C}(m/s^2)}.
\end{eqnarray}
Consequently,
\begin{eqnarray}
\frac{
\sqrt{\left\langle(\delta k)^2\right\rangle}
}{s^2}
\sim
\frac{1}{\sqrt{d\beta s}}\,
\frac{1}{\sqrt{{\cal C}(m/s^2)}}.
\end{eqnarray}
The static auxiliary fluctuations are therefore suppressed at large
\(d\), with an additional suppression in the low--temperature regime
\(\beta s\gg1\).

\medskip
\noindent
Consequently, at leading order in the combined large--\(d\) and
low--temperature saddle expansion, one freezes the auxiliary field at
\(k=k_0\).  The effective action governing the coordinate matrices then
reduces to
\begin{eqnarray}
S_{\rm eff}[X;\theta]\big|_{\rm leading\ saddle}
&=&
S_{\rm MHO}[X;\theta,m_{\rm eff}]
\nonumber\\
&=&
N\int_{0}^{\beta}\!dt\;
\mathrm{Tr}
\left[
\frac{1}{2}(D_tX_a)^2
+
\frac{m_{\rm eff}}{2}X_a^2
\right].
\end{eqnarray}

\medskip
\noindent
In particular, the \emph{dominant large--\(d\) contribution} is given
by a gauged matrix harmonic oscillator with the dynamically generated
effective mass
\begin{eqnarray}
m_{\rm eff}
=
s^2
=
\tilde{k}_0
=
m+k_0.
\end{eqnarray}
Thus, at leading order in the large--\(d\) expansion, the original
commutator--squared interaction is encoded in a self--consistent mass
term, while fluctuations of \(\delta k\) generate the subleading
non--Gaussian corrections.

\subsection{Free energy}

At the large--$d$ saddle, the localized theory reduces to a gauged matrix harmonic
oscillator with effective mass
\begin{eqnarray}
s^2 \equiv m_{\rm eff} = m + k_0,
\end{eqnarray}
where $k_0$ is determined by the gap equation. Integrating out the Gaussian matrices $X_a$ yields an effective action for the
Polyakov loop eigenvalues $\{\theta_i\}$.

We start from the action 
\begin{eqnarray}
S_{\rm MHO}[X;\theta,m_{\rm eff}]&=&
\frac{N}{2}\sum_{a=1}^{d}\sum_{n\in\mathbb{Z}}\sum_{i,j=1}^N
X_a^{ij}(-n)\,
\Bigg[
\Big(\omega_n+\frac{\phi_{ij}}{\beta}\Big)^2\,P_{ij,kl}
\;+\;s^2\,P_{ij,kl}
\Bigg]\,
X_a^{kl}(n),\nonumber\\
\end{eqnarray}
with
\begin{eqnarray}
\phi_{ij}=\theta_i-\theta_j.
\end{eqnarray}
Each adjoint mode $(i,j)$ experiences shifted Matsubara frequencies \(\omega_n \;\longrightarrow\; \omega_n+{\phi_{ij}}/{\beta}\), and thus for each matrix element this produces the Gaussian determinant
\begin{eqnarray}
\prod_{n\in\mathbb{Z}}
\Bigg[
\Big(\omega_n+\frac{\phi_{ij}}{\beta}\Big)^2+s^2
\Bigg]\;\propto\;
\cosh(\beta s)-\cos\phi_{ij},\qquad \omega_n=\frac{2\pi n}{\beta}.
\end{eqnarray}
The result is derived by expressing the determinant in logarithmic form and performing the Matsubara sum \eqref{Matsu}.  Explicitly, set
\begin{eqnarray}
x=\beta s,
\qquad
\phi=\phi_{ij}.
\end{eqnarray}
Up to an overall \(\beta\)-dependent normalization independent of \(s\) and \(\phi\),
we have
\begin{eqnarray}
\prod_{n\in\mathbb Z}
\left[
\left(\omega_n+\frac{\phi}{\beta}\right)^2+s^2
\right]
\;\propto\;
\prod_{n\in\mathbb Z}
\left[
(2\pi n+\phi)^2+x^2
\right].
\end{eqnarray}
Let
\begin{eqnarray}
{\cal D}(x,\phi)
=
\prod_{n\in\mathbb Z}
\left[
(2\pi n+\phi)^2+x^2
\right].
\end{eqnarray}
Then
\begin{eqnarray}
\frac{\partial}{\partial x}\log{\cal D}(x,\phi)
&=&
2x\sum_{n\in\mathbb Z}
\frac{1}{(2\pi n+\phi)^2+x^2}.
\end{eqnarray}
Using the standard Matsubara sum \eqref{Matsu} in the form 
\begin{eqnarray}
\sum_{n\in\mathbb Z}
\frac{1}{(2\pi n+\phi)^2+x^2}
=
\frac{1}{2x}\,
\frac{\sinh x}{\cosh x-\cos\phi},
\end{eqnarray}
we obtain
\begin{eqnarray}
\frac{\partial}{\partial x}\log{\cal D}(x,\phi)
=
\frac{\sinh x}{\cosh x-\cos\phi}
=
\frac{\partial}{\partial x}
\log\!\big(\cosh x-\cos\phi\big).
\end{eqnarray}
Hence
\begin{eqnarray}
{\cal D}(x,\phi)
\;\propto\;
\cosh x-\cos\phi.
\end{eqnarray}
Restoring \(x=\beta s\), this gives
\begin{eqnarray}
\prod_{n\in\mathbb{Z}}
\Bigg[
\Big(\omega_n+\frac{\phi_{ij}}{\beta}\Big)^2+s^2
\Bigg]
\;\propto\;
\cosh(\beta s)-\cos\phi_{ij}.
\end{eqnarray}
The proportionality constant is independent of \(s\) and \(\phi_{ij}\), and
therefore contributes only an irrelevant additive constant to the free energy.

\medskip
\noindent
Therefore, after integrating out the coordinate modes \(X_a^{ij}(n)\), the bosonic
matrix harmonic--oscillator partition function is, up to an overall normalization,

\begin{eqnarray}
\log Z_{\rm MHO}(\theta;s)
&=&
-\frac{d}{2}\sum_{i,j=1}^{N}
P_{ij,ji}\,
\log\!\Big(\cosh(\beta s)-\cos\phi_{ij}\Big)
\;+\;\text{const}.
\end{eqnarray}
The corresponding free energy is
\begin{eqnarray}
F_{\rm MHO}(\theta;s)
&=&
-\frac{1}{\beta}\log Z_{\rm MHO}
=
\frac{d}{2\beta}\sum_{i,j=1}^{N}
P_{ij,ji}\,
\log\!\Big(\cosh(\beta s)-\cos(\theta_i-\theta_j)\Big)
\;+\;\text{const}.\nonumber\\
\end{eqnarray}
\subsection{Extent of space}

The extent of space is defined as
\begin{eqnarray}
R^2
=
\frac{1}{N\beta}
\Big\langle \int_0^{\beta}\! dt\ \mathrm{Tr}X_a(t)\,X_a(t)\Big\rangle.
\end{eqnarray}
For the Gaussian theory obtained at the large--$d$ saddle point, the momentum--space
propagator in the presence of holonomy is
\begin{eqnarray}
\Big\langle X_a^{ij}(n)\,X_a^{kl}(-n)\Big\rangle
=
\frac{1}{N}\,
\frac{P_{ij,kl}}
{\Big(\omega_n+\tfrac{\phi_{ij}}{\beta}\Big)^2+s^2}.
\end{eqnarray}
The equal--time (coincident) correlator follows again from the standard Matsubara sum
\begin{eqnarray}
\sum_{n\in\mathbb{Z}}
\frac{1}{\Big(\omega_n+\tfrac{\phi_{ij}}{\beta}\Big)^2+s^2}
=
\frac{\beta}{2s}\,
\frac{\sinh(\beta s)}{\cosh(\beta s)-\cos\phi_{ij}}.
\end{eqnarray}
Therefore, the extent of space is given explicitly by 
\begin{eqnarray}
R^2(\theta;s)
&=&
\frac{d}{N^2}\,
\sum_{i,j}
P_{ij,ji}\;
\frac{1}{2s}\,
\frac{\sinh(\beta s)}
{\cosh(\beta s)-\cos\phi_{ij}} .
\label{R2-holonomy}
\end{eqnarray}
\medskip
\noindent
As a consistency check, the same quantity can be extracted from the free energy
of the matrix harmonic oscillator via
\begin{eqnarray}
\frac{\partial F_{\rm MHO}}{\partial s^2}
=
\frac{N^2}{2}\,R^2,
\end{eqnarray}
up to the same normalization conventions used in the action.

\subsection{Consistency between the saddle equation and the extent of space}

At the large--\(d\) saddle point, the auxiliary effective action satisfies
\eqref{saddle-point-cond}, namely
\begin{eqnarray}
\left\langle Y_{\mu\rho}\right\rangle_{k_0}
=
-\frac{1}{\alpha}\,
\mu_{\mu\rho\nu\sigma}\,
k_{0\,\nu\sigma}
=
\frac{k_0}{4\alpha N}\,
\delta_{\mu\rho}.
\label{Y-saddle}
\end{eqnarray}
This relation follows from the stationarity of the auxiliary-field
effective action, or equivalently from the vanishing expectation value
of the coefficient of the term linear in \(\delta k\).

Evaluating the same expectation value in the Gaussian saddle and invoking
rotational symmetry in adjoint space, we obtain
\begin{eqnarray}
\left\langle Y_{\mu\rho}\right\rangle_{k_0}
=
\delta_{\mu\rho}\,
\frac{1}{N^2-1}
\sum_{\lambda}
\left\langle Y_{\lambda\lambda}\right\rangle_{k_0}
=
\delta_{\mu\rho}\,
\frac{N}{2}\,
\frac{R^2}{N^2-1}.
\label{Y-expect}
\end{eqnarray}
Equating \eqref{Y-saddle} and \eqref{Y-expect} yields
\begin{eqnarray}
R^2
=
\left(1-\frac{1}{N^2}\right)
\frac{k_0}{2\alpha}.
\label{R2-k0}
\end{eqnarray}
Using \(\sum_{i,j}P_{ij,ji}=N^2-1\), the zero--temperature Gaussian
result (see below)
\begin{eqnarray}
R^2
=
\left(1-\frac{1}{N^2}\right)
\frac{d}{2s}
\end{eqnarray}
immediately implies
\begin{eqnarray}
\frac{k_0}{\alpha}
=
\frac{d}{s}.
\end{eqnarray}
Upon using \(k_0=s^2-m\), this is precisely equivalent to the
saddle--point gap equation \eqref{eq:zeroT}.

Thus \eqref{Y-saddle} is not an additional assumption.  Rather, it
expresses the consistency between the expectation-value saddle equation
of the auxiliary effective action and the extent of space computed in
the emergent Gaussian theory.  At zero temperature, the two evaluations
agree at leading order in the large--\(d\) saddle expansion, while in the
low--temperature regime the thermal corrections are exponentially
suppressed.  This is consistent with the gauged matrix harmonic
oscillator providing the leading large--\(d\), low--temperature
description of the original BFSS model.

\subsection{Low--temperature evaluation and mass regimes}

\medskip
\noindent
In the low--temperature confined phase ($\beta s\gg1$), the dominant
holonomy saddle is uniform, $\rho(\theta)=1/2\pi$.  For a general function
\(f(\phi_{ij})\), the adjoint projector
\(P_{ij,ji}=1-\delta_{ij}/N\) gives
\begin{eqnarray}
\frac{1}{N^2}\sum_{i,j}P_{ij,ji}\,f(\phi_{ij})
&=&
\frac{1}{N^2}\sum_{i,j}f(\phi_{ij})
-\frac{1}{N^2}f(0)
\nonumber\\
&\longrightarrow&
\int_{-\pi}^{\pi}\frac{d\phi}{2\pi}\,f(\phi)
-\frac{1}{N^2}f(0).
\label{uniform-projector-average}
\end{eqnarray}

\medskip
\noindent
For the matrix harmonic oscillator determinant, we take
\begin{eqnarray}
f(\phi)=\log\!\big(\cosh x-\cos\phi\big),
\qquad
x=\beta s.
\end{eqnarray}
The required quantities are
\begin{eqnarray}
\int_{-\pi}^{\pi}\frac{d\phi}{2\pi}\,
\log\!\big(\cosh x-\cos\phi\big)
&=&
x-\log 2,
\qquad x>0,
\\
f(0)
&=&
\log(\cosh x-1)
\nonumber\\
&=&
x-\log 2+2\log(1-e^{-x}).
\end{eqnarray}
Hence, Eq.~\eqref{uniform-projector-average} becomes
\begin{eqnarray}
\frac{1}{N^2}\sum_{i,j}P_{ij,ji}\,
\log\!\big(\cosh x-\cos\phi_{ij}\big)
&\longrightarrow&
\Big(1-\frac{1}{N^2}\Big)(x-\log 2)
-\frac{2}{N^2}\log(1-e^{-x})
\nonumber\\
&=&
\Big(1-\frac{1}{N^2}\Big)(x-\log 2)
+\mathcal{O}\!\left(\frac{e^{-x}}{N^2}\right).
\end{eqnarray}

\medskip
\noindent
Since the total free energy contains an overall factor \(N^2\), its leading
low--temperature form is therefore
\begin{eqnarray}
F_{\rm MHO}
&=&
(N^2-1)\,
\frac{d}{2}\,s
+\text{const}
+\mathcal{O}\!\big(\beta^{-1}e^{-\beta s}\big),
\qquad
(\beta s\gg1).
\label{F-lowT}
\end{eqnarray}
The additive constant is independent of \(s\).

\medskip
\noindent
Similarly, in the same uniform low--temperature saddle, the extent of space
is
\begin{eqnarray}
R^2
&=&
\Big(1-\frac{1}{N^2}\Big)\,
\frac{d}{2s}
+\mathcal{O}\!\big(e^{-\beta s}\big),
\qquad
(\beta s\gg1).
\label{R2-lowT}
\end{eqnarray}
To the displayed order, Eqs.~\eqref{F-lowT} and \eqref{R2-lowT} satisfy
\begin{eqnarray}
\frac{\partial F_{\rm MHO}}{\partial s^2}
=
\frac{N^2}{2}\,R^2,
\end{eqnarray}
as expected.

\medskip
\noindent
The dependence on the bare mass \(m\) enters entirely through the
zero--temperature gap equation \eqref{eq:zeroT}, which yields two distinct
regimes:

\medskip
\noindent\emph{(i) Small mass, $m\ll(\alpha d)^{2/3}$.}
Writing $\varepsilon=m/(\alpha d)^{2/3}\ll1$, one finds
\begin{eqnarray}
s_0
&=&
(\alpha d)^{1/3}
\left(1+\frac{\varepsilon}{3}+\mathcal{O}(\varepsilon^2)\right).
\end{eqnarray}
Hence,
\begin{eqnarray}
F_{\rm MHO}
&=&
(N^2-1)\,
\frac{d}{2}\,(\alpha d)^{1/3}
\left(1+\frac{\varepsilon}{3}+\cdots\right),
\\
R^2
&=&
\Big(1-\frac{1}{N^2}\Big)\,
\frac{d^{2/3}}{2\alpha^{1/3}}
\left(1-\frac{\varepsilon}{3}+\cdots\right).
\end{eqnarray}
In this regime the extent of space scales as $R^2\sim d^{2/3}$ and the free energy as
$F\sim d^{4/3}$, reflecting a genuinely interacting large--$d$ saddle.

\medskip
\noindent\emph{(ii) Large mass, $m\gg(\alpha d)^{2/3}$.}
The positive root admits
\begin{eqnarray}
s_0
&=&
\sqrt{m}+\frac{\alpha d}{2m}
+\mathcal{O}(m^{-5/2}),
\end{eqnarray}
leading to
\begin{eqnarray}
F_{\rm MHO}
&=&
(N^2-1)\,
\frac{d}{2}\,\sqrt{m}
\left(1+\frac{\alpha d}{2m^{3/2}}+\cdots\right),
\\
R^2
&=&
\Big(1-\frac{1}{N^2}\Big)\,
\frac{d}{2\sqrt{m}}
\left(1-\frac{\alpha d}{2m^{3/2}}+\cdots\right).
\end{eqnarray}
Here the matrices are tightly localized near the origin, $R^2\sim d/\sqrt{m}$, and the theory
approaches a truly Gaussian regime dominated by the quadratic term.

\medskip
\noindent
Thus, at low temperature the free energy and the extent of space are entirely controlled by the
gap $s_0(m,d)$: the small--$m$ regime retains strong BFSS interactions with $d^{4/3}$ scaling,
while large $m$ smoothly suppresses both $R^2$ and interaction effects.

\section{Holonomy physics at low--\(T\)}\label{section5}

\subsection{Holonomy effective action}

\medskip
\noindent
After the reduction to the gauged matrix harmonic oscillator, the remaining gauge
dynamics is encoded in the holonomy eigenvalues \(\theta_i\), or equivalently in
the Polyakov moments
\begin{eqnarray}
u_n=\frac{1}{N}\sum_{j=1}^{N}e^{in\theta_j},
\qquad
u_{-n}=u_n^\ast,
\qquad
q=e^{-\beta s}.
\end{eqnarray}
Integrating out the \(d\) Gaussian coordinate matrices gives an attractive
contribution to the holonomy potential, while the Vandermonde determinant from
the Haar measure gives a universal repulsive contribution. Dropping
\(\theta\)-independent constants, these are
\begin{eqnarray}
&&S_X[\theta]
=
-\,d\,N^2\sum_{n\ge1}\frac{q^n}{n}\,|u_n|^2,
\label{SX}
\\
&&S_{\rm Vdm}[\theta]
=
N^2\sum_{n\ge1}\frac{1}{n}\,|u_n|^2.
\label{SVdm}
\end{eqnarray}
Hence the holonomy effective action is
\begin{eqnarray}
S_{\rm hol}[\theta]
&=&
N^2\sum_{n\ge1}
\frac{1-dq^n}{n}\,|u_n|^2
+\hbox{const},
\qquad
q=e^{-\beta s}.
\label{Seff-hol}
\end{eqnarray}
The Fourier derivation of \eqref{SX}, \eqref{SVdm}, and \eqref{Seff-hol} is given
in Appendix~\ref{appendix-holonomy-effective-action}.

\subsection{Confined phase and critical point}

\noindent
The coefficient \(1-dq^n\) displays the competition between the repulsive
Vandermonde measure and the attractive coordinate determinant. The first Polyakov
mode becomes marginal when \(dq=1\), giving the Gaussian large--\(d\)
deconfinement, or Hagedorn, criterion.

\noindent
Indeed, the holonomy effective action \eqref{Seff-hol} is positive definite provided
\begin{eqnarray}
1-d\,q^n>0 \qquad \forall\,n\ge1 .
\end{eqnarray}
In this regime the unique minimum is
\begin{eqnarray}
u_n=0 \qquad (n\ge1),
\end{eqnarray}
corresponding to the uniform (center--symmetric) eigenvalue density
\begin{eqnarray}
\rho(\theta)=\frac{1}{2\pi},\qquad \theta\in[-\pi,\pi]. \label{rho-eq1}
\end{eqnarray}
Since the most stringent condition comes from the lowest mode, confinement reduces to
\begin{eqnarray}
d\,q<1.
\end{eqnarray}

\medskip
\noindent
The confined saddle becomes marginal when the coefficient of the $n=1$ mode vanishes, namely
\begin{eqnarray}
1-d\,q=0.
\end{eqnarray}
Recall that
\begin{eqnarray}
&&q=e^{-\beta s},\qquad 
s^2=m_{\rm eff}=m+k_0,
\nonumber\\
&&s=s_0\bigg[1+\frac{2}{3}\,
\frac{s_0^2-m}{3s_0^2-m}\,
\langle\cos\phi\rangle\,e^{-\beta s_0}
+\cdots\bigg],
\nonumber\\
&&s_0=(\alpha d)^{1/3}
\bigg[1+\frac{\epsilon}{3}
-\frac{\epsilon^3}{81}
+\frac{\epsilon^4}{243}
+\cdots\bigg],
\qquad
\epsilon=\frac{m}{(\alpha d)^{2/3}} .
\end{eqnarray}
Here $s_0$ is determined by the zero--temperature gap equation \eqref{eq:zeroT}.

\noindent
In the center--symmetric phase one has $\langle\cos\phi\rangle=0$, so the leading thermal correction to $s$
vanishes and $s\simeq s_0$. Setting $\alpha=1$ at the end of the calculation, the critical temperature is therefore fixed by
\begin{eqnarray}
\beta_c\,s_0=\log d
\qquad\Longleftrightarrow\qquad
T_c=\frac{s_0}{\log d}.\label{Tc-general}
\end{eqnarray}
Explicitly, we have
\begin{eqnarray}
T_c
=\frac{d^{1/3}}{\log d}
\bigg[1+\frac{\epsilon}{3}
-\frac{\epsilon^3}{81}
+\frac{\epsilon^4}{243}
+\cdots\bigg].
\end{eqnarray}
\medskip
\noindent
Thus, at large $d$ with fixed $m$, one finds
\begin{eqnarray}
T_c\to\infty \qquad (d=1,\ {\bf BFSS}_2),
\end{eqnarray}
whereas for $d>1$,
\begin{eqnarray}
T_c\sim\frac{d^{1/3}}{\log d}\gg1 \qquad (d>1,\ {\bf BFSS}_{d+1}).
\end{eqnarray}
This explains why the low--temperature dynamics of the BFSS$_d$ model is accurately captured by the gauged matrix
harmonic oscillator: for a wide range $T<T_c$ the system remains in the confined phase.

\subsection{Effect of a mass deformation on the transition temperature}

The previous results apply in the regime of small mass,
\begin{eqnarray}
m \ll (\alpha d)^{2/3},
\end{eqnarray}
where the dynamically generated scale is controlled by
$s_0\sim(\alpha d)^{1/3}$.
In the opposite, large--mass regime,
\begin{eqnarray}
m \gg (\alpha d)^{2/3},
\end{eqnarray}
the positive root of the zero--temperature gap equation \eqref{eq:zeroT}
admits the asymptotic expansion
\begin{eqnarray}
s_0
=
\sqrt{m}
+\frac{\alpha d}{2m}
+\mathcal{O}(m^{-5/2}),
\qquad
m \gg (\alpha d)^{2/3}.
\label{large-m-s0}
\end{eqnarray}
Substituting this result into the general expression \eqref{Tc-general} for the critical temperature,
the latter becomes
\begin{eqnarray}
T_c
=
\frac{\sqrt{m}}{\log d}
\left[
1+\frac{\alpha d}{2m^{3/2}}
+\mathcal{O}(m^{-3})
\right],
\qquad d>1.
\label{Tc-large-m}
\end{eqnarray}
Thus, a positive mass deformation increases the gap $s$ and pushes the
deconfinement instability to parametrically higher temperatures.
In particular, for moderately large values of $d$
(e.g.\ $d=2$ in BFSS$_3$),
one may keep the system in the uniform, center--symmetric
(stringy / Gaussian) phase over all temperatures of interest
by choosing $m$ sufficiently large,
while still retaining quantitative control within the large--$d$
saddle--point framework.

\subsection{Deconfined phase and eigenvalue distributions}

\noindent
For $d\,q>1$ (equivalently $T>T_c$), the uniform distribution becomes unstable.
Introducing a continuous eigenvalue density $\rho(\theta)$ and its moments
\begin{eqnarray}
u_n=\int_{-\pi}^{\pi}\!d\theta\,\rho(\theta)\,e^{in\theta},
\qquad
\rho(\theta)\ge0,\quad \int_{-\pi}^{\pi}\rho(\theta)\,d\theta=1,
\end{eqnarray}
\medskip
\noindent At leading order, the holonomy effective action (\ref{Seff-hol}) can be written as a functional of $\rho$:
\begin{eqnarray}
S_{\rm hol}[\rho]
=
-\frac{N^2}{2}\int_{-\pi}^{\pi}\!d\theta\!\int_{-\pi}^{\pi}\!d\phi\;
\rho(\theta)\rho(\phi)\,
\log\!\Big(2\sin\frac{\theta-\phi}{2}\Big)^2
\;-\;
N^2 d\sum_{n\ge1}\frac{q^n}{n}\,u_n u_{-n}
\;+\;\text{const}.\nonumber\\
\label{Shol-rho}
\end{eqnarray}
We extremize $S_{\rm hol}$ subject to $\int\rho=1$ by introducing a Lagrange multiplier $\lambda$:
\begin{eqnarray}
0=\frac{\delta}{\delta\rho(\theta)}\Big(S_{\rm hol}[\rho]+\lambda\int_{-\pi}^{\pi}\!d\theta\,\rho(\theta)\Big).
\label{var-rho}
\end{eqnarray}
Varying \eqref{Shol-rho} gives
\begin{eqnarray}
0
=
-\int_{-\pi}^{\pi}\!d\phi\;\rho(\phi)\,
\log\!\Big(2\sin\frac{\theta-\phi}{2}\Big)^2
\;-\;
d\sum_{n\ge1}\frac{q^n}{n}\Big(u_{-n}e^{in\theta}+u_ne^{-in\theta}\Big)
\;+\;\lambda,
\label{EL-1}
\end{eqnarray}
where we used
\begin{eqnarray}
\frac{\delta u_n}{\delta\rho(\theta)}=e^{in\theta},
\qquad
\frac{\delta u_{-n}}{\delta\rho(\theta)}=e^{-in\theta}.
\end{eqnarray}
Taking one derivative with respect to $\theta$ removes the constant $\lambda$ and yields
\begin{eqnarray}
-\mathrm{P.V.}\!\int_{-\pi}^{\pi}\!d\phi\;\rho(\phi)\,
\frac{\partial}{\partial\theta}\log\!\Big(2\sin\frac{\theta-\phi}{2}\Big)^2
=
d\sum_{n\ge1}q^n\Big(u_{-n}ie^{in\theta}-u_ni e^{-in\theta}\Big).
\label{EL-2}
\end{eqnarray}
Using
\begin{eqnarray}
\frac{\partial}{\partial\theta}\log\!\Big(2\sin\frac{\theta-\phi}{2}\Big)^2
=
\cot\!\frac{\theta-\phi}{2},
\end{eqnarray}
and writing the right-hand side as $2d\sum_{n\ge1}q^n\,\Im(u_ne^{-in\theta})$, we obtain
\begin{eqnarray}
-\mathrm{P.V.}\!\int_{-\pi}^{\pi}\!d\phi\,
\rho(\phi)\,\cot\!\frac{\theta-\phi}{2}
=
2d\sum_{n\ge1} q^n\,\Im\!\big(u_ne^{-in\theta}\big).
\label{rho-eq-general}
\end{eqnarray}
For the relevant saddles one may take $\rho(\theta)=\rho(-\theta)$, hence $u_n\in\mathbb{R}$ and
$\Im(u_ne^{-in\theta})=-u_n\sin(n\theta)$, giving the large--$N$ saddle--point equation 
\begin{eqnarray}
\mathrm{P.V.}\!\int_{-\pi}^{\pi}\!d\phi\,
\rho(\phi)\,\cot\!\frac{\theta-\phi}{2}
=
2d\sum_{n\ge1} q^n\,u_n\,\sin(n\theta).
\label{rho-eq}
\end{eqnarray}

\medskip
\noindent
Just above $T_c$, only the $n=1$ mode condenses and the density is well approximated by
\begin{eqnarray}
\rho(\theta)
=
\frac{1}{2\pi}\Big(1+2u_1\cos\theta\Big),
\qquad |u_1|\ll1,\label{rho-nonuniform}
\end{eqnarray}
which continuously deforms the uniform distribution.
At higher temperature, positivity of $\rho$ enforces a gapped solution of
\eqref{rho-eq}, corresponding to clumping of eigenvalues around $\theta=0$.

\subsection{Physical interpretation of the holonomy phases}

It is important to stress the correct physical meaning of the holonomy saddle.
When the eigenvalues $\{\theta_i\}$ are \emph{uniformly distributed} on the thermal circle,
\begin{eqnarray}
\rho(\theta)=\frac{1}{2\pi},
\qquad
u_n=0 \ \ (n\ge1),
\end{eqnarray}
the Polyakov loop vanishes and the system is in a \emph{center--symmetric, confined phase}.
In the D0--brane interpretation, this corresponds to a \emph{stringy phase} in which the
D0--branes are delocalized along the Euclidean time circle and no single temporal position
is preferred.

By contrast, a \emph{black--hole phase} requires \emph{localization} of the D0--branes at a
point on the thermal circle. This corresponds to a \emph{non--uniform} eigenvalue
distribution,
\begin{eqnarray}
\rho(\theta)\neq \frac{1}{2\pi},
\qquad
u_1\neq 0,
\end{eqnarray}
with a spectral gap in $\rho(\theta)$.
Such a saddle breaks center symmetry and signals deconfinement in the matrix model,
which is the correct gauge--theory dual of a black zero--brane geometry.

Therefore:
\begin{itemize}
\item \textbf{Uniform holonomy} $\;\Rightarrow\;$ string / Hagedorn--like phase,
\item \textbf{Non--uniform (gapped) holonomy} $\;\Rightarrow\;$ black--hole phase.
\end{itemize}

In BFSS$_2$, the uniform phase persists at all temperatures, consistent with the absence
of a finite--temperature black hole and the emergence of an AdS$_2$ stringy regime.
For $d>1$, the deconfinement transition corresponds to the onset of D0--brane localization
and black--hole formation.

\subsection{Black--string dictionary and the critical--dimension question}
\label{subsection:black-string-critical-dimension}

\medskip
\noindent
The holonomy phases admit a sharper gravitational interpretation when the
one--dimensional model is regarded as the high--temperature reduction of
thermal \(1+1\)-dimensional Yang--Mills theory on a spatial circle
\cite{Aharony:2004igF,Harmark:2004ws}.  Denoting the thermal and spatial
circumferences of the parent theory by \(\widetilde\beta\) and \(L\),
respectively, the limit \(\widetilde\beta\to0\) removes the nonzero temporal
Kaluza--Klein modes and, with antiperiodic thermal boundary conditions, the
fermions.  The surviving one--dimensional coordinate is the original spatial
coordinate.  Consequently,
\begin{eqnarray}
\beta_{\rm 1d}=L,
\qquad
U_{\rm 1d}
=
{\cal P}\exp\left(i\int_0^L dx\,A_x\right)
=
W_x,
\end{eqnarray}
so the Polyakov loop of the reduced model is the spatial Wilson line of the
parent \(1+1\)-dimensional theory.

\medskip
\noindent
After T--duality, the D1--branes are mapped to D0--branes, and the phases
\(\theta_i\) of the spatial Wilson line \(W_x\) are interpreted as the
angular positions of the individual D0--branes on the dual circle
\cite{Hanada2016,Ydri:2018MMST}. The resulting gravitational dictionary is
therefore
\begin{eqnarray}
\begin{array}{ccl}
\rho(\theta)=1/2\pi
&\Longleftrightarrow&
\hbox{uniform black string},
\\[1mm]
\rho(\theta)\ \hbox{non--uniform with full support}
&\Longleftrightarrow&
\hbox{non--uniform black string},
\\[1mm]
\rho(\theta)\ \hbox{gapped and localized}
&\Longleftrightarrow&
\hbox{localized black hole}.
\end{array}
\label{matrix-black-string-dictionary}
\end{eqnarray}
In particular, the weak deformation
\begin{eqnarray}
\rho(\theta)
=
\frac{1}{2\pi}
\left(1+2u_1\cos\theta+\cdots\right)
\end{eqnarray}
is the matrix counterpart of the Gregory--Laflamme deformation of a uniform
black string.  The condition
\begin{eqnarray}
1-de^{-\beta s}=0
\label{matrix-GL-threshold}
\end{eqnarray}
therefore identifies the onset of the corresponding linear instability. In the full two--dimensional phase diagram, the two high--temperature
transitions between uniform, non--uniform, and gapped distributions may join
the direct low--temperature black--string/black--hole transition at a
multicritical region \cite{Aharony:2004igF}. See also\cite{Hanada2016,Ydri:2018MMST}.

\begin{figure}[t]
\centering
\includegraphics[
width=0.72\textwidth,
height=0.76\textheight,
keepaspectratio
]{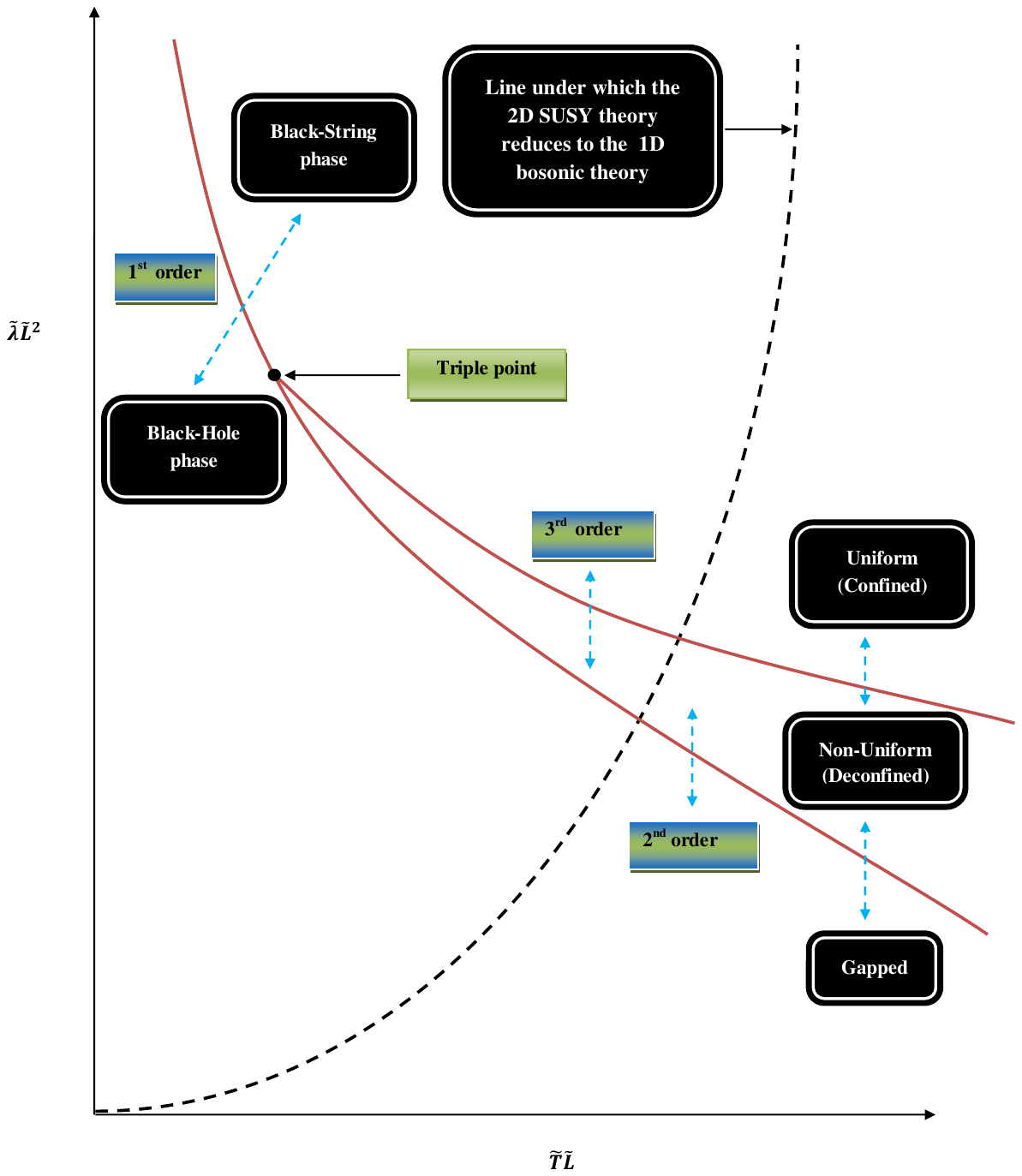}
\caption{Schematic phase diagram of thermal \(1+1\)-dimensional maximally
supersymmetric Yang--Mills theory on a spatial circle and its
high--temperature reduction to the one--dimensional bosonic matrix model.
At low temperature, the uniform black--string and localized black--hole
phases are separated by a first--order transition.  At the putative
multicritical point, this transition splits into a third--order
uniform/non--uniform transition and a second--order
non--uniform/gapped transition.  The dashed curve indicates the boundary
below which dimensional reduction to the one--dimensional bosonic theory is
valid.  The figure is schematic and is adapted from
Refs.~\cite{Aharony:2004igF,Hanada2016,Ydri:2018MMST}.}
\label{fig:2d-1d-black-string-phase}
\end{figure}

\medskip
\noindent
The multicritical organization displayed in
Fig.~\ref{fig:2d-1d-black-string-phase} concerns the global thermal and
coupling phase diagram.  A related but more refined phenomenon occurs for
neutral Kaluza--Klein black strings in ordinary vacuum Einstein gravity.
Consider a uniform black string in \(D\) spacetime dimensions, extended
along a compact coordinate \(z\sim z+L\), and define
\begin{eqnarray}
n=D-4.
\end{eqnarray}
At the Gregory--Laflamme threshold, the uniform string develops a static
linear zero mode with wave number \(k_{\rm GL}\).  Introducing its amplitude
\(\varepsilon\) as a perturbative parameter, one seeks a family of nearby
static solutions of the full nonlinear Einstein equations of the form
\begin{eqnarray}
g_{\mu\nu}(\varepsilon)
=
g_{\mu\nu}^{\rm UBS}
+
\varepsilon h_{\mu\nu}^{(1)}
+
\varepsilon^2 h_{\mu\nu}^{(2)}
+\cdots.
\end{eqnarray}
The wave number and thermodynamic quantities are simultaneously allowed to
shift along this family,
\begin{eqnarray}
k(\varepsilon)
=
k_{\rm GL}+k_2\varepsilon^2+\cdots,
\qquad
M(\varepsilon)
=
M_{\rm GL}+M_2\varepsilon^2+\cdots.
\end{eqnarray}
This perturbative continuation in the space of static solutions generates
the weakly non--uniform black--string branch that bifurcates from the
uniform branch at the Gregory--Laflamme point; and it should not be confused
with the real--time evolution of an unstable string.

In the large-\(D\) effective description, the corresponding mass-density
profile may be written as
\begin{eqnarray}
\frac{m(z)}{\overline m}
=
1+\varepsilon\cos\!\big(k(\varepsilon)z\big)
+\frac{\varepsilon^2}{6}
\cos\!\big(2k(\varepsilon)z\big)
+\cdots,
\label{GL-density-profile}
\end{eqnarray}
where \(z\sim z+L\) is the compact string coordinate and
\(\varepsilon=0\) gives the uniform string.  Since
\(k(\varepsilon)-k_{\rm GL}=O(\varepsilon^2)\), one may equivalently replace
\(k(\varepsilon)\) by \(k_{\rm GL}\) in
\eqref{GL-density-profile} at the order displayed.

The effective density \(m(z)\) is related to the local horizon radius by
\begin{eqnarray}
r_{\rm h}(z)
=
r_0
\left(
\frac{m(z)}{\overline m}
\right)^{1/n},
\qquad
n=D-4,
\end{eqnarray}
and therefore, to linear order in the deformation,
\begin{eqnarray}
\frac{r_{\rm h}(z)}{r_0}
=
1+\frac{\varepsilon}{n}\cos(k_{\rm GL}z)
+O(\varepsilon^2).
\end{eqnarray}
Thus \(\varepsilon\) measures an order-one modulation of the effective mass
density, whereas the corresponding geometrical modulation of the horizon
radius is suppressed by \(1/n\) at large \(D\)
\cite{Sorkin:2004Critical,Emparan:2018NUBS,EmparanHerzog:2020}.

\medskip
\noindent
This construction makes the matrix/gravity correspondence more precise.
Identifying the angular coordinate on the T--dual circle with the normalized
string coordinate,
\begin{eqnarray}
\theta=\frac{2\pi z}{L},
\end{eqnarray}
the normalized eigenvalue density and the normalized gravitational mass
density play analogous roles:
\begin{eqnarray}
2\pi\rho(\theta)
\ \Longleftrightarrow\
\frac{m(z)}{\overline m}.
\label{rho-mass-density-dictionary}
\end{eqnarray}
Hence the weak matrix deformation \(2\pi\rho(\theta)
=
1+2u_1\cos\theta+\cdots\) is the counterpart of \eqref{GL-density-profile}, with
\begin{eqnarray}
\varepsilon
\ \Longleftrightarrow\
2u_1
\end{eqnarray}
at linear order, up to the normalization convention for the deformation
amplitude.  The first Polyakov moment \(u_1\) is therefore the matrix
analogue of the Gregory--Laflamme order parameter.

\medskip
\noindent
Sorkin showed that the branch emerging from the Gregory--Laflamme point
changes character at a critical spacetime dimension \(D_*\simeq13\)
\cite{Sorkin:2004Critical}.  Later large-\(D\) calculations locate the
change more precisely near \(D_*\simeq13.5\), and show that both its
thermodynamic and dynamical stability change there
\cite{Emparan:2018NUBS,EmparanHerzog:2020}.

\medskip
\noindent
For \(D>D_*\), a stable weakly non--uniform string emerges continuously from
the unstable uniform string, consistently with a second--order transition at
\(M_{\rm GL}\).  For \(D<D_*\), the weakly non--uniform branch is not the
stable endpoint near \(M_{\rm GL}\), and a first--order transition to a
finite--amplitude configuration is required.  The latter may be a strongly
non--uniform string or a localized black hole; the critical dimension controls
the local branch near the Gregory--Laflamme point and does not by itself imply
immediate horizon breakup.

\medskip
\noindent
The matrix analogue of this question requires the nonlinear effective action
for the first Polyakov moment,
\begin{eqnarray}
\frac{S_{\rm hol}}{N^2}
=
S_0
+
a_2(d,T)|u_1|^2
+
a_4(d,T)|u_1|^4
+
a_6(d,T)|u_1|^6
+\cdots,
\qquad
a_2(d,T)=1-de^{-\beta s}.\nonumber\\
\label{Landau-matrix-GL}
\end{eqnarray}
For each \(d\), the linear instability temperature \(T_c(d)\) is determined
by
\begin{eqnarray}
a_2\big(d,T_c(d)\big)=0.
\end{eqnarray}
A matrix critical dimension \(d_*\), if it exists, would then be defined by
\begin{eqnarray}
a_4\big(d_*,T_c(d_*)\big)=0,
\qquad
a_6\big(d_*,T_c(d_*)\big)>0.
\end{eqnarray}
The sign of \(a_4\big(d,T_c(d)\big)\) on the two sides of \(d_*\) determines
the character of the branch emerging from the uniform saddle.  For
\(a_4>0\), a locally stable weakly non--uniform branch develops
continuously from \(u_1=0\), whereas for \(a_4<0\) the weak branch is
locally unstable and the transition requires a finite-amplitude
configuration stabilized by the higher-order terms.  The point
\(a_2=a_4=0\), with \(a_6>0\), is therefore the matrix analogue of a
tricritical point.

\medskip
\noindent
The leading maximally symmetric saddle derived here fixes
\(a_2\) and hence the linear instability \eqref{matrix-GL-threshold}, but it
does not yet determine the complete nonlinear coefficients.  Establishing an
analogue of the Sorkin critical dimension would require subleading
\(1/d\) corrections, the holonomy backreaction on the auxiliary saddle, and
possibly more general color--dependent components of the auxiliary kernel.

\medskip
\noindent
Finally, two distinct notions of dimension enter the comparison.  In the
Sorkin problem, \(D\) is the total spacetime dimension of a neutral
Kaluza--Klein black string in vacuum Einstein gravity and is treated as a
variable parameter.  In the physical matrix-model realization considered
here, the dual black objects instead live in the fixed ten-dimensional
bulk spacetime of type-II string theory.

The parameter \(d\) counts the adjoint matrix coordinates of the D0-brane
constituents.  For the maximally supersymmetric system,
\begin{eqnarray}
D_{\rm bulk}=10,
\qquad
d=9,
\end{eqnarray}
and the nine matrices describe the nine spatial directions transverse to
the D0-brane worldline.  Equivalently, the parent D1-brane theory has eight
transverse scalar matrices; after T--duality along its spatial circle, that
circle becomes the ninth transverse D0-brane coordinate
\cite{Hanada2016,Ydri:2018MMST}.

The uniform, non--uniform, and localized phases are therefore phases of
ten-dimensional charged and dilatonic black D0-brane geometries, while
varying \(d\) defines a generalized family of matrix models.  Except for the
physical value \(d=9\), this variation should not be interpreted as varying
the spacetime dimension of a fixed type-II gravitational theory.  The
comparison with the Sorkin critical dimension is consequently structural,
rather than a numerical identification of \(D\) with \(d\).

\section{Yang--Mills observable}\label{section6}

\subsection{Large--$d$ saddle value of the Yang--Mills observable}

\noindent Recall that the bosonic Yang--Mills observable is defined by
\begin{eqnarray}
\Big\langle \frac{N}{4}\int_{0}^{\beta}\! dt\ \mathrm{Tr}[X_a,X_b]^2\Big\rangle
&=&
\left.\frac{\partial \ln Z(\alpha)}{\partial \alpha}\right|_{\alpha=1}
\;=\;
-\beta\,\left.\frac{\partial F(\alpha)}{\partial \alpha}\right|_{\alpha=1}.
\label{YM-def-again}
\end{eqnarray}

\medskip
\noindent At leading order in $1/d$ we evaluate the partition function by the maximally symmetric saddle $k_{\mu\rho}(t)=k_{0\,\mu\rho}=2k_0\,\delta_{\mu\rho}$,
so that the localized theory reduces to the gauged matrix harmonic oscillator with
\begin{eqnarray}
s^2\equiv m_{\rm eff}=m+k_0.
\end{eqnarray}
The corresponding large--$d$ effective action (before integrating over $\theta$) is
\begin{eqnarray}
S_{\rm eff}(\theta;\alpha)
=
S_{\rm Vdm}[\theta]
+S_{\rm MHO}(\theta;s)
+S_k[k_0]
\;+\;\cdots,
\end{eqnarray}
where $S_{\rm Vdm}$ is the Vandermonde contribution, $S_{\rm MHO}(\theta;s)$ is the Gaussian determinant of $X_a$,
and the dots denote subleading corrections (including the $\delta k$ determinant, discussed below).

Using $\mu_{\mu\rho\nu\sigma}\delta_{\nu\sigma}=-(8N)^{-1}\delta_{\mu\rho}$ one finds

\begin{eqnarray}
S_k[k_0]
&=&
\frac{N}{4\alpha}\int_{0}^{\beta}\!dt\;
\mu_{\mu\rho\nu\sigma}\,k_{0\,\mu\rho}\,k_{0\,\nu\sigma}
\;=\;
-\frac{\beta}{8\alpha}\,(N^2-1)\,k_0^2.
\label{Sk-again}
\end{eqnarray}
Thus

\begin{eqnarray}
S_{\rm eff}(\theta;\alpha)
=
S_{\rm Vdm}[\theta]
+S_{\rm MHO}(\theta;s)
-\frac{\beta}{8\alpha}\,(N^2-1)\,k_0^2
\;+\;\cdots.
\label{Seff-leading}
\end{eqnarray}

\medskip
\noindent \noindent At leading order (large--$d$ saddle), we approximate the full free energy by
\begin{eqnarray}
F(\theta;\alpha)&\equiv&
\frac{1}{\beta}\,
S_{\rm eff}(\theta;\alpha)
\simeq F_{\rm Vdm}(\theta)+
F_{\rm MHO}(\theta;s)\;-\;\frac{N^2-1}{8\alpha}\,k_0^2,
\label{F-saddle-again}
\end{eqnarray}
where $k_0\equiv k_0(\theta;\alpha)$ is fixed by the gap equation, i.e.\ by the stationarity of $F$ with respect to $k_0$:
\begin{eqnarray}
0
&=&
\frac{\partial F}{\partial k_0}
=
\frac{\partial F_{\rm MHO}}{\partial s^2}\,\frac{\partial s^2}{\partial k_0}
\;-\;\frac{N^2-1}{4\alpha}\,k_0
=
\frac{\partial F_{\rm MHO}}{\partial s^2}\;-\;\frac{N^2-1}{4\alpha}\,k_0.
\label{saddle-k0}
\end{eqnarray}
(Here $\partial s^2/\partial k_0=1$.)

\medskip
\noindent Now differentiate the saddle free energy $F(\alpha)=F(\theta;\alpha,k_0(\alpha))$ with respect to $\alpha$:
\begin{eqnarray}
\frac{dF}{d\alpha}
&=&
\left.\frac{\partial F}{\partial \alpha}\right|_{k_0}
\;+\;
\left.\frac{\partial F}{\partial k_0}\right|_{\alpha}\,
\frac{dk_0}{d\alpha}.
\label{chain}
\end{eqnarray}
The key point is that the second term vanishes \emph{at the saddle} by \eqref{saddle-k0}:
\begin{eqnarray}
\left.\frac{\partial F}{\partial k_0}\right|_{\alpha}=0
\qquad\Longrightarrow\qquad
\frac{dF}{d\alpha}
=
\left.\frac{\partial F}{\partial \alpha}\right|_{k_0}.
\label{envelope}
\end{eqnarray}
This is the envelope theorem in this context: \emph{implicit} $\alpha$--dependence through the optimized
parameter $k_0(\alpha)$ does not contribute at leading saddle order.

\medskip
\noindent Differentiating then \eqref{F-saddle-again} with respect to $\alpha$ at fixed saddle $k_0$
(envelope theorem at the stationary point of the $k$--effective potential) yields
\begin{eqnarray}
-\beta\,\frac{\partial F}{\partial \alpha}
&\simeq&
-\beta\,\frac{\partial}{\partial \alpha}
\Big(-\frac{N^2-1}{8\alpha}\,k_0^2\Big)
\;=\;
-\frac{\beta}{8\alpha^2}\,(N^2-1)\,k_0^2.
\label{YM-from-k0}
\end{eqnarray}
Therefore, at leading order in the large--$d$ saddle, the Yang--Mills observable becomes
\begin{eqnarray}
\Big\langle \frac{N}{4}\int_{0}^{\beta}\! dt\ \mathrm{Tr}[X_a,X_b]^2\Big\rangle
&=&
-\beta\,\left.\frac{dF}{d\alpha}\right|_{\alpha=1}
\;\simeq\;
-\frac{\beta}{8}\,(N^2-1)\,k_0^2,
\label{YM-envelope-final}
\end{eqnarray}
with $k_0$ evaluated at the $\alpha=1$ saddle. This is manifestly negative, consistent with $\mathrm{Tr}[X_a,X_b]^2\le 0$ for Hermitian $X_a$.

\medskip
\noindent
For completeness, let us also include the contribution of fluctuations around the
maximally symmetric saddle. The quadratic fluctuation of the auxiliary field is
\begin{eqnarray}
S_{\delta k}^{(2)}[\delta k]
=
\frac{N}{4\alpha}\int_{0}^{\beta}\!dt\;
\mu_{\mu\rho\nu\sigma}\,
\delta k_{\mu\rho}(t)\,\delta k_{\nu\sigma}(t).
\label{Sdk2-again}
\end{eqnarray}
Restricting, for simplicity, to the zero mode
\begin{eqnarray}
\delta k_{\mu\rho}(t)=\delta k_{\mu\rho},
\end{eqnarray}
one obtains
\begin{eqnarray}
S_{\delta k,0}^{(2)}
=
\frac{N\beta}{4\alpha}\,
\mu_{\mu\rho\nu\sigma}\,
\delta k_{\mu\rho}\,\delta k_{\nu\sigma}.
\end{eqnarray}
Thus the zero--mode fluctuation produces only a Gaussian determinant. Indeed, the Gaussian integral gives
\begin{eqnarray}
Z_{\delta k,0}
\;\propto\;
\Big[\det\nolimits_{\rm sym}\!\Big(\frac{N\beta}{2\alpha}\,\mu\Big)\Big]^{-1/2},
\label{Zdk0}
\end{eqnarray}
where $\det_{\rm sym}$ is taken over symmetric pairs $(\mu\rho)$ of dimension
$D_{\rm sym}=\frac{(N^2-1)N^2}{2}$. Hence
\begin{eqnarray}
\log Z_{\delta k,0}
=
-\frac{1}{2}\log\det\nolimits_{\rm sym}\mu
-\frac{D_{\rm sym}}{2}\log\Big(\frac{N\beta}{2\alpha}\Big)
+\text{const},
\end{eqnarray}
so the induced shift in the free energy is an $\alpha$--dependent \emph{constant},
\begin{eqnarray}
\Delta F_{\delta k,0}
=
-\frac{1}{\beta}\log Z_{\delta k,0}
=
\frac{1}{2\beta}\log\det\nolimits_{\rm sym}\mu
+\frac{D_{\rm sym}}{2\beta}\log\Big(\frac{N\beta}{2\alpha}\Big)
+\text{const},
\qquad
\frac{\partial}{\partial\alpha}\log Z_{\delta k,0}
=
\frac{D_{\rm sym}}{2\alpha}.\nonumber\\
\end{eqnarray}
In a properly normalized Hubbard--Stratonovich representation this Gaussian factor is absorbed into the
overall HS prefactor: it is independent of $\theta$ and $s$, and thus does not affect the holonomy saddle,
the phase transition, or any $\theta$--dependent observable at leading order.

\subsection{The small--$m$ and large--$m$ behaviors}

\medskip
\noindent By expressing $k_0$ in terms of the gap $s$, the Yang--Mills observable  takes the form 
\begin{eqnarray}
\Big\langle \frac{N}{4}\int_0^\beta\!dt\ \Tr[X_a,X_b]^2\Big\rangle
=
-\frac{\beta}{8}\,(N^2-1)\,(s^2-m)^2.
\label{YM-s0}
\end{eqnarray}

\medskip
\noindent The small--$m$ and large--$m$ regimes are obtained from the $T=0$ gap equation  (zero--temperature saddle) which is given by \eqref{eq:zeroT}.

\noindent\emph{(i) Small mass: $m\ll a^{2/3}$.}
With $\varepsilon\equiv m/a^{2/3}\ll1$ one has 
\begin{eqnarray}
s_0
=
a^{1/3}\left(1+\frac{\varepsilon}{3}-\frac{\varepsilon^3}{81}+\frac{\varepsilon^4}{243}+\cdots\right),
\qquad a=\alpha d,
\end{eqnarray}
so
\begin{eqnarray}
k_0
=s_0^2-m
=
a^{2/3}\left(1-\frac{\varepsilon}{3}+\frac{\varepsilon^2}{9}-\frac{2\varepsilon^3}{81}+\cdots\right),
\qquad (\varepsilon\ll1),
\label{k0-smallm}
\end{eqnarray}
and inserting into \eqref{YM-s0} gives
\begin{eqnarray}
\Big\langle \frac{N}{4}\int_0^\beta\!dt\ \Tr[X_a,X_b]^2\Big\rangle
\;\simeq\;
\frac{\beta}{8}\,(N^2-1)\,(\alpha d)^{4/3}
\left(1-\frac{2\varepsilon}{3}+\frac{\varepsilon^2}{3}+\cdots\right)
\Bigg|_{\alpha=1}.
\label{YM-smallm}
\end{eqnarray}

\noindent\emph{(ii) Large mass: $m\gg a^{2/3}$.}
The positive root admits
\begin{eqnarray}
s_0
=
\sqrt{m}+\frac{a}{2m}-\frac{3a^2}{8m^{5/2}}+\cdots,
\qquad (m\gg a^{2/3}),
\end{eqnarray}
hence
\begin{eqnarray}
k_0
=s_0^2-m
=
\frac{a}{\sqrt{m}}-\frac{a^2}{2m^2}+\cdots,
\qquad (m\gg a^{2/3}),
\label{k0-largem}
\end{eqnarray}
and therefore
\begin{eqnarray}
\Big\langle \frac{N}{4}\int_0^\beta\!dt\ \Tr[X_a,X_b]^2\Big\rangle
\;\simeq\;
\frac{\beta}{8}\,(N^2-1)\,
\left(\frac{(\alpha d)^2}{m}+\mathcal{O}(m^{-5/2})\right)
\Bigg|_{\alpha=1}.
\label{YM-largem}
\end{eqnarray}

\medskip
\noindent Large positive $m$ confines the matrices close to the origin, strongly suppressing commutators and hence
$\Tr[X_a,X_b]^2$. In this regime the quadratic term dominates, and the theory is  truly Gaussian.
By contrast, at small $m$ the quartic interaction dynamically generates a mass gap
$k_0\sim d^{2/3}$, making the Yang--Mills observable parametrically large at large $d$.
The theory is then intrinsically non-perturbative, yet its dynamics are accurately captured by the
Gaussian structure of the maximally symmetric large--$d$ saddle, which encodes the essential physics
of the BFSS model.

\medskip
\noindent  Let us also recall that at finite $T$ the holonomy dependence enters through the MHO determinant,
\begin{eqnarray}
F_{\rm MHO}(\theta;s)
=
\frac{d}{2\beta}\sum_{i,j}P_{ij,ji}\,
\log\!\Big(\cosh(\beta s)-\cos(\theta_i-\theta_j)\Big)+\text{const},
\end{eqnarray}
and the holonomy effective action is obtained by combining $F_{\rm MHO}$ with the Vandermonde term. One finds, at leading order, the total action (see Appendix~\ref{appendix-holonomy-effective-action})
\begin{eqnarray}
S_{\rm hol}[\theta]
=
N^2\sum_{n\ge1}\frac{1-dq^n}{n}\,|u_n|^2+\text{const},
\qquad
q=e^{-\beta s},\qquad
u_n=\frac{1}{N}\Tr\,U^n.
\label{Seff-hol-recall}
\end{eqnarray}
Thus the phase structure is controlled by $dq$:
\begin{eqnarray}
dq<1:\ \ u_n=0\ \ \Rightarrow\ \ \rho(\theta)=\frac{1}{2\pi}
\qquad\text{(uniform / confined)}, 
\end{eqnarray}
whereas for $dq>1$ the uniform saddle is unstable and $u_1$ condenses, eventually yielding a gapped (clumped) distribution.

\medskip
\noindent Across the transition, the holonomy changes from the uniform saddle to a
non--uniform, eventually clumped, saddle. Equivalently, the Polyakov moments
change from
\begin{eqnarray}
u_n=0
\qquad \hbox{(uniform / confined)}
\end{eqnarray}
to
\begin{eqnarray}
u_n\neq 0
\qquad \hbox{(non--uniform / deconfined)}.
\end{eqnarray}
In the gap equation this dependence enters through the projector--weighted
holonomy moments
\begin{eqnarray}
{\cal C}_p(U)
=
\frac{|\Tr U^p|^2-1}{N^2-1}
=
\frac{N^2|u_p|^2-1}{N^2-1},
\end{eqnarray}
or equivalently through the kernel \({\cal R}(s;U)\) in
\eqref{eq:gap-projector-averaged}--\eqref{Rkernel_projected}. Therefore
\eqref{YM-s0} should be understood as
\begin{eqnarray}
\Big\langle \frac{N}{4}\int_0^\beta\!dt\ \Tr[X_a,X_b]^2\Big\rangle
\;\simeq\;
-\frac{\beta}{8}(N^2-1)\,\Big(s^2(T;U)-m\Big)^2.
\end{eqnarray}
Here \(s(T;U)\) is evaluated in the dominant holonomy saddle: uniform in the
confined phase and non--uniform in the deconfined phase. In practice, at low
temperature the uniform saddle gives \(s\simeq s_0\), up to exponentially small
thermal corrections, so the small--\(m\) and large--\(m\) formulas
\eqref{YM-smallm} and \eqref{YM-largem} apply directly. In the deconfined phase,
the condensed moments \(u_n\), or equivalently the nonzero
\({\cal C}_p(U)\), correct the gap and hence the Yang--Mills observable.

\section{Double--scaling saddle and high--temperature branch}\label{section7}

\subsection{The Yang--Mills--mass-deformation balance}

\noindent
At the large--$d$ saddle, the quartic interaction is encoded entirely in the dynamical mass
shift $k_0$, or equivalently in
\begin{eqnarray}
s^2=m+k_0 .
\end{eqnarray}
The Yang--Mills and quadratic contributions to the energy density scale parametrically as

\begin{eqnarray}
{\cal O}_{\rm YM}
&\equiv&
\frac{1}{N\beta}\Big\langle \int_0^\beta\!dt\ \frac{N}{4}\Tr[X_a,X_b]^2\Big\rangle
\;\sim\;
-N\,k_0^2,\\
{\cal O}_{\rm quad}
&\equiv&
\frac{1}{N\beta}\Big\langle \int_0^\beta\!dt\ \frac{Nm}{2}\Tr X_a^2\Big\rangle
\;\sim\;
Nm\,R^2
\;\sim\;
Nm\,\frac{d}{2s}.
\end{eqnarray}

\medskip
\noindent
Our goal is to identify a regime in which the quartic interaction is neither negligible nor
parametrically dominant over the quadratic term. Schematically, this ``sweet spot'' is defined by
\begin{eqnarray}
{\cal O}_{\rm YM}\ \sim\ {\cal O}_{\rm quad}.
\label{sweet-def}
\end{eqnarray}
This condition precisely marks the crossover between the interaction--dominated regime,
$m\ll(\alpha d)^{2/3}$, and the mass--dominated regime, $m\gg(\alpha d)^{2/3}$.

\medskip
\noindent
It therefore suggests choosing the mass deformation to scale as
\begin{eqnarray}
m\sim d^{2/3}.
\end{eqnarray}

\medskip
\noindent
In this regime:
\begin{enumerate}
\item The gap \(s\) is enhanced, and hence the deconfinement instability is pushed
to higher temperatures, even for moderate fixed \(d\), such as \(d=2\) in
\(\mathrm{BFSS}_3\). Equivalently, the uniform holonomy regime is enlarged. In
this sense the theory behaves, over a parametrically wider range, as \(d\) effectively decoupled BFSS$_2$--type gauged matrix harmonic oscillators,
since the holonomy--mediated interaction between the different matrix directions
becomes weak at large \(d\).

\item The Yang--Mills observable remains sizable, but no longer dominates
parametrically over the quadratic term. Thus the double--scaling limit
interpolates smoothly between the interaction--dominated regime and the
mass--dominated Gaussian regime. In this sense the original \(\mathrm{BFSS}_{d+1}\) interaction is retained
self--consistently in the Gaussian saddle through the dynamically generated
mass shift \(k_0\).

The BFSS$_2$--type factorization described above should therefore be understood
as an effective bulk approximation. The endpoint dynamics and the full holonomy
structure still retain information about the original \(\mathrm{BFSS}_{d+1}\)
theory.

\item In this double--scaling limit, the theory is therefore neither reduced to a
purely free massive model nor driven by uncontrolled strong commutator dynamics.
The matrices are in fact localized near the origin and weakly non--commuting: the
commutator contribution per matrix pair is parametrically suppressed. This is the
IKKT--like aspect of the double--scaled saddle, reproducing the characteristic
Yang--Mills, or almost--commuting, phase familiar from the IKKT reduction.

\end{enumerate}

\medskip
\noindent 
We now impose the double--scaling condition
\begin{eqnarray}
\kappa \;\equiv\;\frac{m^{3/2}}{\alpha d}\quad \Longleftrightarrow\quad
m \;=\;(\kappa \alpha d)^{2/3},
\label{def-kappa}
\end{eqnarray}
and solve the $T=0$ gap equation \eqref{eq:zeroT}.
Since $m\sim (\alpha d)^{2/3}$ in this limit, it is natural to scale
\begin{eqnarray}
s_0\;=\;(\alpha d)^{1/3}\,y,
\qquad
m\;=\;(\alpha d)^{2/3}\,\kappa^{2/3}.
\label{scaling-y}
\end{eqnarray}
Substituting into \eqref{eq:zeroT} gives the \emph{exact} reduced cubic equation for $y$,
\begin{eqnarray}
y^3-\kappa^{2/3}\,y-1=0,
\label{cubic-y}
\end{eqnarray}
which is independent of $d$. Hence the double--scaled saddle is
\begin{eqnarray}
s(\kappa)=(\alpha d)^{1/3}\,y(\kappa),
\qquad
k_0(\kappa)=s^2(\kappa)-m=(\alpha d)^{2/3}\Big(y(\kappa)^2-\kappa^{2/3}\Big).
\label{s-k0-kappa}
\end{eqnarray}

\medskip
\noindent Writing \eqref{cubic-y} as $y^3+py+q=0$ with \(p=-\kappa^{2/3}\) and \(q=-1\), the discriminant is
\begin{eqnarray}
\Delta_\kappa
=\Big(\frac{q}{2}\Big)^2+\Big(\frac{p}{3}\Big)^3
=
\frac{1}{4}-\frac{\kappa^2}{27}.
\label{Delta-kappa}
\end{eqnarray}
A convenient real root is
\begin{eqnarray}
y(\kappa)
=
\Big(\frac{1}{2}+\sqrt{\Delta_\kappa}\Big)^{\!1/3}
+
\Big(\frac{1}{2}-\sqrt{\Delta_\kappa}\Big)^{\!1/3}.
\label{Cardano-y}
\end{eqnarray}

\subsection{Observables at fixed large $\kappa$: nearly commuting, localized matrices}
\noindent 
The critical temperature, extent of space and Yang-Mills observable are given respectively by

\begin{eqnarray}
T_c(\kappa)
=
\frac{s(\kappa)}{\log d}
=
\frac{(\alpha d)^{1/3}}{\log d}\,y(\kappa).
\label{Tc-kappa}
\end{eqnarray}

\begin{eqnarray}
R^2(\kappa)
=
\Big(1-\frac{1}{N^2}\Big)\frac{d}{2\,s(\kappa)}
+\mathcal{O}(e^{-\beta s})
=
\Big(1-\frac{1}{N^2}\Big)\frac{(\alpha d)^{2/3}}{2\alpha\,y(\kappa)}
+\mathcal{O}(e^{-\beta d^{1/3}y}).
\label{R2-kappa}
\end{eqnarray}

\begin{eqnarray}
\Big\langle -\frac{N}{4}\int_0^\beta dt\ \Tr[X_a,X_b]^2\Big\rangle_{\kappa}
=
\frac{N^2-1}{8\alpha^2}\,k_0(\kappa)^2
=
\frac{N^2-1}{8\alpha^2}\,(\alpha d)^{4/3}\,
\Big(y(\kappa)^2-\kappa^{2/3}\Big)^2.
\label{YM-kappa}
\end{eqnarray}

\medskip
\noindent The entire dependence on the double--scaling parameter is encoded in the single
dimensionless function $y(\kappa)$ solving \eqref{cubic-y}, and the leading large--$d$
scalings are
\begin{eqnarray}
  s\sim d^{1/3},
\end{eqnarray}
and
\begin{eqnarray}
T_c\sim \frac{d^{1/3}}{\log d},\qquad
R^2\sim d^{2/3},\qquad
\langle -\Tr[X,X]^2\rangle\sim d^{4/3},
\end{eqnarray}
with $\kappa$--dependent prefactors given explicitly by
Eqs.~(\ref{Tc-kappa}), (\ref{R2-kappa}), and (\ref{YM-kappa}),
where the function $y(\kappa)$ is determined in closed form by the
Cardano solution~(\ref{Cardano-y}). In particular, for sufficiently large fixed $\kappa$ one has the asymptotic behavior
\begin{eqnarray}
y(\kappa)
=
\kappa^{1/3}
+\frac{1}{2\kappa^{2/3}}
+\mathcal{O}(\kappa^{-5/3}),
\qquad (\kappa\to\infty),
\label{y-large-kappa-text}
\end{eqnarray}
so that $y(\kappa)$ is large and close to $\kappa^{1/3}$. Consequently, the gap $s\propto d^{1/3}y(\kappa)$ increases with $\kappa$, which drives the critical temperature $T_c\propto s/\log d$ upward even for moderately small $d$.

\medskip 
\noindent
In the double--scaling limit (we set $\alpha=1$)
\begin{eqnarray}
\kappa \;\equiv\; \frac{m^{3/2}}{d}=\text{fixed},
\qquad
m\sim d^{2/3},
\qquad
s\sim d^{1/3},
\qquad
k_0=s^2-m\sim d^{2/3}.
\label{ds-limit}
\end{eqnarray}
Thus, the quadratic and quartic terms in the BFSS action are indeed parametrically of the \emph{same}
order at large~$d$, which means that this regime interpolates between the interaction--dominated and
mass--dominated saddles and realizes the desired ``sweet spot'' discussed above.

\medskip
\noindent
The structure emerging in this limit is that of \emph{localized, almost commuting}
matrices. To make this explicit, it is convenient to pass from observables summed over
$a,b$ to quantities defined per matrix and per matrix pair, using rotational invariance
in the $SO(d)$ index~$a$.

\medskip
\noindent
By $SO(d)$ symmetry, each matrix contributes equally to the extent of space. Thus
\begin{eqnarray}
\frac{1}{N\beta}\Big\langle \int_0^\beta\!dt\ \Tr X_1^2\Big\rangle
=
\frac{R^2}{d}
=
\Big(1-\frac{1}{N^2}\Big)\frac{1}{2s}
+\mathcal{O}(e^{-\beta s})
\;\sim\; d^{-1/3}.
\label{X1-localization}
\end{eqnarray}
Each individual matrix $X_a$ is therefore localized near the origin, with a typical width
that shrinks as $d^{-1/3}$, even though the total extent grows as $R^2\sim d^{2/3}$ due to
the sum over $d$ matrices.

\medskip
\noindent
Similarly, the Yang--Mills observable sums over $\sim d^2$ ordered pairs $(a,b)$. By
$SO(d)$ invariance, the typical contribution per pair scales as
\begin{eqnarray}
-\frac{1}{N\beta}\Big\langle \int_0^\beta\!dt\ \Tr[X_1,X_2]^2\Big\rangle
\;\sim\;
-\frac{1}{d^2}\,
\frac{1}{N\beta}\Big\langle \int_0^\beta\!dt\ \Tr[X_a,X_b]^2\Big\rangle
\;\sim\;
d^{-2/3}.
\label{comm-small}
\end{eqnarray}
Hence
\begin{eqnarray}
-\Big\langle \Tr[X_1,X_2]^2 \Big\rangle\;\propto\; d^{-2/3}\ \longrightarrow\ 0
\qquad (d\to\infty),
\label{almost-commuting}
\end{eqnarray}
so the matrices become asymptotically commuting in the large--$d$ limit, despite the
presence of a non--trivial self--consistent mass gap.

\medskip
\noindent
Taken together, Eqs.~\eqref{X1-localization} and \eqref{almost-commuting} describe a regime
of \emph{nearly commuting, localized matrices}: each $X_a$ is clumped near the origin,
while commutators are parametrically suppressed. This is precisely the qualitative
behavior expected in the Yang--Mills (``commuting'') phase of the IKKT reduction, where
geometry is encoded in an almost diagonal matrix configuration with small,
controlled noncommutative fluctuations generated here by the double--scaled saddle.

\subsection{The high--temperature limit}

At sufficiently high temperature the theory is in the deconfined phase, and the holonomy eigenvalues
cluster so that $\phi=\theta_i-\theta_j\simeq 0$. The exact gap equation
\begin{eqnarray}
\frac{d}{s}\,
\frac{\sinh(\beta s)}{\cosh(\beta s)-\cos\phi}
=\frac{s^{2}-m}{\alpha},
\qquad s=\sqrt{\tilde k_0}>0,
\label{eq:gap_exact}
\end{eqnarray}
reduces to
\begin{eqnarray}
\frac{d}{s}\,
\frac{\sinh(\beta s)}{\cosh(\beta s)-1}
=
\frac{d}{s}\,\coth\!\Big(\frac{\beta s}{2}\Big)
=\frac{s^{2}-m}{\alpha}.
\label{eq:gap_phi0}
\end{eqnarray}
For $\beta\to 0$ one has $\coth(\beta s/2)=2/(\beta s)+\mathcal{O}(\beta s)$, and keeping the
leading term gives
\begin{eqnarray}
\frac{2d}{\beta s^{2}}=\frac{s^{2}-m}{\alpha}.
\label{eq:gap_highT_leading}
\end{eqnarray}
Writing $y=s^{2}$, this yields a quadratic equation
\begin{eqnarray}
y(y-m)=\frac{2\alpha d}{\beta},
\label{eq:y_quadratic}
\end{eqnarray}
whose physical (positive) solution is
\begin{eqnarray}
s^{2}
=
\frac{1}{2}\left(
m+\sqrt{m^{2}+\frac{8\alpha d}{\beta}}
\right),
\qquad
k_{0}=s^{2}-m
=
\frac{1}{2}\left(
-\,m+\sqrt{m^{2}+\frac{8\alpha d}{\beta}}
\right).
\label{eq:s2_k0_highT}
\end{eqnarray}
In the strict high--temperature limit $\beta\to 0$ this behaves as
\begin{eqnarray}
s^{2}\sim\sqrt{\frac{2\alpha d}{\beta}},
\qquad
s\sim\left(\frac{2\alpha d}{\beta}\right)^{1/4},
\qquad
k_{0}\sim\sqrt{\frac{2\alpha d}{\beta}}.
\label{eq:s_scaling_highT}
\end{eqnarray}

\medskip
\noindent In this regime the saddle point remains Gaussian, and the adjoint matrices are again governed by
a matrix harmonic oscillator (M.H.O.) with a self--consistent frequency $s(\beta)$ determined by
\eqref{eq:s2_k0_highT}. This contrasts with the low--temperature (uniform--holonomy) branch, where the M.H.O. frequency
saturates to a temperature--independent value $s_0$ determined by the large--$d$ gap equation
$s_0^3-m s_0=\alpha d$, implying
\begin{eqnarray}
m_{\rm eff}\equiv s_0\sim(\alpha d)^{1/3}
\qquad (T\to 0).
\end{eqnarray}
By contrast, the present high--temperature (deconfined) branch exhibits a genuinely temperature--dependent
mass scale. In particular,

\begin{eqnarray}
m_{\rm eff}^2 \equiv s^2 \;\sim\; \sqrt{\frac{2\alpha d}{\beta}}
\;\propto\; \sqrt{\alpha d\,T},
\qquad
m_{\rm eff}\equiv s \;\sim\; (2\alpha d\,T)^{1/4}\qquad (T\to \infty).
\label{eq:meff_highT}
\end{eqnarray}
Thus, while both phases are described by the same M.H.O. structure, the low--$T$ dynamics is
controlled by a $T$--independent mass scale set by $d$ (and $m$), whereas the high--$T$ phase exhibits a
qualitatively new scaling in which the oscillator frequency grows also with temperature.

\medskip
\noindent
In the M.H.O. approximation, integrating out $d$ adjoint bosons of frequency $s$ generates for the
first holonomy Fourier mode $u_1$ a quadratic coefficient of the schematic form
\begin{eqnarray}
a_1(\beta)\;=\;1-d\,e^{-\beta s(\beta)}+\cdots.
\label{eq:a1_form}
\end{eqnarray}
Thus the confinement/deconfinement transition is still controlled by the condition
\begin{eqnarray}
a_1(\beta_c)=0
\qquad\Longrightarrow\qquad
d\,e^{-\beta_c s(\beta_c)}=1
\qquad\Longrightarrow\qquad
\beta_c\,s(\beta_c)=\ln d.
\label{eq:tc_condition}
\end{eqnarray}
This is the same criticality condition as in the low--temperature (uniform--holonomy) phase;
however, while at low $T$ the frequency $s$ approaches a constant $s_0\sim(\alpha d)^{1/3}$,
here $s$ itself depends on temperature through the high--$T$ saddle.
Using the high--$T$ scaling $s(\beta)\sim (2\alpha d/\beta)^{1/4}$ from \eqref{eq:s_scaling_highT},
the criticality condition becomes
\begin{eqnarray}
\beta_c\left(\frac{2\alpha d}{\beta_c}\right)^{1/4}=\ln d
\qquad\Longrightarrow\qquad
(2\alpha d)^{1/4}\,\beta_c^{3/4}=\ln d,
\label{eq:tc_algebra}
\end{eqnarray}
hence
\begin{eqnarray}
\beta_c
\;\sim\;
\frac{(\ln d)^{4/3}}{(2\alpha d)^{1/3}},
\qquad\Rightarrow\qquad
T_c^{\rm high}=\frac{1}{\beta_c}
\;\sim\;
\frac{(2\alpha d)^{1/3}}{(\ln d)^{4/3}}.
\label{eq:Tc_highT_scaling}
\end{eqnarray}
Therefore, although the M.H.O. structure and the condition $\beta_c s(\beta_c)=\ln d$ are unchanged,
the temperature dependence of the effective mass scale on the high--$T$ branch leads to a
parametrically different scaling of $T_c$ compared to the low--$T$ result
\begin{eqnarray}
  T_c^{\rm low}\sim \frac{s_0}{\ln d}\sim \frac{(\alpha d)^{1/3}}{\ln d}.
  \end{eqnarray}

\medskip
\noindent In summary, the low--temperature analysis dictates a regime of validity for
\(T < T_c^{\rm low},\) where the M.H.O. frequency saturates to the temperature--independent value
$s_0\sim(\alpha d)^{1/3}$.  Conversely, the high--temperature analysis applies for
\(T > T_c^{\rm high},\) where the effective frequency grows with temperature as
$s(\beta)\sim(2\alpha d/\beta)^{1/4}$. Since parametrically \(
T_c^{\rm high} < T_c^{\rm low},\)  there exists an intermediate window
\begin{eqnarray}
T_c^{\rm high} \;<\; T \;<\; T_c^{\rm low},
\end{eqnarray}
in which both approximations overlap. In this regime the dynamics is consistently governed
by a Gaussian saddle with a self--consistent mass, and is therefore dominated by
approximately commuting matrices.

\medskip
\noindent This can be seen more clearly by analyzing basic observables at the large--$d$ Gaussian (M.H.O.)
saddle, using their exact saddle--point expressions and comparing the low-- and high--temperature
branches.

At the large--$d$ saddle point, the extent of space and the Yang--Mills term are given respectively by
\begin{eqnarray}
R^2 
\;\equiv\;
\frac{1}{N\beta}\Big\langle \int_0^{\beta} dt \Tr X_a^2 \Big\rangle
\;=\;
\frac{d}{2s},
\end{eqnarray}
\begin{eqnarray}
\Big\langle \frac{N}{4}\int_0^{\beta} dt\Tr[X_a,X_b]^2 \Big\rangle
\;=\;
-\,\frac{N^2-1}{8\alpha^2}\,k_0^2,
\qquad
k_0=s^2-m.
\end{eqnarray}
On the high--temperature (deconfined) branch one has
$s\sim(2\alpha d T)^{1/4}$ and $k_0\sim\sqrt{2\alpha d T}$, which implies
\begin{eqnarray}
\frac{R^2}{d} 
\;\sim\;
(2\alpha d T)^{-1/4},
\qquad
\frac{1}{d^2} \Big\langle \frac{N}{4}\int_0^{\beta} dt\Tr[X_a,X_b]^2 \Big\rangle
\;\sim\;-T (\alpha d)^{-1}.
\end{eqnarray}
At low temperature, the M.H.O. frequency saturates to
$s_0\sim(\alpha d)^{1/3}$ with $k_0\sim (\alpha d)^{2/3}$, yielding
\begin{eqnarray}
\frac{R^2}{d} 
\;\sim\;
(\alpha d)^{-1/3},
\qquad
\frac{1}{d^2} \Big\langle \frac{N}{4}\int_0^{\beta} dt\Tr[X_a,X_b]^2 \Big\rangle
\;\sim\;
- (\alpha d)^{-2/3}.
\end{eqnarray}
Thus, in both the low--$T$ and high--$T$ limits, the basic objects per matrix and per matrix pair
are parametrically suppressed at large $d$ (and, on the high--temperature branch, also with
increasing $T$).

These results imply that, in both limits, the dynamics is dominated by approximately commuting
matrices. Combined with the existence of the overlap window
$T_c^{\rm high}<T<T_c^{\rm low}$, this establishes an extended temperature range in which the
Gaussian/M.H.O. description is self--consistent and commuting--matrix configurations control
the dynamics. Importantly, this window widens with increasing $d$, providing a clear signal for
the emergence of IKKT--type physics in the large--$d$ limit.

\subsection{Size of the auxiliary-field fluctuations at high temperature}

\noindent
We now verify directly that the matrix harmonic oscillator description
remains valid on the high--temperature branch.  As in Sec.~4.3,
the quadratic kernel governing the auxiliary-field fluctuations contains
two contributions: the bare Hubbard--Stratonovich kernel and the
polarization kernel induced by the connected fluctuations of the
coordinate matrices.

\medskip
\noindent
Consider a time--independent fluctuation in the maximally symmetric
channel,
\begin{eqnarray}
k=k_0+\delta k,
\qquad
s^2=m+k_0.
\end{eqnarray}
Suppressing the holonomy shifts and the common adjoint multiplicity, the
effective action in this channel takes the schematic form
\begin{eqnarray}
S_{\rm eff}(k)
=
-\frac{\beta}{8\alpha}k^2
+
\frac{d}{2}
\sum_{n\in\mathbb Z}
\log\!\left(\omega_n^2+m+k\right).
\end{eqnarray}
The term linear in \(\delta k\) vanishes by the finite--temperature
auxiliary saddle equation
\begin{eqnarray}
-\frac{\beta k_0}{4\alpha}
+
\frac{d}{2}
\sum_{n\in\mathbb Z}
\frac{1}{\omega_n^2+s^2}
=
0\qquad \Longleftrightarrow\qquad
\frac{k_0}{\alpha}
=
\frac{d}{s}
\coth\!\left(\frac{\beta s}{2}\right).
\end{eqnarray}
The quadratic term generated by the coordinate determinant is governed
by the static polarization sum which at high temperature is given by 
\begin{eqnarray}
\sum_{n\in\mathbb Z}
\frac{1}{(\omega_n^2+s^2)^2}
&=&
\frac{\beta}{4s^3}
\coth\!\left(\frac{\beta s}{2}\right)
+
\frac{\beta^2}{8s^2}
\operatorname{csch}^2\!\left(\frac{\beta s}{2}\right)
\nonumber\\
&=&
\frac{1}{s^4}
\left[
1+\mathcal{O}\!\left((\beta s)^4\right)
\right],
\qquad
\beta s\ll1.
\end{eqnarray}
The coordinate contribution to a static eigenvalue of the auxiliary
Hessian therefore scales as
\begin{eqnarray}
\Gamma_{{\rm stat},X}^{(2)}
\sim
\frac{d}{s^4},
\end{eqnarray}
up to its numerical coefficient, sign, and adjoint tensor eigenvalue.

\medskip
\noindent
The bare contribution is proportional to \(\beta/\alpha\).  Expanding
the finite--temperature saddle equation for high temperature \(\beta s\ll1\), and using
\(k_0=s^2-m\), gives
\begin{eqnarray}
\frac{\beta}{\alpha}
&=&
\frac{2d}{s^2(s^2-m)}
\left[
1+\mathcal{O}\!\left((\beta s)^2\right)
\right]
\nonumber\\
&=&
\frac{2d}{s^4}
\frac{s^2}{s^2-m}
\left[
1+\mathcal{O}\!\left((\beta s)^2\right)
\right].
\end{eqnarray}
Thus, provided \(s^2-m\) is of order \(s^2\), the bare and polarization
contributions have the same parametric scaling.  For a stable eigenmode
of the complete tensor kernel, understood along the appropriate
steepest--descent contour, one may therefore write
\begin{eqnarray}
\Gamma_{\rm stat}^{(2)}
\sim
\frac{d}{s^4}\,
{\cal C}_{\rm HT}\!\left(\frac{m}{s^2}\right)
\left[
1+\mathcal{O}\!\left((\beta s)^2\right)
\right],
\end{eqnarray}
where \({\cal C}_{\rm HT}\) combines the bare and coordinate
polarization contributions.

\medskip
\noindent
Provided that \({\cal C}_{\rm HT}\neq0\) and that no parametrically soft
auxiliary mode is present, the variance of the corresponding static
fluctuation is controlled parametrically by the inverse Hessian:
\begin{eqnarray}
\left\langle(\delta k)^2\right\rangle
\sim
\frac{s^4}{d}\,
\frac{1}{{\cal C}_{\rm HT}(m/s^2)}.
\end{eqnarray}
Consequently,
\begin{eqnarray}
\frac{
\sqrt{\left\langle(\delta k)^2\right\rangle}
}{s^2}
\sim
\frac{1}{\sqrt{d}}\,
\frac{1}{
\sqrt{{\cal C}_{\rm HT}(m/s^2)}
}.
\end{eqnarray}
The auxiliary-field fluctuations on the high--temperature branch are
therefore suppressed as \(d^{-1/2}\).  Unlike the low--temperature
regime, there is no additional suppression by \(\beta s\), since the
static Matsubara mode dominates the high--temperature polarization.
The large--\(d\) limit nevertheless provides the required control of
the Gaussian saddle, while auxiliary-field fluctuations generate the
subleading non--Gaussian corrections to the matrix harmonic oscillator
description.

\medskip
\noindent
Thus the matrix harmonic oscillator description remains valid on the
high--temperature branch at leading order in the large--\(d\) expansion.
In contrast with the low--temperature regime, high temperature does not
provide an additional suppression of the auxiliary-field fluctuations;
rather, their relative size remains of order \(d^{-1/2}\).  The control
of the Gaussian saddle at high temperature is therefore supplied by the
large number of coordinate matrices, while fluctuations of the
auxiliary field generate the subleading non--Gaussian corrections.

\section{On supersymmetric extensions and  Molien--Weyl approximations}\label{section8}

\subsection{Saddle point equation with split masses}

In many supersymmetric BFSS$_{d+1}$ models with mass deformations (BMN$_{d+1}$),
the bosonic mass term breaks the $SO(d)$ rotational symmetry to
$SO(n_1)\times SO(n_2)$ with $n_1+n_2=d$.  
The quadratic potential then takes the form
\begin{eqnarray}
m_1\sum_{a_1=1}^{n_1}\Tr X_{a_1}^2
\;+\;
m_2\sum_{a_2=1}^{n_2}\Tr X_{a_2}^2.
\end{eqnarray}
\medskip
\noindent
Let us then split the transverse index as $a=(a_1,a_2)$ with

\begin{eqnarray}
a_1=1,\ldots,n_1,\qquad a_2=1,\ldots,n_2,\qquad n_1+n_2=d,
\end{eqnarray}
and assign masses $m_1$ and $m_2$ to the two flavors.

The Hubbard--Stratonovich localization proceeds exactly as in the $SO(d)$--symmetric
case; the only modification is that, after integrating out the bosonic matrices,
the one--loop determinants appear with multiplicities $n_1$ and $n_2$.

\medskip
\noindent Indeed, in the constant-$k$ sector, after integrating out all bosonic matrices, we obtain the effective action
\begin{eqnarray}
S_{\rm eff}[k]
&=&
\frac{n_1}{2}\sum_{n\in\mathbb{Z}}
\operatorname{Tr}\!\log\!\big(P{\cal W}^{(1)}(n)P\big)
\;+\;
\frac{n_2}{2}\sum_{n\in\mathbb{Z}}
\operatorname{Tr}\!\log\!\big(P{\cal W}^{(2)}(n)P\big)
\;+\;
\frac{N\beta}{4\alpha}\,\mu_{\mu\rho\nu\sigma}\,k_{\mu\rho}k_{\nu\sigma},\nonumber\\
\label{Seff-split-general}
\end{eqnarray}
with
\begin{eqnarray}
{\cal W}^{(s)}_{ij,kl}(n)
&=&
\Big(\omega_n+\frac{\theta_i-\theta_j}{\beta}\Big)^2\,P_{ij,kl}
\;+\;m_s\,P_{ij,kl}
\;+\;k_{ij,kl},
\qquad s=1,2.
\end{eqnarray}
Imposing the commuting--symmetric ansatz

\begin{eqnarray}
k_{ij,kl}\;\equiv\;k_{ij}\,P_{ij,kl}\;=\;P_{ij,kl}\,k_{kl},
\label{ansatz-k-split}
\end{eqnarray}
the quadratic kernels reduce to
\begin{eqnarray}
{\cal W}^{(s)}(n)_{ij,kl}
&=&
\Delta^{(s)}_{ij}(n)\,P_{ij,kl}
\;=\;
P_{ij,kl}\,\Delta^{(s)}_{kl}(n),
\label{W-Delta-split}
\\[1mm]
\Delta^{(s)}_{ij}(n)
&\equiv&
\Big(\omega_n+\frac{\theta_i-\theta_j}{\beta}\Big)^2+m_s+k_{ij},
\qquad s=1,2.
\label{Delta-split}
\end{eqnarray}
Therefore the effective action simplifies exactly as before to
\begin{eqnarray}
S_{\rm eff}[k]
&=&
\frac{1}{2}\sum_{s=1}^{2}n_s
\sum_{n\in\mathbb{Z}}\sum_{i,j=1}^{N}
P_{ij,ji}\,\log\!\Big(\Delta^{(s)}_{ij}(n)\Big)
\;+\;
\frac{N\beta}{4\alpha}\,\mu_{\mu\rho\nu\sigma}\,k_{\mu\rho}k_{\nu\sigma}.
\label{Seff-split-simplified}
\end{eqnarray}

\medskip
\noindent Varying now \eqref{Seff-split-simplified} with respect to the variables $k_{ij}$ gives
\begin{eqnarray}
0=\frac{\partial S_{\rm eff}}{\partial k_{ij}}
&=&
\frac{1}{2}\sum_{s=1}^{2}n_s
\sum_{n\in\mathbb{Z}}\,P_{ij,ji}\,
\frac{1}{\Delta^{(s)}_{ij}(n)}
\;+\;\frac{N\beta}{2\alpha}\,
\mu_{\mu\rho\nu\sigma}\,k_{\nu\sigma}\,
\frac{\partial k_{\mu\rho}}{\partial k_{ij}},
\label{var-kij-split}
\end{eqnarray}
which simplifies, following the same steps as before, to 

\begin{eqnarray}
0
&=&
\frac{1}{2}\sum_{s=1}^{2}n_s
\sum_{n\in\mathbb{Z}}\,P_{ij,ji}\,
\frac{1}{\Delta^{(s)}_{ij}(n)}
\;+\;\frac{N\beta}{2\alpha}\,
\mu_{\mu\rho\nu\sigma}\,k_{\nu\sigma}\,
\Lambda_{\mu}^{\,ij}\,\Lambda_{\rho}^{\,ji}.
\label{gap-pre-sum-split}
\end{eqnarray}

\medskip
\noindent Performing the Matsubara sums with
\begin{eqnarray}
\tilde{k}^{(s)}_{ij}\equiv k_{ij}+m_s,\qquad s=1,2,
\qquad (m_1\ \hbox{multiplicity }n_1,\;\; m_2\ \hbox{multiplicity }n_2),
\end{eqnarray}
yields

\begin{eqnarray}
\sum_{n\in\mathbb{Z}}
\frac{1}{\Big(\omega_n+\frac{\theta_i-\theta_j}{\beta}\Big)^2+\tilde{k}^{(s)}_{ij}}
&=&
\beta\,\frac{1}{2\sqrt{\tilde{k}^{(s)}_{ij}}}\,
\frac{\sinh\!\Big(\beta\sqrt{\tilde{k}^{(s)}_{ij}}\Big)}
{\cosh\!\Big(\beta\sqrt{\tilde{k}^{(s)}_{ij}}\Big)-\cos(\theta_i-\theta_j)}.
\label{Matsu-split}
\end{eqnarray}
Substituting \eqref{Matsu-split} into \eqref{gap-pre-sum-split} gives the split-mass saddle equation:
\begin{eqnarray}
0
&=&
\frac{\beta}{4}\,P_{ij,ji}\,
\sum_{s=1}^{2}n_s\,
\frac{1}{\sqrt{\tilde{k}^{(s)}_{ij}}}\,
\frac{\sinh\!\Big(\beta\sqrt{\tilde{k}^{(s)}_{ij}}\Big)}
{\cosh\!\Big(\beta\sqrt{\tilde{k}^{(s)}_{ij}}\Big)-\cos(\theta_i-\theta_j)}
\;+\;
\frac{N\beta}{2\alpha}\,
\mu_{\mu\rho\nu\sigma}\,k_{\nu\sigma}\,
\Lambda_{\mu}^{\,ij}\,\Lambda_{\rho}^{\,ji}.\nonumber\\
\label{gap-final-split}
\end{eqnarray}

\medskip
\noindent Next, we impose the maximally symmetric ansatz:

\begin{eqnarray}
k_{ij}=k_0\quad \forall\,i,j
\qquad\Longrightarrow\qquad
k_{ij,kl}=k_0\,P_{ij,kl}.
\label{max-sym-ansatz}
\end{eqnarray}

\medskip
\noindent With this maximally symmetric ansatz, the second term of \eqref{gap-final-split} simplifies as before while the first term  becomes a function of
\begin{eqnarray}
\tilde{k}^{(s)}_{ij}\;\to\;\tilde{k}^{(s)}_0\equiv k_0+m_s,
\qquad s_s\equiv \sqrt{\tilde{k}^{(s)}_0}=\sqrt{k_0+m_s}.
\label{defs-s-split}
\end{eqnarray}
Thus \eqref{gap-final-split} reduces to
\begin{eqnarray}
0
&=&
\frac{\beta}{4}\,P_{ij,ji}\,
\sum_{s=1}^{2}n_s\,
\frac{1}{s_s}\,
\frac{\sinh(\beta s_s)}{\cosh(\beta s_s)-\cos\phi}
\;-\;\frac{\beta k_0}{4\alpha}\,P_{ij,ji},
\qquad \phi\equiv\theta_i-\theta_j.
\label{gap-max-pre-sum}
\end{eqnarray}
Now sum over \(i,j\) and divide by
\(\sum_{i,j}P_{ij,ji}=N^2-1\). This gives the scalar maximally symmetric gap
equation
\begin{eqnarray}
\sum_{s=1}^{2}n_s\,
\frac{1}{s_s}\,
{\cal R}(s_s;U)
=
\frac{k_0}{\alpha},
\qquad
s_s=\sqrt{k_0+m_s},
\label{gap-exact-max-split}
\end{eqnarray}
where the projector--weighted holonomy kernel is
\begin{eqnarray}
{\cal R}(s_s;U)
&=&
\frac{1}{N^2-1}
\sum_{i,j=1}^{N}
P_{ij,ji}\,
\frac{\sinh(\beta s_s)}
{\cosh(\beta s_s)-\cos(\theta_i-\theta_j)} .
\label{Rkernel_split}
\end{eqnarray}
Equations~\eqref{gap-exact-max-split} and \eqref{Rkernel_split} are the
split--mass analogues of the original projector--averaged gap equation
\eqref{eq:gap-projector-averaged} and holonomy kernel
\eqref{Rkernel_projected}. The crucial point is that it is a
\emph{sum of two thermal factors} with two different gaps $s_1,s_2$. 

\medskip
\noindent The gap equation at low temperature is obtained following the same steps as before. We expand the $T$-dependent factor of \eqref{gap-exact-max-split} in the limit $\beta\to\infty$ as
\begin{eqnarray}
\frac{\sinh(\beta s_s)}{\cosh(\beta s_s)-\cos\phi}
&=&
1+2\cos\phi\,e^{-\beta s_s}+\mathcal{O}(e^{-2\beta s_s}).
\label{ratio-expansion-split}
\end{eqnarray}
Insert \eqref{ratio-expansion-split} into \eqref{gap-exact-max-split}:
\begin{eqnarray}
\sum_{s=1}^{2}n_s\,\frac{1}{s_s}
\Big[1+2\cos\phi\,e^{-\beta s_s}+\mathcal{O}(e^{-2\beta s_s})\Big]
&=&
\frac{k_0}{\alpha}.
\label{gap-expand-split}
\end{eqnarray}
At strictly $T=0$ (drop all $e^{-\beta s_s}$ terms) we get the \emph{split-mass $T=0$ gap equation}:
\begin{eqnarray}
\frac{n_1}{s_1}+\frac{n_2}{s_2}
\;=\;\frac{n_1}{\sqrt{k_0+m_1}}+\frac{n_2}{\sqrt{k_0+m_2}}
\;=\;
\frac{k_0}{\alpha}.
\label{gap-zeroT-split}
\end{eqnarray}
This replaces our cubic $s_0^3-ms_0=\alpha d$ and, in general, does \emph{not} collapse to a single
effective mass parameter.

\medskip
\noindent
The $T=0$ maximally symmetric gap equation  \eqref{gap-zeroT-split} can also be put in the following equivalent form 
\begin{eqnarray}
\frac{n_2}{\sqrt{s_1^2+\Delta m}}
=\frac{s_1^2-m_1}{\alpha}-\frac{n_1}{s_1},
\qquad
\Delta m:=m_2-m_1.
\label{isolate}
\end{eqnarray}
Squaring once yields an eighth--degree polynomial equation for $s_1$,
\begin{eqnarray}
\big(s_1^3-m_1 s_1-\alpha n_1\big)^2\,(s_1^2+\Delta m)
-\alpha^2 n_2^{\,2}\,s_1^{\,2}=0,
\label{gap_T0_split_polynomial}
\end{eqnarray}
supplemented by the sign condition
\begin{eqnarray}
\frac{s_1^2-m_1}{\alpha}-\frac{n_1}{s_1}>0,
\label{sign_constraint}
\end{eqnarray}
which discards spurious solutions introduced by squaring.

\medskip
\noindent
Equivalently, the two--flavor saddle point equation, or more precisely the gap equation \eqref{gap-zeroT-split} or \eqref{gap_T0_split_polynomial}, is a \emph{single} self--consistency condition for
the Hubbard--Stratonovich mass parameter $k_0$:
\begin{eqnarray}
s_1^2=m_1+k_0,\qquad s_2^2=m_2+k_0
\qquad\Longrightarrow\qquad
s_2^2-s_1^2=m_2-m_1,
\label{eq:s1s2_def}
\end{eqnarray}
or, solving in terms of $s_1$,
\begin{eqnarray}
k_0=s_1^2-m_1,\qquad
s_2=\sqrt{s_1^2+\Delta m}.
\label{reconstruct}
\end{eqnarray}
Thus the two gaps are not independent; both are determined by the same saddle $k_0$. The cubic equation of the single--flavor case is simply replaced by an 
eighth--degree equation in the split--flavor case.

\medskip
\noindent
Now introduce the fractions $r_s=n_s/d$ (with $r_1+r_2=1$) and the large--$d$ scaling variables
\begin{eqnarray}
s_*=(\alpha d)^{1/3},\qquad k_*=(\alpha d)^{2/3}.
\end{eqnarray}
Solving \eqref{gap-zeroT-split} at large $d$ yields
\begin{eqnarray}
k_0
&=&
k_*-\frac{1}{3}\,\bar m+\mathcal{O}(d^{-2/3}),
\qquad
\bar m=r_1 m_1+r_2 m_2,
\label{eq:k0_leading}
\\[2mm]
s_{1,0}
&=&
s_*+\frac{m_1}{2s_*}-\frac{\bar m}{6s_*}
+\mathcal{O}(d^{-1}),
\label{eq:s10_leading}
\\[2mm]
s_{2,0}
&=&
s_*+\frac{m_2}{2s_*}-\frac{\bar m}{6s_*}
+\mathcal{O}(d^{-1}).
\label{eq:s20_leading}
\end{eqnarray}
In particular,
\begin{eqnarray}
s_{1,0}\sim s_{2,0}\sim(\alpha d)^{1/3},
\qquad
s_{2,0}^2-s_{1,0}^2=m_2-m_1,
\qquad
\frac{s_{2,0}^2-s_{1,0}^2}{s_*^{\,2}}=\mathcal{O}(d^{-2/3}),
\end{eqnarray}
so the mass splitting is parametrically subleading in the large--$d$ expansion.
At leading order, both flavors share the same dynamically generated gap which swamps their bare masses.

\medskip
\noindent
Finally, by expanding the localized action about $k_0$, the dominant bosonic theory is therefore
a sum of two gauged matrix harmonic oscillators,
\begin{eqnarray}
S_{\rm MHO}^{\rm split}[X;U]
=
N\int_{0}^{\beta}\!dt\ \Tr\Bigg[
\frac{1}{2}\sum_{a=1}^{d}(D_tX_a)^2
+\frac{s_1^2}{2}\sum_{a_1=1}^{n_1}X_{a_1}^2
+\frac{s_2^2}{2}\sum_{a_2=1}^{n_2}X_{a_2}^2
\Bigg].
\label{eq:MHO_split}
\end{eqnarray}

\subsection{Gaussian Molien--Weyl approximation}
The basic model of interest is the bosonic BFSS$_{d+1}$ theory with a mass deformation,
namely the BMN$_{d+1}$ model, defined by

\begin{eqnarray}
S
=
N\int_{0}^{\beta}\! dt\ \mathrm{Tr}\bigg[\frac{1}{2}(D_tX_a)^2+\frac{m}{2}X_a^2-\frac{1}{4}[X_a,X_b]^2\bigg].
\end{eqnarray}
In the large--$d$ limit, this interacting theory is well approximated by a gauged
matrix harmonic oscillator,

\begin{eqnarray}
   S=N\int dt \mathrm{Tr}\bigg[\frac{1}{2}(D_tX_a)^2+\frac{1}{2}s^2X_a^2\bigg],\label{singlemass}
\end{eqnarray}
where the dynamically generated mass $s^2=k_0+m$  is determined by the cubic gap equation
$s^3-ms-\alpha d=0$. Here, $k_0$  is the large--$d$ saddle point.

\medskip
\noindent
In the presence of a split mass deformation, the quadratic and interaction terms are
effectively replaced as

\begin{eqnarray}
\frac{m_1}{2}\sum_{a_1=1}^{n_1}X_a^2+\frac{m_2}{2}\sum_{a_2=1}^{n_2}X_a^2-\frac{1}{4}[X_a,X_b]^2\longrightarrow \frac{1}{2}s_1^2\sum_{a_1=1}^{n_1}X_{a_1}^2+\frac{1}{2}s_2^2\sum_{a_2=1}^{d}X_{a_2}^2,
\end{eqnarray}
leading to the effective action
\begin{eqnarray}
   S=N\int dt \mathrm{Tr}\bigg[\frac{1}{2}(D_tX_a)^2+\frac{1}{2}s_1^2X_{a_1}^2+\frac{1}{2}s_2^2X_{a_2}^2\bigg].\label{splitmass}
\end{eqnarray}
The two gaps are related by $s_2=\sqrt{s_1^2+\Delta m}=k_0+m_2$, while $s_1=k_0+m_1$ is fixed by the
eighth--degree equation
\begin{eqnarray}
\big(s_1^3-m_1 s_1-n_1\big)^2\,(s_1^2+\Delta m)-n_2^{\,2}\,s_1^{\,2}=0.
\end{eqnarray}

\medskip
\noindent
In the large--$d$ limit the single--flavor gap behaves as $s\sim d^{1/3}$, and the
two--flavor gaps become degenerate,
$s_1\sim s_2\sim d^{1/3}$.
Consequently, even in the split--mass case, the dynamics is well approximated at leading
order by the single--mass action \eqref{singlemass}.

Moreover, in the absence of mass deformation, corresponding to the genuine
BFSS$_{d+1}$ model, the large--$d$ gap equation yields the exact relation
\(
s^2 = d^{1/3}.
\)

\medskip
\noindent
Since the resulting theory is Gaussian, the integration over the matrices $X_a$ can be
performed exactly. Fixing the gauge on the thermal circle reduces the gauge field to a
constant holonomy
\(
g=\mathcal{P}\exp\big(i\int_0^\beta dt\,A_0\big),
\)
and the path integral reduces to a group integral over $g$.
Diagonalizing $g=\mathrm{diag}(z_1,\ldots,z_N)$, the normal--ordered $SU(N)$ partition function
can be written as a Molien--Weyl integral~\cite{OConnor:2023mss,OConnor:2024udv}
\begin{eqnarray}
  Z_{N,d}(x)&=&\frac{1}{N!}\frac{1}{(1-x_1)^{n_1(N-1)}(1-x_2)^{n_2(N-1)}}\oint \prod_{i=1}^N\frac{dz_i}{2\pi i z_i}\Delta_A(-1,z)\frac{1}{\Delta_b^{n_2}(-x_2,z)\Delta_b^{n_1}(-x_1,z)},\nonumber\\
\end{eqnarray}
where the Vandermonde--Faddeev--Popov determinant $\Delta_A$ and the bosonic determinant
$\Delta_B$ are built from
\begin{eqnarray}
\Delta(x,z)
=
\prod_{i<j}\Big(1+x\,\frac{z_i}{z_j}\Big)
\prod_{i<j}\Big(1+x\,\frac{z_j}{z_i}\Big),
\end{eqnarray}
and with the fugacities $x_1$ and $x_2$ defined in terms of the masses $m_1$ and $m_2$ by the equation
\begin{eqnarray}
x_i=e^{-\beta m_i}.
\end{eqnarray}

\medskip
\noindent
These holonomy models are closely related to multitrace matrix models, in which the
dynamics is entirely encoded in the eigenvalue distribution~\cite{Ydri:2021cam,Ydri:2015vba}.

In this language, the
bosonic determinants (and, in supersymmetric extensions, the combined bosonic and
fermionic determinants) play, for the holonomy angles, the role that the classical
scalar potential
\(\Tr (b M^2 + c M^4 + \cdots)\)
plays for the eigenvalues of $M$ in multitrace matrix models.
More generally, one may introduce an explicit eigenvalue potential $V(\theta)$ through
a Boltzmann weight $\exp[-V(\theta)]$, leading to a broad class of scalar--gauge matrix
models governed by hybrid scalar--holonomy dynamics.

\subsection{Supersymmetric completion and Monte Carlo simulations}

\noindent
To construct the supersymmetric completion of the Gaussian models
discussed above, it is instructive to work out explicitly the case of
\(\mathrm{BMN}_3\), and then extend the same construction to the allowed
supersymmetric \(\mathrm{BFSS}_{d+1}\) and \(\mathrm{BMN}_{d+1}\) models
listed in Sec.~1.1.

\medskip
\noindent
However, the meaning of the large--\(d\) construction in the supersymmetric
case should be first stated carefully.  Supersymmetry restricts the admissible
values of \(d\) through the corresponding Fierz identities, and hence
there is no continuous family of fully supersymmetric
\(\mathrm{BFSS}_{d+1}\) or \(\mathrm{BMN}_{d+1}\) theories for arbitrary
\(d\).  Nevertheless, the large--\(d\) limit used here is literal at the
level of the bosonic sector: one first determines the self--consistent
large--\(d\) bosonic gap mass \(s^2\), or the corresponding set of gap
masses in the split--mass case.  The resulting Gaussian bosonic theory
is then completed supersymmetrically, by analogy with the large--mass
Gaussian limit of the corresponding supersymmetric model, through the
addition of the fermionic sector and the imposition of the
boson--fermion mass relations dictated by supersymmetry.

\medskip
\noindent
The resulting theory is therefore the supersymmetric completion of the
large--\(d\) bosonic Gaussian saddle.  The qualifier ``large--\(d\)''
refers to the origin of the dynamically generated bosonic masses,
whereas the supersymmetric completion is performed at each admissible
fixed value of \(d\). This construction is implemented explicitly by the three--step
procedure \ref{step:gap}, \ref{step:fermionic-sector}, and
\ref{step:completion} given below.

\medskip
\noindent
The starting point is then BFSS$_3$ matrix quantum mechanics with mass deformation defined by the action

 \begin{eqnarray}
   S_{\rm BFSS_3}&=&N\int dt \mathrm{Tr}\bigg[\frac{1}{2}(D_tX_a)^2-\frac{1}{2}[X_1,X_2]^2+\frac{i}{2}\bar{\Psi} \gamma^0_ED_t\Psi+\frac{1}{2}\bar{\Psi}\gamma^a_E[X_a,\Psi]\nonumber\\
     &+&\frac{\mu^2}{72}X_a^2-\frac{i\mu}{8}\bar{\Psi}\Psi \bigg].
 \end{eqnarray}
 The simplest approximation is obtained by dropping the interaction terms, leading to the
Gaussian theory

 \begin{eqnarray}
   S_{\rm BFSS_3}^{\rm Gaussian}&=&N\int dt \mathrm{Tr}\bigg[\frac{1}{2}(D_tX_a)^2+\frac{i}{2}\bar{\Psi} \gamma^0_ED_t\Psi+\frac{\mu^2}{72}X_a^2-\frac{i\mu}{8}\bar{\Psi}\Psi \bigg],
 \end{eqnarray}
 which is still supersymmetric. Writing the fermions in terms of a complex spinor
$\psi$, this action takes the form
 
 \begin{eqnarray}
   S_{\rm BFSS_3}^{\rm Gaussian}
   =N\int dt \mathrm{Tr}\bigg[\frac{1}{2}(D_tX_a)^2+\frac{1}{2}{\psi}^{\dagger}D_t\psi+\frac{\mu^2}{72}X_a^2+\frac{\mu}{8}{\psi}^{\dagger}\psi \bigg].
 \end{eqnarray}
 
\medskip
\noindent
This action should in fact be compared with the BFSS$_2$ model defined by the action 

       \begin{eqnarray}
        S_{\rm BFSS_2}=N\int dt \mathrm{Tr}\bigg(\frac{1}{2}(D_tX)^2+\frac{1}{2}{\psi} D_t\psi-\frac{1}{2}\Lambda(t)X^2-\rho(t) X \bigg),
       \end{eqnarray}
       with $\rho(t)=0$ and $-\Lambda=\mu^2/36$. Identifying $m=m_b^2$, one finds the mass
relation
\begin{eqnarray}
m_b=\frac{\mu}{6},
\qquad
m_f=\frac{\mu}{4}=\frac{3}{2}\,m_b .
\end{eqnarray}
\medskip
\noindent
It follows that the supersymmetric analogue of the large--$d$ single--mass Gaussian
model \eqref{singlemass} is
 
 \begin{eqnarray}
   S_{\rm BFSS_3}^{\rm Gaussian/SUSY}=N\int dt \mathrm{Tr}\bigg[\frac{1}{2}(D_tX_a)^2+\frac{1}{2}s_b^2X_a^2+\frac{i}{2}\bar{\Psi} \gamma^0_ED_t\Psi-\frac{is_f}{2}\bar{\Psi}\Psi \bigg],\label{singlemass-susy}
 \end{eqnarray}
 with
 \begin{eqnarray}
  s_b^2=k_0+m=k_0+m_b^2,\qquad s_f=\frac{3}{2}s_b.
 \end{eqnarray}
 \medskip
\noindent
An intermediate model between BFSS$_3$ and BFSS$_2$ consists of a single bosonic matrix
(as in BFSS$_2$) but a complex fermion (as in BFSS$_3$):
        \begin{eqnarray}
   S_{\rm BFSS_{2/3}}^{\rm Gaussian}
   &=&N\int dt  \mathrm{Tr} \bigg[\frac{1}{2}(D_tX)^2+\frac{1}{2}{\psi}^{\dagger}D_t\psi-\frac{1}{2}\Lambda(t) X^2-\rho(t) X+\frac{\mu}{8}{\psi}^{\dagger}\psi \bigg].
        \end{eqnarray}
        \medskip
\noindent
Introducing the fugacities
\begin{eqnarray}
x_b=e^{-\beta m_b},
\qquad
x_f=e^{-\beta m_f},
\end{eqnarray}
the quadratic bosonic and fermionic degrees of freedom can be integrated out exactly
using the Molien--Weyl formula~\cite{OConnor:2023mss,OConnor:2024udv}. For BFSS$_3$ one obtains

        \begin{eqnarray}
          Z_N^{(3)}(x_b,x_f)&=&\frac{1}{N!}\frac{1}{2^{N-1}}\frac{(1+x_f)^{N-1}}{(1-x_b)^{N-1}}\oint \prod_{i=1}^N\frac{dz_i}{2\pi i z_i}\Delta_A(-1,z)\frac{\Delta_f(x_f,z)}{\big[\Delta_b(-x_b,z)\big]^2}.\nonumber\\\label{5N}
        \end{eqnarray}
        Here,  $(m_b,m_f)=(\tfrac{\mu}{6},\tfrac{\mu}{4})$ or $(m_b,m_f)=(s_b,s_f)$.

        \medskip
\noindent
BFSS$_2$ is special in that it contains a massless real fermion, yet its Hamiltonian formulation still involves a complex spinor (like BFSS$_3$). Thus, its Molien--Weyl partition function reads

        \begin{eqnarray}
          Z_N^{(2)}(x_b,x_f)&=&\frac{1}{N!}\frac{1}{2^{N-1}}\frac{(1+x_f)^{N-1}}{(1-x_b)^{N-1}}\oint \prod_{i=1}^N\frac{dz_i}{2\pi i z_i}\Delta_A(-1,z)\frac{\Delta_f(x_f,z)}{\Delta_b(-x_b,z)},\nonumber\\
          \label{4N}
        \end{eqnarray}
        with $(m_b,m_f)=(\sqrt{-\Lambda},0)$. In fact, this partition functions holds also for the BFSS$_{2/3}$ model with 
$(m_b,m_f)=(\sqrt{-\Lambda},\mu/4)$.

\medskip
\noindent
For a general BMN$_{d+1}$ model, the construction of its supersymmetric
large--$d$ Gaussian approximation follows a simple three--step procedure:
\begin{enumerate}

\item\label{step:gap}
determine the large--$d$ bosonic gap mass $s_1$, or the pair $s_1,s_2$
in the split--mass case;

\item\label{step:fermionic-sector}
analyze the large--mass limit of the supersymmetric theory to fix the
form of the fermionic sector and the boson--fermion mass relation that
preserve supersymmetry at the Gaussian level;

\item\label{step:completion}
replace the bare bosonic masses by their large--$d$ gap values and add
the fermionic completion dictated by the large--mass Gaussian theory,
while keeping the boson--fermion mass relation fixed.

\end{enumerate}

\medskip
\noindent
Thus, for bosonic masses $m_1,m_2$ with multiplicities $n_1,n_2$ and $n_f$ complex
fermions of mass $m_f$, the normal--ordered $SU(N)$ partition function takes the
Molien--Weyl form~\cite{OConnor:2023mss,OConnor:2024udv}

\begin{eqnarray}
  Z_{N,d}(x)&=&\frac{1}{N!}\frac{(1+x_f)^{n_f(N-1)}}{(1-x_1)^{n_1(N-1)}(1-x_2)^{n_2(N-1)}}\oint \prod_{i=1}^N\frac{dz_i}{2\pi i z_i}\Delta_A(1,z)\frac{\Delta_f^{n_f}(x_f,z)}{\Delta_b^{n_2}(-x_2,z)\Delta_b^{n_1}(-x_1,z)}.\nonumber\\
\end{eqnarray}
Equivalently, this can be written as a purely holonomy model,
\begin{eqnarray}
Z_{N,d}(x)
=
\oint \prod_{i=1}^N\frac{d\theta_i}{2\pi}\,
e^{-S_{\rm eff}(\theta)} ,
\label{pureholonomy}
\end{eqnarray}
which defines the gauged supersymmetric matrix harmonic oscillator (SMHO), generalizing the gauged matrix harmonic oscillator (MHO) defined by equation  \eqref{splitmass}. The single--flavor case is recovered by setting $s_1=s_2$.

Alternatively, one may integrate out only the Gaussian fermionic degrees of freedom using the
Molien--Weyl formula, leading to a purely bosonic holonomy model, which we refer to as
the \emph{Bosonic Molien--Weyl} model. In this current formulation the bosonic sector is
Gaussian, while the full bosonic action—including the commutator interaction terms,
both Yang--Mills and possible Chern--Simons contributions—may also be retained explicitly yielding a genuine  \emph{Bosonic Molien--Weyl} model~\cite{Ydri2025}.

Accordingly, the coordinate and gauge fields are treated as in the standard lattice
formulation of BFSS$_{d+1}$ matrix models~\cite{Ydri2025}. In particular, we
non--perturbatively fix the local $U(N)$ gauge symmetry on the lattice by assigning the
holonomy matrix
\(
D=\mathrm{diag}(e^{i\theta_1},\ldots,e^{i\theta_N})
\)
to the final temporal link. This procedure leads to a bosonized matrix quantum mechanics
of the form

\begin{eqnarray}
  S_{\rm eff}&=&N\sum_{n=0}^{\Lambda-1}\bigg[\frac{1}{a}\mathrm{Tr}{X}_a^{ 2}(n)+\frac{as_1^2}{2}\sum_{a_1=1}^{n_1}\mathrm{Tr}{X}_{a_1}^{ 2}(n)+\frac{as_2^2}{2}\sum_{a_2=1}^{n_2}\mathrm{Tr}{X}_{a_2}^{ 2}(n)\bigg]\nonumber\\
  &-&\frac{1}{a}\sum_{n=0}^{\Lambda-2}\mathrm{Tr}{X}_a^{}(n){X}_a^{}(n+1)-\frac{N}{a}\sum_{i,j}e^{-i(\theta_i-\theta_j)}({X}_a^{}(\Lambda-1))_{ij}({X}_a^{}(0))_{ji}\nonumber\\
  &-&\frac{1}{2}\sum_{i\ne j}\ln\sin^2\frac{\theta_i-\theta_j}{2}~,~{\rm Vandermonde~determinant}\nonumber\\
     &-&\frac{n_f}{2}\sum_{i\ne j}\ln\bigg[(1+x_f)^2-4x_f\sin^2\frac{\theta_i-\theta_j}{2}\bigg]~,~{\rm fermion~determinant}.\label{bosonized}
\end{eqnarray}
\medskip
\noindent
The lattice has spacing $a$, with the inverse temperature defined by \(\beta = \Lambda\, a\). Again, the single--flavor case is recovered by setting $s_1=s_2$, which indeed coincide
as $s_1\sim s_2\sim d^{1/3}$ in the large--$d$ limit. In the absence of mass deformation,
corresponding to the true BFSS$_{d+1}$ model, we have exactly
$s=d^{1/3}$. The values of $n_i$, $s_i$, and $n_f$ for the models considered here
are summarized in Table~(\ref{so3II}). Monte Carlo simulations of these models, either in the purely holonomic
form~\eqref{pureholonomy} or in the bosonized lattice form~\eqref{bosonized},
will be reported in detail elsewhere~\cite{Ydri2025}.

\begin{table}[h]
\centering
\renewcommand{\arraystretch}{1.5}
\begin{tabular}{|p{2.0cm}|p{5.0cm}|p{5.9cm}|p{3.2cm}|}
\hline
\textbf{BFSS\(_{d+1}\) QM} & \textbf{BPS configuration} & \textbf{Spectrum} & \textbf{Deformation Parameter} \\
\hline
\textbf{BMN\(_{10}\)} & \(\begin{aligned}
&&X_a = 0~,~a=1,...,6 \\
  &&D_t X^p = 0~,~p=7,8,9\\
  &&[X^p, X^q] = i\,\frac{1}{3} \mu\, \epsilon^{pqr} X_r
\end{aligned}
 \) & \(
\begin{aligned}
  &6B~:~ m_{b1}=\mu/6~,~ n_{b1}=6\\
  &3B~:~m_{b2}=\mu/3~,~n_{b2}=3\\
  &8F~:~m_f=\mu/4~,~n_f=8.
\end{aligned}
  \) & \( \mu \) \\
\hline
\textbf{BMN\(_{6}\)} type I & \(\begin{aligned}
&&X_1 = R \cos(\frac{1}{6} t \mu), \\
&&X_2 = R \sin( \frac{1}{6} t \mu), \\
  &&D_t X^p = 0~,~p=3,4,5\\
  &&[X^p, X^q] = i\,\frac{1}{3} \mu\, \epsilon^{pqr} X_r
\end{aligned}
\) & \(
\begin{aligned}
  &2B~:~ m_{b1}=\mu/6~,~n_{b1}=2\\
  &3B~:~m_{b2}=\mu/3~,~n_{b2}=3\\
  &4F~:~m_f=\mu/4~,~n_f=4.
\end{aligned}
  \) & \( \mu \) \\
\hline
\textbf{BMN\(_{6}\)} type II & \text{None} & \(
\begin{aligned}
  &1B~:~ m_{b1}=\mu/2~,~n_{b1}=1\\
  &4B~:~m_{b2}=\mu/6~,~n_{b2}=4\\
  &4F~:~m_f=\mu/4~,~n_f=4.
\end{aligned}
  \) & \( \mu \) \\
\hline
\textbf{BMN\(_{4}\)} type I & \(
\begin{aligned}
  &&D_t X^a = 0~,~a=1,2,3\\
  &&[X^a, X^b] = i\,\frac{1}{3} \mu_2\, \epsilon^{abc} X_c
\end{aligned}
  \) & \(
\begin{aligned}
  &3B~:~ m_{b}=\sqrt{\mu_1^2+\mu_2^2}/3~,~n_b=3\\
  &2F~:~m_f=\sqrt{\mu_1^2+\mu_2^2}/2~,~n_f=2.
\end{aligned}
  \)  & \( \mu_1 \ne 0,\, \mu_2 \) \\
\hline
\textbf{BMN\(_{4}\)} type I & \(\begin{aligned}
  &&D_t X^a = 0,\\
  &&[X^a, X^b] = i\,\frac{1}{3} \mu_2\, \epsilon^{abc} X^c
\end{aligned}
 \) & \(
\begin{aligned}
  &3B~:~ m_{b}=\mu_2/3~,~n_{b}=3\\
  &2F~:~m_f=\mu_3/2~,~n_f=2.
\end{aligned}
  \) & \( \mu_2,\, \mu_3~ \text{is fictitious} \) \\
\hline
\textbf{BMN\(_{4}\)} type II & 
\(
\begin{aligned}
&&X_1 = R \cos(\frac{1}{6} t \mu_1), \\
&&X_2 = R \sin( \frac{1}{6} t \mu_1), \\
&&X_3 = 0
\end{aligned}
\) & \(
\begin{aligned}
  &2B~:~ m_{b1}=\mu/6~,~n_{b1}=2\\
   &1B~:~ m_{b2}=\mu/3~,~n_{b2}=1\\
  &2F~:~m_f=\mu/4~,~n_f=2.
\end{aligned}
  \) & \( \mu \) \\
\hline
\textbf{BMN\(_{3}\)} & \(\begin{aligned}
&&X_1 = R \cos(\frac{1}{6} t \mu), \\
&&X_2 = R \sin( \frac{1}{6} t \mu).
\end{aligned}
\) & \(
\begin{aligned}
  &2B~:~ m_{b1}=\mu/6~,~n_b=2\\
  &1F~:~m_f=\mu/4~,~n_f=1.
\end{aligned}
  \) & \( \mu \) \\
\hline
\textbf{BMN\(_{2}\)} & \( {\bf AdS}^2 \) & \(
\begin{aligned}
  &1B~:~ m_{b1}=\Lambda(t)~,~n_b=1\\
  &1/2F~:~m_f=0~,~n_f=1.
\end{aligned}
  \) & \( \Lambda(t),\, \rho(t) \) \\
\hline
\end{tabular}
\caption{Summary of deformation parameters, spectra and classical BPS solutions of the BFSS/BMN models.}
\label{so3II}
\end{table}

\section{Conclusion}\label{conclusion}
\medskip
\noindent
We have analyzed the large--\(d\) dynamics of the mass--deformed bosonic
\(\mathrm{BFSS}_{d+1}\) matrix model in a correlated double--scaling limit in
which
\begin{eqnarray}
d\longrightarrow \infty,
\qquad
m\longrightarrow \infty,
\qquad
\kappa\equiv\frac{m^{3/2}}{\alpha d}
\quad \hbox{fixed}.
\end{eqnarray}
Equivalently,
\begin{eqnarray}
m=(\kappa\alpha d)^{2/3},
\end{eqnarray}
so that, after setting \(\alpha=1\), the ratio \(m/d^{2/3}\) is held fixed. This
is precisely the scaling in which the explicit mass deformation and the
dynamically generated Yang--Mills mass shift remain parametrically balanced.

\medskip
\noindent
The analysis is controlled by a Gaussian saddle, equivalently a gauged matrix
harmonic oscillator, whose self--consistent frequency \(s\) is determined by the
large--\(d\) gap equation. In the double--scaling limit this saddle remains valid
over a parametrically wide range of temperatures, covering both the enlarged
uniform--holonomy regime and the high--temperature Gaussian branch.

\medskip
\noindent
Our first main result is that, in this double--scaling limit, the mass--deformed
\(\mathrm{BFSS}_{d+1}\) model is genuinely dominated by the uniform,
center--symmetric holonomy phase. In particular, the effective holonomy action
takes the universal Gross--Witten--Wadia form with coefficients \(1-dq^n\), where
\begin{eqnarray}
q=e^{-\beta s}.
\end{eqnarray}
Since the saddle frequency scales as \(s\sim d^{1/3}\), the deconfinement
instability is pushed to parametrically higher temperature,
\begin{eqnarray}
T_c\sim \frac{d^{1/3}}{\log d}.
\end{eqnarray}
Thus the uniform holonomy region expands parametrically with \(d\). This is the
only regime observed in \(\mathrm{BFSS}_2\) matrix quantum mechanics, where no
finite--temperature deconfinement transition exists. As a consequence, the
double--scaled \(\mathrm{BFSS}_{d+1}\) saddle does not merely produce a Gaussian
approximation: it opens a parametrically large \(\mathrm{BFSS}_2\)--like, or
AdS\(_2\)--like, stringy regime.

\medskip
\noindent
In this regime the bulk dynamics reduces to \(d\) effectively independent gauged
matrix harmonic oscillators. The holonomy-mediated interaction between different
matrix directions is weak because the system lies deep in the uniform holonomy
region at large \(d\), where
\begin{eqnarray}
dq=d\,e^{-\beta s}\ll 1.
\end{eqnarray}

\medskip
\noindent
Our second main result concerns the structure of the theory in the overlap window
\begin{eqnarray}
T_c^{\rm high}<T<T_c^{\rm low},
\end{eqnarray}
where both the low--temperature and high--temperature saddle--point analyses are
simultaneously valid. In this window, explicit evaluation of the extent of space
and of the Yang--Mills observable shows that the basic quantities per matrix and
per matrix pair are parametrically suppressed at large \(d\). In particular, the
commutator contribution between two fixed matrix directions satisfies
\begin{eqnarray}
-\Big\langle \Tr[X_1,X_2]^2\Big\rangle
\sim d^{-2/3}
\longrightarrow 0
\qquad (d\to\infty).
\end{eqnarray}
Thus noncommutativity is weak at the saddle point, and the dynamics is dominated
by approximately commuting matrices. The resulting physics is therefore
IKKT--like: spacetime emerges from nearly commuting matrices, while interactions
are encoded in collective large--\(N\) effects rather than in strong matrix
noncommutativity.

\medskip
\noindent
Importantly, this overlap window widens with increasing \(d\). Together with the
parametrically enlarged uniform-holonomy regime described above, this provides a
clear signal that the double--scaling limit of the mass--deformed
\(\mathrm{BFSS}_{d+1}\) model interpolates between two complementary descriptions:
a \(\mathrm{BFSS}_2\)--like, AdS\(_2\)--like stringy regime governed by weakly
coupled single--matrix gauged harmonic-oscillator sectors, and an IKKT--type
commuting--matrix regime in which the matrices are localized and almost
commuting. This identifies the double--scaled large--\(d\) model as a natural
arena in which both descriptions coexist and can be treated in a unified and
controlled framework.

\medskip
\noindent
Finally, we explained how the large--\(d\) Gaussian framework extends to
mass--deformed supersymmetric BFSS/BMN models and their Molien--Weyl approximations.
In the split--mass case, different bosonic sectors acquire different oscillator
gaps, but these gaps are tied together by a single self--consistent
Hubbard--Stratonovich saddle and become degenerate at leading order in large
\(d\). The resulting Gaussian theory admits two useful formulations: a purely
holonomic Molien--Weyl model, obtained by integrating out all Gaussian bosonic and
fermionic degrees of freedom, and a bosonized lattice model, in which only the
fermions are integrated out while the bosonic matrices remain explicit. These formulations provide practical Gaussian benchmarks for Monte Carlo
simulations of the full interacting BFSS/BMN theory.

\section{Acknowledgments}

\medskip
\noindent
The author would like to acknowledge helpful discussions with Denjoe O'Connor from the Dublin Institute for Advanced Studies. The author is especially grateful for Denjoe O'Connor's continued institutional hosting and generous support over the years, including travel, accommodation, and living expenses.

\medskip
\noindent
The author also acknowledges the use of ChatGPT-5.5, as well as previous versions, in several auxiliary capacities: 
(1) as a language editor; 
(2) as a LaTeX generator; 
(3) as a Mathematica-like symbolic tool; 
(4) as an assistant in searching for and reviewing references; 
and, more importantly, 
(5) as an ``artificial'' sounding board for testing, organizing, and refining ideas, effectively replacing in this role the function often played by human collaborators. However, the scientific vision, concept, design, direction, final scientific and mathematical editing, and all intellectual responsibility for this work remain solely with the author.

\appendix 
\section{Fourier derivation of the holonomic effective action}\label{appendix-holonomy-effective-action}

\subsection{Coordinate contribution}
\noindent
The $X_a$ determinant contributes the action 
\begin{eqnarray}
S_X[\theta]
&=&
\frac{d}{2}\sum_{i,j=1}^N P_{ij,ji}\,
\log\!\Big(\cosh(\beta s)-\cos(\theta_i-\theta_j)\Big),
\end{eqnarray}
up to $\theta$--independent constants.

For $x>0$ one has the identity
\begin{eqnarray}
\log(\cosh x-\cos\phi)
=
\text{const}
-2\sum_{n\ge 1}\frac{e^{-nx}}{n}\cos(n\phi),
\qquad x>0,
\label{log-Fourier}
\end{eqnarray}
where again ``const'' is independent of $\phi$. To derive this relation, we start from the elementary factorization

\begin{eqnarray}
\cosh x-\cos\phi
=
\frac{1}{2}\Big(e^{x}+e^{-x}-e^{i\phi}-e^{-i\phi}\Big)=
\frac{1}{2}e^{x}\Big(1-e^{-x}e^{i\phi}\Big)\Big(1-e^{-x}e^{-i\phi}\Big),
\qquad x>0.\nonumber\\
\end{eqnarray}
Taking the logarithm yields
\begin{eqnarray}
\log(\cosh x-\cos\phi)
&=&
x-\log 2
+\log\!\big(1-qe^{i\phi}\big)
+\log\!\big(1-qe^{-i\phi}\big),
\qquad q\equiv e^{-x}\in(0,1).\nonumber\\
\end{eqnarray}
Using the Taylor expansion
\begin{eqnarray}
\log(1-z)=-\sum_{n\ge1}\frac{z^n}{n},
\qquad |z|<1,
\end{eqnarray}
we then obtain
\begin{eqnarray}
\log\!\big(1-qe^{i\phi}\big)
=
-\sum_{n\ge1}\frac{q^n}{n}e^{in\phi},\qquad
\log\!\big(1-qe^{-i\phi}\big)
=
-\sum_{n\ge1}\frac{q^n}{n}e^{-in\phi}.
\end{eqnarray}
Adding these two series gives immediately our desired identity (\ref{log-Fourier}).

Apply now this identity \eqref{log-Fourier} with $x=\beta s$ and $\phi=\theta_i-\theta_j$ we get the action 
\begin{eqnarray}
S_X[\theta]
&=&
\frac{d}{2}\sum_{i,j}P_{ij,ji}
\left[
\text{const}
-2\sum_{n\ge1}\frac{e^{-n\beta s}}{n}\cos\!\big(n(\theta_i-\theta_j)\big)
\right]\nonumber\\
&=&
\text{const}
\;-\;
d\sum_{n\ge1}\frac{q^n}{n}\,
\sum_{i,j=1}^N P_{ij,ji}\,\cos\!\big(n(\theta_i-\theta_j)\big),
\qquad q\equiv e^{-\beta s}.
\label{SX-Fourier-step}
\end{eqnarray}
Next, we introduce the Fourier modes of the holonomy defined by the Polyakov moments
\begin{eqnarray}
u_n\equiv \frac{1}{N}\sum_{j=1}^N e^{in\theta_j},
\qquad
u_{-n}=u_n^\ast.
\end{eqnarray}
Then
\begin{eqnarray}
\sum_{i,j=1}^N e^{in(\theta_i-\theta_j)}
=
\Big(\sum_i e^{in\theta_i}\Big)\Big(\sum_j e^{-in\theta_j}\Big)
=
N^2\,u_n\,u_{-n}
=
N^2\,|u_n|^2,
\end{eqnarray}
and since the result is real we also have
\begin{eqnarray}
\sum_{i,j=1}^N \cos\!\big(n(\theta_i-\theta_j)\big)
=\sum_{i,j=1}^N\Re\Big(e^{in\theta_i}e^{-in\theta_j}\Big)=
N^2\,|u_n|^2.
\label{cos-sum}
\end{eqnarray}
We also include the adjoint projector factor:
\begin{eqnarray}
\sum_{i,j}P_{ij,ji}\cos\!\big(n(\theta_i-\theta_j)\big)
&=&
\sum_{i,j}\cos\!\big(n(\theta_i-\theta_j)\big)
-\frac{1}{N}\sum_i \cos(0),\qquad P_{ij,ji}=
1-\frac{1}{N}\delta_{ij}\nonumber\\
&=&
N^2|u_n|^2 - 1.
\label{proj-cos-sum}
\end{eqnarray}
Dropping the $\theta$--independent constants in \eqref{SX-Fourier-step} using \eqref{proj-cos-sum}, one obtains
\begin{eqnarray}
S_X[\theta]
=
-\,d\,N^2\sum_{n\ge1}\frac{q^n}{n}\,|u_n|^2,
\qquad
q=e^{-\beta s}.\label{contr1}
\end{eqnarray}
This term is attractive as it favors eigenvalue clumping.

\subsection{Vandermonde contribution}
After fixing to the static diagonal (Polyakov) gauge, the Haar measure of $SU(N)$
induces the Vandermonde determinant, which can be written as the effective action
\begin{eqnarray}
S_{\rm Vdm}[\theta]
&=&
-\sum_{i<j}\log\!\left[4\sin^2\!\Big(\frac{\theta_i-\theta_j}{2}\Big)\right].
\end{eqnarray}
This contribution is purely entropic and universal, as it arises solely from the
group measure and is present even in the absence of matter fields.

\medskip
\noindent
To analyze this term, we use the Fourier identity
\begin{equation}
-\log\!\left(2\sin\frac{\phi}{2}\right)
=
\sum_{n\ge1}\frac{\cos(n\phi)}{n}
+\text{const},
\qquad 0<\phi<2\pi,
\label{Fourier2}
\end{equation}
This identity follows from a standard Fourier expansion. For $|z|<1$ one has
\begin{equation}
-\log(1-z)=\sum_{n\ge1}\frac{z^n}{n}.
\end{equation}
Setting $z=re^{i\phi}$ with $0<r<1$ and taking the real part yields
\begin{equation}
-\Re\log(1-re^{i\phi})
=
\sum_{n\ge1}\frac{r^n}{n}\cos(n\phi).
\end{equation}
Since $\Re\log w=\log|w|$, this can be written as
\begin{equation}
-\log|1-re^{i\phi}|
=
-\frac12\log\!\big(1-2r\cos\phi+r^2\big)
=
\sum_{n\ge1}\frac{r^n}{n}\cos(n\phi),
\qquad 0<r<1.
\end{equation}
Finally, taking the Abel limit $r\to1^{-}$ gives, for $0<\phi<2\pi$, the desired Fourier identity~(\ref{Fourier2}).

Using now the Fourier identity (\ref{Fourier2}), one finds
\begin{eqnarray}
S_{\rm Vdm}[\theta]
&=&
\sum_{i<j}2\sum_{n\ge1}\frac{\cos\big(n(\theta_i-\theta_j)\big)}{n}+\text{const}.
\end{eqnarray}
Using
\[
\sum_{i<j}\cos\big(n(\theta_i-\theta_j)\big)
=\frac{N^2}{2}|u_n|^2-\frac{N}{2},
\]
we obtain, up to $\theta$–independent constants,
\begin{eqnarray}
S_{\rm Vdm}[\theta]
&=&
N^2\sum_{n\ge1}\frac{1}{n}\,|u_n|^2.\label{contr2}
\end{eqnarray}
\medskip
Adding both contributions (\ref{contr1}) and (\ref{contr2}), the leading large-$d$ holonomy effective action is
\begin{eqnarray}
S_{\rm hol}[\theta]
&=&
N^2\sum_{n\ge1}\frac{1-d\,q^n}{n}\,|u_n|^2
\;+\;\text{const}.
\end{eqnarray}

\end{document}